\documentclass[review]{elsarticle}

\usepackage[colorlinks,citecolor=blue,linktoc=all,linkcolor=cyan]{hyperref}
\usepackage{graphicx}

\usepackage[T1]{fontenc}
\usepackage{dsfont}               
\usepackage{mathrsfs}             
\usepackage{slashed}              
\usepackage{amsmath}
\usepackage{amssymb}
\usepackage{amsbsy}
\usepackage{amsfonts}
\usepackage{bm}
\usepackage{xcolor}

\newcommand{\bA}{{\bf A}}
\newcommand{\bb}{{\bf b}}
\newcommand{\bbe}{{\bf e}}
\newcommand{\bj}{{\bf j}}
\newcommand{\bk}{{\bf k}}
\newcommand{\bK}{{\bf K}}
\newcommand{\bL}{{\bf L}}
\newcommand{\bn}{{\bf n}}

\newcommand{\bp}{{\bf p}}
\newcommand{\br}{{\bf r}}
\newcommand{\bR}{{\bf R}}
\newcommand{\bs}{{\bf s}}
\newcommand{\bv}{{\bf v}}

\newcommand{\mmmatrix}[9]{ \left(\! \begin{array}{ccc}#1 & #2 & #3\\[1mm] #4 & #5 & #6\\[1mm] #7 & #8 & #9\\ \end{array}\!\right) }
\newcommand{\doublet}[2]{ \left( \begin{array}{c}#1 \\ #2 \end{array}\right) }
\newcommand{\triplet}[3]{ \left( \begin{array}{c}#1 \\ #2 \\ #3\end{array}\right) }
\newcommand{\lr}[1]{ \langle #1 \rangle}
\newcommand{\bra}[1]{\langle #1|}
\newcommand{\ket}[1]{|#1 \rangle}
\renewcommand{\Re}{\mbox{Re}}
\renewcommand{\Im}{\mbox{Im}}

\def\lsim{\mathrel{\rlap{\lower4pt\hbox{\hskip1pt$\sim$}}
		\raise1pt\hbox{$<$}}}         
\def\gsim{\mathrel{\rlap{\lower4pt\hbox{\hskip1pt$\sim$}}
		\raise1pt\hbox{$>$}}}         

\numberwithin{equation}{section}
\numberwithin{table}{section}
\numberwithin{figure}{section}

\journal{Progress in Particle and Nuclear Physics}

\usepackage{titlesec}
\usepackage{sectsty}
\titleformat{\section}{\normalfont\Large\bfseries}{\thesection}{1em}{}
\titleformat{\subsection}{\normalfont\large\bfseries}{\thesubsection}{1em}{}
\titleformat{\subsubsection}{\normalfont\normalsize\bfseries}{\thesubsubsection}{1em}{}

\begin{document}
	
	\begin{frontmatter}
		
		\title{Promises and challenges of high-energy vortex states collisions}

		\author[myaddress]{Igor P. Ivanov}
		\ead{ivanov@mail.sysu.edu.cn}
		\address[myaddress]{School of Physics and Astronomy, Sun Yat-sen University, 519082 Zhuhai, China}
		
		\begin{abstract}
			Vortex states of photons, electrons, and other particles are non--plane-wave solutions 
			of the corresponding wave equation with helicoidal wave fronts. 
			These states possess an intrinsic orbital angular momentum with respect to the average propagation direction,
			which represents a new degree of freedom, previously unexplored in particle or nuclear collisions.
			Vortex states of photons, electrons, neutrons, and neutral atoms have been experimentally produced, albeit at low energies, 
			and are being intensively explored.
			Anticipating future experimental progress, one can ask what additional insights on nuclei and particles 
			one can gain once collisions of high-energy vortex states become possible.
			This review describes the present-day landscape of physics opportunities, experimental progress
			and suggestions relevant to vortex states in high energy collisions.
			The aim is to familiarize the community with this emergent cross-disciplinary topic
			and to provide a sufficiently complete literature coverage, highlighting some results and calculational techniques.
		\end{abstract}
		
		
	\end{frontmatter}
	
	\newpage
	
	\thispagestyle{empty}
	\tableofcontents
	

	\newpage
	\section{Vortex states in nutshell}\label{section-introduction}

\begin{figure}[!h]
	\centering
	\includegraphics[width=\textwidth]{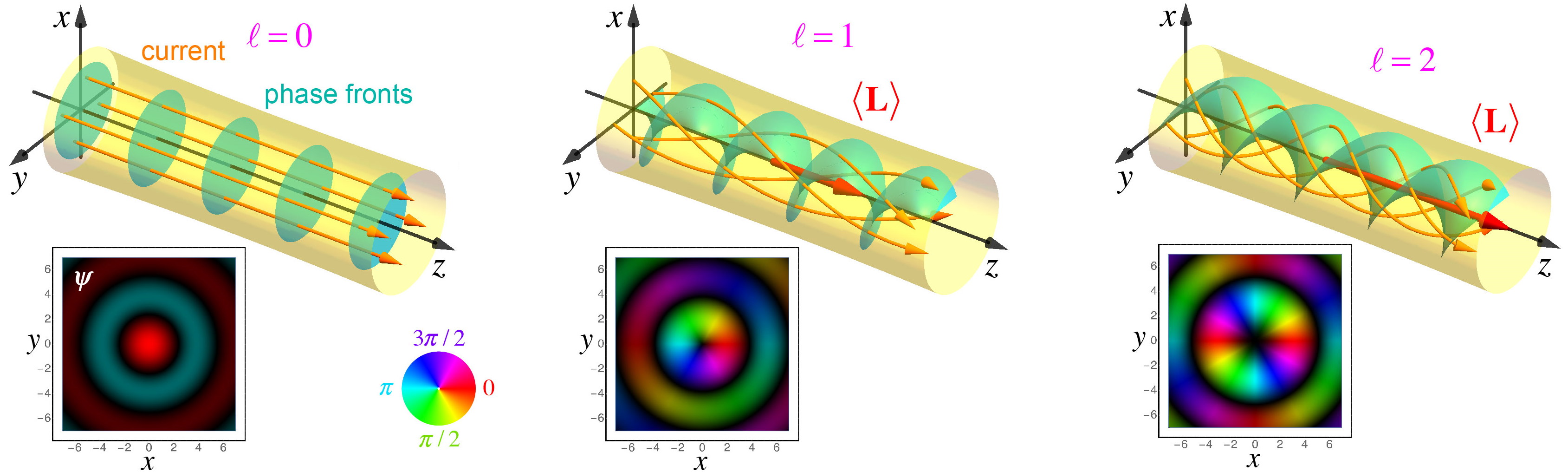}
	\caption{Vortex beams are cylindrical solutions of the wave equation. 
		They propagate along axis $z$ and carry an intrinsic $z$-projection of the orbital angular
		momentum (OAM). The 3D illustrations show the phase fronts and probability current
		streamlines for beams with different topological vortex charges $\ell$. 
		The 2D plots show the wave function distribution in the transverse plane for the Bessel vortex states,
		with the phase encoded in color.
		Reproduced from \cite{Bliokh:2017uvr} with permission.}\label{fig-vortex-1}
\end{figure}

This review deals with a cross-disciplinary topic which, in the past decade, grew into a remarkably vibrant 
field of research in optics, electromagnetism, microscopy, beam physics, atomic and condensed matter physics,
regularly making headlines in physics news and deserving cover pages in {\em Nature} and {\em Science}.
It is much less known within the particle physics and nuclear physics communities,
as the main experimental achievements reported in this field are limited to low energies. 
However the field is ripe with ideas and techniques which, to say the least,
already stimulate outside-of-the-box thinking in high energy physics.
Numerous proposals have been published of how to bring vortex states of photons and massive particles
to higher energies, in the MeV or even GeV domains.
When demonstrated experimentally, they will open up a wealth of new opportunities
in nuclear physics, hadron structure, quantum electrodynamics (QED) phenomena
in regimes which were unthinkable before.

The past few years witnessed a burst of activity which already led to about hundred publications
on possible applications of vortex states in particle and nuclear physics. 
This rapidly growing literature deserves a dedicated review, which will complement
the existing books \cite{Torres-applications,andrews2012angular} 
and reviews on vortex photons \cite{Allen:1999,Molina-Terriza:2007,Paggett:2017,photonics-review-2017,Knyazev-Serbo:2018,babiker2018atoms}
and vortex electrons \cite{harris2015structured,Bliokh:2017uvr,Lloyd:2017,larocque2018twisted}.
Although this review introduces an inevitable personal bias in description of the results,
the author hopes that the literature coverage is sufficiently complete and representative.

It is appropriate to begin this review with quick answers to some 
immediate questions which are frequently asked within the particle physics community upon first learning about vortex states.
\begin{itemize}
	\item What are vortex states?
	
	A vortex, or twisted, state of a field is a non--plane-wave solution of the corresponding wave equation 
	which, on average, propagates along certain direction (axis $z$) and, at the same time, possesses 
	a nonzero $z$-projection of the intrinsic orbital angular momentum (OAM).
	This is not the relative OAM of constituents in a composite particle, such as an atom or a proton.
	Nor is it the OAM of a particle moving in external fields, such as given by Landau levels of the electron in 
	homogeneous magnetic field.
	Here, we talk about the OAM of the center of mass wave function of a free propagating particle.
	
	For a scalar field, this feature comes from the azimuthally-dependent phase factor of the wave function: 
	$\psi(\br)\propto \exp(i\ell\varphi_r)$, where $\varphi_r$ is the azimuthal angle in the transverse plane. 
	Axis $z$, corresponding to the transverse distance $\rho=0$, represents a phase singularity line
	of zero intensity. Fig.~\ref{fig-vortex-1} should help visualize this state with the aid of wave fronts, local current streamlines,
	the intensity profiles, and phase evolution across the transverse plane.
	For a vector or spinor field, the description requires care and will be elaborated below, 
	but a vortex state of a photon or electron still possesses a phase singularity line and an azimuthally-dependent phase factor.
	
	One must clearly distinguish OAM-dominated beams and vortex states.
	An OAM-dominated beam can be described semiclassically as a beam of pointlike particles,
	each of them moving at an angle and possessing a classical OAM with respect to the overall beam axis.
	This is literally a cloud of particle, which propagates and rotates as a whole. 
	The mechanical OAM of this beam can be large and is a collective classical effect.
	
	In contrast, the OAM associated with a vortex state is not a collective effect. 
	After quantization, each particle (a scalar, a photon, an electron and so on)
	possesses the same azimuthally varying phase factor and, therefore, carries a nonzero OAM $z$-projection. For a scalar particle,
	one gets $L_z = \hbar \ell$, which can be arbitrarily large. Electrons and photons, strictly speaking, do not possess a definite OAM, 
	but the values of $L_z$ involved in the definition of a vortex electron and photon can be large, too.
	This new degree of freedom, which bypasses the limitations of spin state space, is already routinely used in optics, electron microscopy
	and atomic physics. If imposed on initial particles in high-energy collisions, 
	it will, hopefully, lead to new insights and applications in nuclear and particle physics.

	\item Vortex states are just peculiar superpositions of plane waves. Can they offer any new insights in particle physics?
	
	Yes, they can, and the keyword is coherence. Vortex states are coherent superposition of plane waves, 
	and their wave fronts bear a non-trivial, topologically protected phase structure.
	Scattering of vortex states in a fixed potential leads to different angular distribution and polarization preferences of the process
	if compared with plane wave scattering. Collisions of vortex states with atoms, nuclei, and hadrons can modify selection rules
	in a position-dependent way. 
	Free space collision of two vortex states leads to controlled interference between plane wave amplitudes with difference initial momenta
	leading to the same final state, which has not been studied before. The resulting cross sections
	offer access to new observables which are absent in plane wave collisions or in collisions of broad Gaussian wave packets.
	
	A vortex state can have an azimuthally symmetric intensity profile, and yet it explicitly breaks the left-right reflection symmetry.
	Indeed, consider the transverse plane $\br_\perp = (x, y)$ at some position along axis $z$ and write the wave function $\psi(\br_\perp)$,
	centered at the origin $\lr{\br_\perp} = 0$. 
	Then, $|\psi(\br_\perp)|^2$ is invariant under the reflection $(x, y) \mapsto (x, -y)$, but the wave function itself is not.
	A vortex state can also be put in a spin-orbital coupled states which exhibits a remarkable, spatially-dependent 
	polarization impossible for plane waves, which offers additional opportunities for spin or parity related observables.
	
	\item Do vortex states offer new insights in the spin content of the proton?
	
	They may, but no theoretical work has yet thoroughly studied the implications.
	The relevance of vortex state collisions for the proton spin puzzle \cite{Aidala:2012mv,Leader:2013jra} is self-evident; 
	it was mentioned in the early paper on vortex particle scattering \cite{Ivanov:2011kk} and
	reiterated afterwards \cite{larocque2018twisting,vanacore2019ultrafast,madan2020quantum}.
	It is somewhat surprising that this interest has not yet materialized in a systematic treatment
	of deep inelastic scattering in the vortex regime. One of the objectives of this review is 
	to stimulate efforts in this direction.
	
	\item How are processes with vortex states calculated?
	
	Since vortex states are superpositions of plane waves,
	the scattering matrix element for vortex state collision can be written as a superposition
	of plane wave scattering matrix elements, weighted with momentum space wave functions,
	or via the Wigner function formalism. Thus, there is no need to quantize the fields in the basis of vortex states,
	and no new Feynman diagrams are involved. 
	What makes the work laborious is performing integrations in generic kinematics
	keeping the dependence of the matrix element and the energy and momentum delta functions.
	
	\item Can vortex state collisions be studied at present day colliders?
	
	In principle, they can, provided one generates vortex states of initial particles 
	and injects them in an existing linear accelerator or a collider without destroying their vortex state.
	Preparation of vortex states requires delicate engineering of their wave fronts. 
	Evolution of these quantum states in the beam optics systems of linear accelerators or in storage rings
	is not well known, as one needs to go beyond the traditional semiclassical treatment,
	but it can be studied numerically and experimentally. 
	This work requires dedicated simulation and experimental efforts to understand the limits of the existing instrumentation
	and to set the goals for realizing vortex state collisions.
	Stimulating these efforts is another objective of this review.
	
	\item What has been done so far in experiment?
	
	Vortex states of optical photons were produced back in 1990's \cite{Allen:1992zz} and have already found numerous applications 
	\cite{Torres-applications,andrews2012angular,Allen:1999,Molina-Terriza:2007,Paggett:2017,Knyazev-Serbo:2018,babiker2018atoms}.
	Vortex states of X rays were also experimentally demonstrated \cite{Bahrdt:2013eoa,Nature-Phot-2019,photonics-review-2017}.
	Many proposals exist on how to generate hard vortex photons with MeV energies and beyond, 
	but they have not yet been experimentally realized.
	
	Following the suggestion of \cite{Bliokh:2007ec}, moderately relativistic vortex electrons with the kinetic energy up to 300 keV
	were experimentally demonstrated in 2010--2011 \cite{Uchida:2010,Verbeeck:2010,McMorran:2011},
	with the focal spot size down to 1 angstrom \cite{Verbeeck:2011-atomic}.
	They were produced in commercially available electron microscopes and cannot be easily scaled to achieve higher energies.
	There are proposals to put high-energy electrons in vortex states or to accelerate low-energy vortex electrons
	to higher energies, but none of them has been tried in experiment.
	
	Finally, cold neutrons \cite{clark2015controlling,sarenac2018methods,sarenac2019generation,Sarenac:2022} 
	and, very recently, slow atoms \cite{luski2021vortex} were put in vortex states.
	Vortex protons or ions have not yet been reported, but ideas on their generation exist.
	
	In short, although high-energy particles in vortex states have not yet been produced, 
	there is no fundamental show-stopper to twisting any sort of particles. 
	This milestone achievement will require novel experimental set-ups and dedicated efforts.
	Nevertheless, certain collision experiments with twisted photons, electrons, and neutrons 
	can be performed with the existing technology.
\end{itemize}
The author hopes that these answers will convince the reader that the topic is promising and worth investigating further. 

The outline of the review is the following. 
Section~\ref{section-describing} contains technical details on how vortex states are described.
Various regimes of vortex particle collisions with their characteristic features are
presented in Section~\ref{section-scattering-general} followed in Section~\ref{section-processes} by
a list of concrete processes which have been studied in literature.
An overview of the experimental situation and prospects are given in Section~\ref{section-experimental}.
The current status of the problem and an outlook presented in Section~\ref{section-summary} summarize the discussion.

Throughout the paper, we usually use the natural units $\hbar = c = 1$, although in certain cases 
we restore these fundamental constants for clarity.
Three-dimensional vectors will be denoted by bold symbols, 
and the transverse momenta will carry the subscript $\perp$.

	\newpage
	\section{Describing vortex states}\label{section-describing}

\subsection{Vortex state of a scalar field}

In quantum field theory, the wave function of a plane wave one-particle state of a scalar field 
with three-momentum $\bk$ and energy $E=\sqrt{M^2 + \bk^2}$ can be written as
\begin{equation}
\Psi_{\rm PW}(\br,t) = \frac{1}{\sqrt{2E}}\psi_{\rm PW}(\br,t) = \frac{1}{\sqrt{2E}}N_{\rm PW} e^{-iEt + i\bk\cdot \br}\,.\label{PW-def}
\end{equation}
We assume that the state is normalized by one particle per large but finite volume $V$:
$\int_V d^3 r\, j^0 = \int_V d^3r\, |\psi_{\rm PW}|^2 = 1$, where $j^\mu = i \Psi^* \partial_\mu \Psi + c.c.$ is the conserved current.
This leads to the plane-wave normalization factor $N_{\rm PW} = 1/\sqrt{V}$. 
It should be mentioned that there exist well-known limitations of the one-particle wave function 
for extremely localized states as well as other related subtleties to the proper normalization of states, \cite{Karlovets:2018iww}.
We do not discuss them here, as they will not affect the main physics points we cover in this review,
although they may be quantitatively important in processes with non--plane-wave states.
Also, the cross sections calculations will not depend on the normalization coefficients
if one uses the same normalization for scattering matrix elements, the flux, and the phase space density of the final states.

The simplest type of a vortex state is the so-called Bessel state \cite{Jentschura:2010ap,Jentschura:2011ih}.
This is a monochromatic solution of the wave equation constructed from plane waves with the same longitudinal momentum $k_z$,
the same modulus of the transverse momentum $|\bk_\perp| = \varkappa$ (sometimes called the conical transverse momentum), 
and, as a result, the same energy $E$.
Its wave function in the cylindrical coordinates $(\rho,\varphi_r,z)$ can be split into the trivial time and $z$-dependent part
and a non-trivial transverse wave function:
\begin{equation}
\Psi_{k_z,\varkappa,\ell}(\br, t) = \frac{1}{\sqrt{2E}} N_{\rm Bes} e^{-iE t + i k_z z} \psi_{\varkappa, \ell}(\br_\perp)\,.
\end{equation}
where $N_{\rm Bes}$ is the new normalization coefficient. 
Following \cite{Jentschura:2010ap,Jentschura:2011ih}, we write the transverse part of the wave function as
\begin{equation}
\psi_{\varkappa, \ell}(\br_\perp) = \int \frac{d^2\bk_\perp}{(2\pi)^2} a_{\varkappa, \ell}(\bk_\perp) e^{i \bk_\perp\!\cdot \br_\perp}\,,
\quad \mbox{where} \quad a_{\varkappa, \ell}(\bk_\perp) = (-i)^\ell e^{i\ell \varphi_k}\sqrt{\frac{2\pi}{\varkappa}}\delta(|\bk_\perp|-\varkappa)\,,
\label{twisted-scalar-1}
\end{equation}
Here $\ell$ is an integer number defining the phase with which every plane wave enters this superposition.
Performing the integration, we get
\begin{equation}
\psi_{\varkappa, \ell}(\br_\perp) = e^{i\ell\varphi_r}\sqrt{\frac{\varkappa}{2\pi}}\, J_{\ell}(\varkappa \rho)\,.\label{twisted-scalar-2}
\end{equation}
Fig.~\ref{fig-vortex-1} offers a visualization of the Bessel vortex state in coordinate space.

The all-important phase factor $\exp(i\ell\varphi_r)$ highlights the presence of a phase vortex around the phase singularity line.
The ``order'' of the phase vortex is quantified with the topological charge. 
To define it, take the phase of the wave function $\arg\psi(\br)$,
compute its gradient ${\bm \nabla}[\arg\psi(\br)]$, integrate it along a closed path around the singularity line, and divide by $2\pi$.
The result, $\oint d\br \cdot {\bm \nabla}[\arg\psi(\br)]/2\pi$, is an integer number called the topological charge.
This integral does not depend on the path (provided the path encircles the singularity line exactly once)
and is equal to $\ell$ for the wave function in Eq.~\eqref{twisted-scalar-2}.
Moreover, even if one ``disturbs'' the vortex state by adding an smooth wave function without a phase singularity,
the topological charge remains the same: it is said that the phase vortex is topologically protected.

Since the vortex state is an eigenstate of the OAM $z$-projection operator $\hat{L}_z = -i\hbar \partial/\partial\varphi_r$,
it carries the OAM $z$-projection $L_z = \hbar \ell$.
Due to the phase factor $\exp(i k_z z + i\ell \varphi_r)$, the wave fronts are represented by helicoidal surfaces.
The streamlines orthogonal to these surfaces show the local current density ${\bf j}(\br)$.
The concentric rings of intensity, defined by the Bessel function $J_{\ell}(\varkappa \rho)$ and shown in Fig.~\ref{fig-vortex-1},
are also a characteristic feature of the vortex state. For $\ell\not = 0$, $J_{\ell}(0) = 0$, 
so that the intensity is exactly zero on the axis;
near the axis, $\psi_{\varkappa, \ell} \propto (\varkappa \rho)^\ell$, 
and the first maximum of the intensity is located at $\rho \approx |\ell|/\varkappa$.

\begin{figure}[!h]
	\centering
	\includegraphics[width=0.4\textwidth]{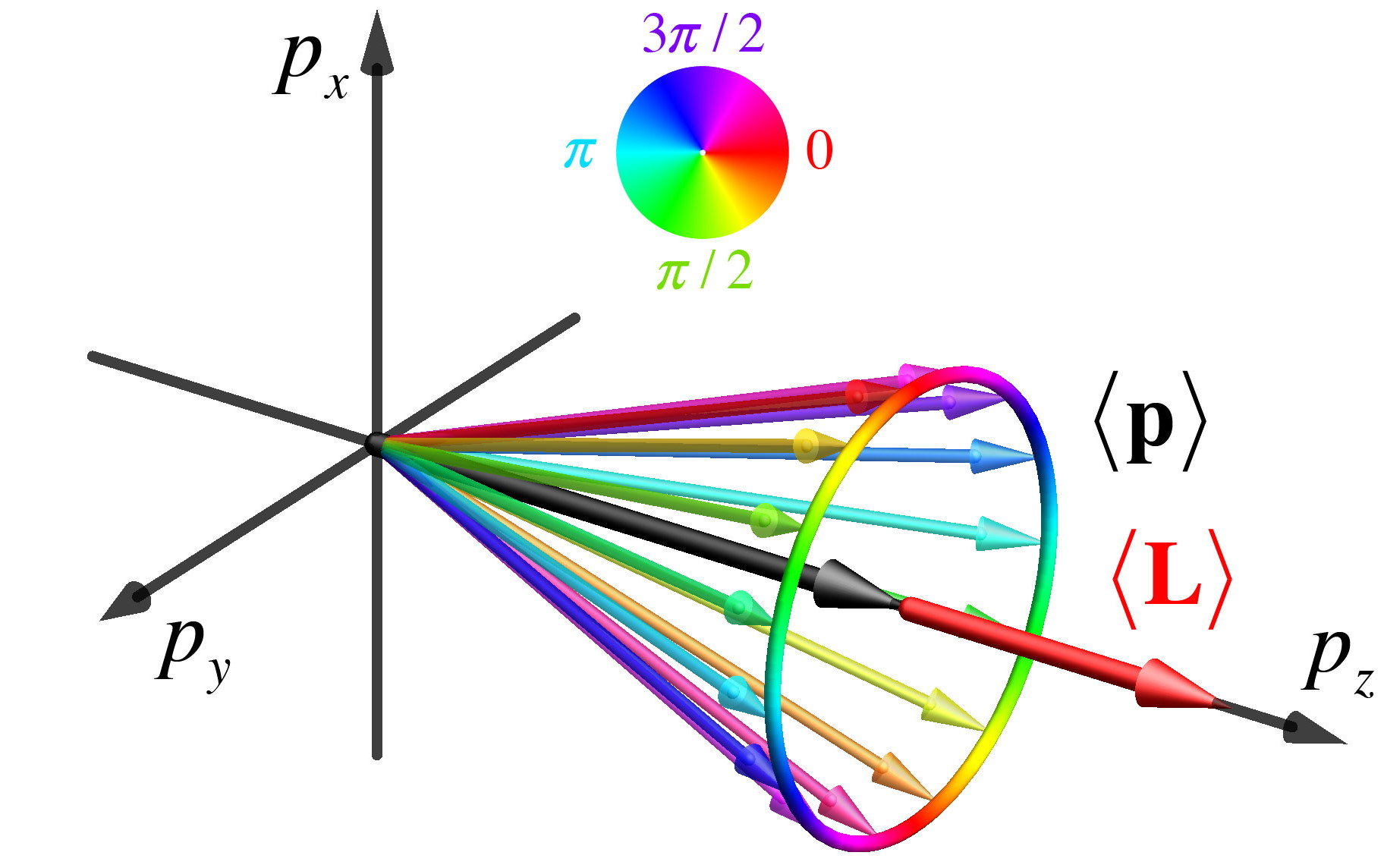}
	\caption{Bessel vortex state in momentum space is represented by an infinitely thin circle with radius $\varkappa$
		located in the plane of fixed $p_z$. Different plane wave components bear the phase factor $\exp(i\ell\varphi_p)$,
		with the phase encoded in color. The average OAM is parallel to the average momentum.
		Reproduced from \cite{Bliokh:2017uvr} with permission.}\label{fig-vortex-2}
\end{figure}

In momentum space, the Bessel state is represented by an infinitely thin circle with radius $\varkappa$
located in the transverse plane at $\lr{\bk} = (0,0, k_z)$, see Fig.~\ref{fig-vortex-2}.
The momenta of the plane wave components of the Bessel state cover the surface of a cone
with the opening angle $\theta$ defined as $\tan\theta = \varkappa/|k_z|$.
The conical transverse momentum $\varkappa$ shows the transverse momentum content 
hidden inside the Bessel vortex state.
The paraxial regime, to which we will often refer below, corresponds to $\theta \ll 1$.

Although the height of the intensity rings given by $J^2_\ell(\varkappa \rho)$ decreases as $\rho$ grows, 
the exact Bessel state cannot be viewed as localized near the axis.
Without a proper radial regularization, the Bessel state is non-normalizable in the transverse plane.
Using the same normalization condition, one particle per large cylindrical volume of length $d_z$ and radius $R$, we get
\begin{equation}
1 = N_{\rm Bes}^2 \int d^3\br\, |\psi_{\varkappa, \ell}(\br_\perp)|^2 = N_{\rm Bes}^2\, d_z\varkappa \, \int_0^R \rho d\rho\, J^2_\ell(\varkappa \rho)
= N_{\rm Bes}^2\, \frac{d_z R}{\pi}\,.
\end{equation}
Thus, the Bessel state normalization coefficient is $N_{\rm Bes} = \sqrt{\pi/(d_zR)}$. 
For the extreme case of $|\ell|/(\varkappa R)$ close to 1, the normalization condition is modified, see \cite{Karlovets:2012eu}.

Just like plane waves, the Bessel states form an orthogonal basis in the transverse plane:
\begin{equation}
\int d^2\br_\perp \,\psi^*_{\varkappa, \ell}(\br_\perp) \psi_{\varkappa', \ell'}(\br_\perp)
= \delta_{\ell\,\ell'}\, \delta(\varkappa-\varkappa')\,.
\end{equation}
By inserting \eqref{twisted-scalar-1} in this condition, one can deduce the regularizing prescription for the square of the 
radial momentum delta function \cite{Jentschura:2010ap,Jentschura:2011ih}:
\begin{equation}
[\delta(|\bk_\perp|-\varkappa)]^2 = \delta(|\bk_\perp|-\varkappa) \cdot \delta(0) \to \delta(|\bk_\perp|-\varkappa) \, \frac{R}{\pi}\,.
\label{radial-reg}
\end{equation}
The fact that Bessel states are non-normalizable in the transverse plane may lead to singularities and other artifacts 
in cross section calculations. A possible way to render the expressions regular is to introduce smearing in 
$\varkappa$ by the amount $\sigma_\varkappa$ \cite{Ivanov:2011bv}. This can be done with the aid of a properly normalized function $f(\varkappa)$ 
which is peaked at $\varkappa = \varkappa_0$ and quickly decreases at $|\varkappa-\varkappa_0| > \sigma_\varkappa$.
One then defines the smeared vortex state 
\begin{equation}
\tilde \Psi_{\varkappa_0, \sigma_\varkappa; \ell}(\br, t) = \int d\varkappa \, f(\varkappa)\, \Psi_{k_z,\varkappa, \ell}(\br, t)\,,
\label{smearing}
\end{equation}
where $k_z$ can either be kept fixed or vary together with $\varkappa$ if one insists on a fixed $E = \sqrt{M^2 + k_z^2 + \varkappa^2}$.
In this way, one effectively regularizes the vortex state in the transverse plane;
the wave function now exhibits only a few (of the order of $\varkappa_0/\sigma_\varkappa$) intensity rings and then quickly decreases.
A drawback of this approach is that results are sensitive to the choice of the function $f(\varkappa)$.

As for the final state particles, one has freedom to choose a basis for their description.
The usual choice for outgoing particles is the plane wave basis, 
which conveniently implements momentum conservation.
Whenever one discusses the final state angular distribution, one automatically assumes the plane wave basis.
However, one could also describe one or several final states particles in the basis of Bessel states.
When doing so, one first must decide with respect to which axis one defines the final particle vortex states.
A natural choice is the same axis $z$ as for the initial state, together with the assumption that they reside 
in the same large normalization volume of the initial Bessel state.
Then, one just needs to replace the plane wave number of states with its Bessel state counterpart according to the following rule
\cite{Jentschura:2011ih,Ivanov:2011kk}:
\begin{equation}
dn = \pi R^2 d_z\,\frac{d^3\bk'}{(2\pi)^3} \to \frac{R d\varkappa'}{\pi}\, \frac{d_z dk'_z}{2\pi}\, \Delta \ell'\,.
\end{equation} 
Here, $\Delta \ell'$ is a reminder that all final state OAM values must be counted.
If one chooses a different axis $z'$ with respect to which the final Bessel states are to be defined,
the expression for the number of states remains the same but much care must be taken
in order to account for two mismatching axes.
In this case, as described in \cite{Ivanov:2011bv,PhysRevA.84.065802}, the exact Bessel states lead to non-physical artifacts,
which can be eliminated once smearing \eqref{smearing} is taken into account.

Although the exact Bessel states are convenient for scattering amplitude calculations, they lead to pathologies 
when integrating the cross section of a two Bessel states collision over the final phase space, see Section~\ref{subsection-double-vortex}.
Regularization of these artifacts with the aid of smearing procedure such as \eqref{smearing} 
introduces certain arbitrariness. A more physically appealing choice is to use 
other vortex states such as the Laguerre-Gaussian (LG) modes,
which contain the all-important $\exp(i\ell\varphi_r)$ factors and are normalized in the transverse plane.
One can also impose the Gaussian profile on the longitudinal distribution, 
which renders the LG vortex state well localized in three dimensions.
Such a state cannot be monochromatic and, therefore, evolves in time, which is also a realistic feature.

A LG wave packet takes the simplest form in the paraxial approximation,
when the typical transverse momenta inside the wave packet are much smaller than the average longitudinal momentum and the mass.
A description of the LG wave packets in a form convenient for high-energy collisions of massive particles
was developed in \cite{Karlovets:2018iww}. 
Adapting the Lorentz-covariant formalism developed previously for neutrino wave packets \cite{Naumov:2010um},
Ref.~\cite{Karlovets:2018iww} begins with a relativistic Gaussian wave packet in momentum space:
\begin{equation}
\psi(k) = 2\sqrt{2}\pi\sigma \frac{e^{-M^2/\sigma^2}}{\sqrt{K_1(2M^2/\sigma^2)}}\exp[(k-\bar k)^2/\sigma^2]\,.
\label{GaussWP-1}
\end{equation}
Here, the four-vector $k$ satisfies the on-mass-shell condition $k^2 = M^2$, which implies $E = \sqrt{M^2 + \bk^2}$,
$\bar k$ is the mean four-momentum, which also satisfies $\bar k^2 = M^2$,
the parameter $1/\sigma$ shows the spatial extent of the wave packet,
and $K_1$ is the modified Bessel function.
This wave packet satisfied the Klein-Fock-Gordon equation and is normalized with the Lorentz-invariant normalization measure
\begin{equation}
\int \frac{d^3 \bk}{(2\pi)^3 2E} |\psi(\bk)|^2 = 1\,.\label{norm-lorentz}
\end{equation}
Its expression in coordinate space and various properties were also presented in \cite{Karlovets:2018iww}.
The exact relativistic Gaussian wave packet can be simplified in the paraxial approximation,
which is expected to apply to most situations and which, as argued in \cite{Karlovets:2018iww}, 
should be defined as $1/\sigma \gg \lambda_c$, the Compton wavelength of the particle with mass $M$.
In this approximation, we again choose axis $z$ along the mean three-momentum $\bk$, so that $\bar \bk = (0,0,\bar k_z)$,
and represent the momentum space wave function as
\begin{eqnarray}
\psi_G(\bk)=\frac{(4\pi)^{3/4}}{\sigma_{\perp}\sqrt{\sigma_{z}}} \sqrt{2E}\,   
\exp\left[-\frac{\bk_{\perp}^2}{2\sigma_{\perp}^2}-\frac{(k_{z}-\bar k_{z})^2}{2\sigma_{z}^2}\right]\,,
\label{GaussWP-2}
\end{eqnarray}
which is again normalized according to \eqref{norm-lorentz}.
Here, we $1/\sigma_{\perp}$ and $1/\sigma_{z}$ show the transverse and longitudinal spatial extents of the wave function.
For the wave packet which is spherically symmetric in its rest frame as implied by \eqref{GaussWP-1}, 
one finds $\sigma_{z} = \bar\gamma\sigma_{\perp}$, where 
$\bar\gamma = \bar E/M$, $\bar E = \sqrt{\bar \bk^2 + M^2}$.
This result agrees with the intuition of the relativistic contraction of longitudinal distances.
However one could also imagine the situation where the longitudinal and transverse extents 
are independent parameters determined by different elements of an experimental setting.
In particular, by making $\sigma_z$ arbitrarily small, one can obtain a close approximation to the 
Gaussian beam instead of a wave packet.

Next, Ref.~\cite{Karlovets:2018iww} presents the LG wave packets, which were first constructed in the Lorentz-covariant form, 
then in the paraxial approximation, and finally nonparaxial corrections were investigated.
In particular, in the paraxial approximation, the principal mode LG wave packet can be written as 
\begin{equation}
\psi_{LG}(\bk) = \psi_G(\bk)\, \left(\frac{k_\perp}{\sigma_\perp}\right)^{|\ell|} \frac{1}{\sqrt{\ell!}} e^{i\ell \varphi_k}\,,
\label{LGWP-1}
\end{equation}
which is also normalized according to \eqref{norm-lorentz}. Notice that one recovers the Gaussian wave packet by setting $\ell = 0$.

The coordinate space wave functions for the Gaussian and LG wave packets can also be explicitly found.
As expected, they exhibit a characteristic time evolution. At early times the wave packet is broad, 
but as it moves along axis $z$ with the average velocity $u=\bar k_z/\bar E$
it shrinks and at some instant reaches the minimal dimensions of $1/\sigma_{\perp}$ and $1/\sigma_{z}$.
After this momentary focusing, it broaden again. 
Notice that, when considering collision of two such wave packets, 
one must take into account not only their shapes but also the time difference $\Delta\tau$ between the instants of their focusing 
and the distance $\bb$ between the focal points.
All these offset parameters can be taken into account with the additional factor $\exp(i b^\mu k_\mu)$ in one of the wave functions,
where the four-vector $b^\mu = (\Delta\tau, \bb)$.

The full set of LG solutions contains the second parameter, $n$, so that the radial wave function exhibits $n+1$ radial
intensity maxima. The expression \eqref{LGWP-1} represents only the principal LG modes with $n=0$.
The paraxial expressions for the general LG wave packets of relativistic massive particles were also obtained in \cite{Karlovets:2018iww},
and their properties were studied.  
Description of the Gaussian and LG vortex wave packets was further developed in \cite{Karlovets:2018fof,Karlovets:2020odl}
within the relativistic Wigner function representation. 
For a pedagogical discussion of the Wigner functions applied to particle collisions, see \cite{Karlovets:2016jrd}.
Wigner functions for the Gaussian and LG vortex wave packets were given in the paraxial approximation 
and then used to derive a universal expression for generic $2\to 2$ cross sections in collisions
of two Gaussian wave packets or a LG state with a Gaussian state, which we will review in Section~\ref{subsection-Wigner-formalism}.

\subsection{A qualitative discussion}\label{section-qualitative}

Before proceeding further, it is instructive to discuss qualitative features of vortex states
which may at first appear confusing.

A convenient property of a plane wave that it remains a plane wave in any inertial reference frame,
albeit with transformed energy and momentum. Bessel vortex states do not follow this pattern.
To construct a vortex state, one first needs to fix a reference frame and then to select an axis  
which will be the phase singularity axis. For the Bessel vortex state, a longitudinal boost 
will keep it a monochromatic Bessel state with a the same $\varkappa$ and $\ell$ but different $k_z$ and $E$.
For a Laguerre-Gaussian state, a longitudinal boost makes the vortex beam non-monochromatic,
as the focal point will itself be moving; 
however, for paraxial beams, this effect is minor. 
Performing a transverse boost changes the space-time evolution significantly even for Bessel vortex states. 
Still, the momentum space wave function retains the ring structure with a non-trivial phase profile.
Such states were dubbed spatiotemporal vortices \cite{sukhorukov2005spatio,bliokh2012spatiotemporal}; 
they will be briefly discussed in Section~\ref{section-beyond-simple}.
Thus, when dealing with vortex states and especially with their collision, 
we unavoidably break the Lorentz covariant description of the initial state.

The dispersion relation $E^2 = \varkappa^2 + k_z^2 + M^2$ of course holds 
for every individual plane wave component of a vortex state.
However, the relation between the energy and the average momentum $\bar \bk = \lr{\bk} = (0,0,k_z)$ is modified.
For the monochromatic Bessel state with energy $E$, we obtain 
$E^2 \not = \bar\bk^2 + M^2$. One can discuss this modification in terms of dynamical correction 
to the mass of the particle \cite{Karlovets:2018iww,Silenko:2019okz}.

Next, focusing on the orbital angular momentum,
one must distinguish the extrinsic and intrinsic OAM. Consider a freely propagating localized wave packet
with an average momentum $\bar \bk$. In a chosen coordinate frame, the motion of its centroid is described
with $\lr{\br}$. The two vectors can be used to construct the quasiclassical OAM $\bL_{\rm ext} = \lr{\br}\times \bar \bk$.
This is the extrinsic OAM, and it can be set to zero by a shift of the origin of the coordinate frame.
By construction, $\bL_{\rm ext} \perp \bar \bk$.

The OAM associated with the vortex state is intrinsic. It is generated by the phase vortex and 
should be defined with respect to the built-in phase singularity axis, 
which is not related to the coordinate frame choice. For the scalar field, by construction, it is parallel to the average momentum:
$\lr{\bL}_{\rm int}\, \|\, \bar \bk$, see Fig.~\ref{fig-vortex-2}. 
For an optical field, the full picture is more involved and allows for several forms of the OAM \cite{Bliokh:2015doa}.

A parallel shift of the reference axis does not shift the average intrinsic OAM of a vortex state 
but it broadens its OAM spectrum.
Indeed, select, in addition to axis $z$, another axis $z'$, which is parallel to $z$ but shifted in the transverse plane 
by the impact parameter ${\bf a}_\perp = (a\cos\varphi_a, a\sin\varphi_a)$.
Consider a scalar Bessel state with parameters $\varkappa$ and $\ell$ defined with respect to $z'$.
Then it can be expanded in the basis of Bessel state defined with respect to axis $z$ as (see e.g. Appendix C of \cite{Ivanov:2016oue})
\begin{equation}
\psi^{({\bf a}_\perp)}_{\varkappa, \ell}(\br_\perp) = 
\psi_{\varkappa, \ell}(\br_\perp-{\bf a}_\perp) = \sum_{\ell'=-\infty}^{+\infty} e^{i(\ell'-\ell)\varphi_a}J_{\ell'-\ell}(\varkappa a)
\, \psi_{\varkappa, \ell'}(\br_\perp)\,.\label{shifted}
\end{equation}
The value of $\varkappa$ is preserved, while the values of $\ell'$ differ from $\ell$ by about $\varkappa a$.
If the shift parameter $a$ is not too large, this OAM range stays narrow.
However, if $z$ and $z'$ are separated by a macroscopic distance, this distribution will be extremely broad.

One can also define yet another axis $z''$, which is tilted by angle $\alpha$ with respect to the axis $z$,
and define a Bessel state around $z''$. The expansion of the new Bessel state via the old basis of Bessel state
around $z$ will now involve an infinite sum of all $\ell'$ \cite{Ivanov:2011bv}.
This is hardly surprising; even a plane wave at oblique incidence can be expanded in the Bessel state basis as \cite{Landau3}
\begin{equation}
e^{i\bk_\perp\!\cdot\br_\perp} = \sqrt{\frac{2\pi}{\varkappa}}\sum_{\ell=-\infty}^{\infty} i^\ell e^{-i\ell\varphi_k} \psi_{\varkappa = |\bk_\perp|, \ell}(\br_\perp)\,.
\label{PW-expansion}
\end{equation}
This extreme broadening of the OAM spectrum from a single $\ell$ to an infinitely broad $\ell'$ range
is an artifact of the Bessel states and can be regularized with smeared-Bessel or with LG states \cite{Ivanov:2011bv}.
However the important lesson is that if one tries to reveal the intrinsic OAM of a vortex state,
one must use the appropriate axis with respect to which to measure it, the phase singularity axis.

This has important consequences for experiment. Suppose we know that a vortex photon is produced 
but we do not know where its phase singularity axis points to.
When a vortex photon hits a traditional pixelized detector, it collapses, in the usual quantum-mechanical sense,
to a localized state, which erases the information about its wave function and a possible phase vortex it may have had.
No OAM can be detected in this way, even if the photon was in a vortex state.
These arguments were recently developed in \cite{Karlovets:2022evc,Karlovets:2022mhb} in a more formal way
in terms of the strong postsection protocol which is always implicitly applied when we 
discuss measurements with traditional local detectors, see more discussion in Section~\ref{subsection-schemes-scattering}.

One can, however, detect the photon in a way which coherently probes a spatial patch of its wavefront of a certain transverse extent.
This can be done, for example, by letting it pass through a customized diffraction grating before hitting the detector.
This is how OAM sorters work for optical photons and electrons \cite{Bliokh:2017uvr,Lloyd:2017,larocque2018twisted,rosales2018review,Knyazev-Serbo:2018}.
However this device will be able to detect the phase vortex of the incident light only if the phase singularity axis
directly passes through the device.
If it does not, a small local patch of the vortex state wavefront far away from the phase singularity axis 
will resemble a plane wave with an additional transverse momentum $\ell/a$. Again, no OAM can be detected in this way.

To illustrate the point, suppose a pointlike source emits a spherical wave
with $L=1, L_z=1$ state with respect to axis $z$. The outgoing wave function $Y_{1}^1(\theta,\varphi)\,e^{ikr}/r$
contains the spherical harmonic $Y_{1}^1(\theta,\varphi) \propto \sin\theta e^{i\varphi}$,
which exhibits a phase vortex with respect to axis $z$. 
When the outgoing wave reaches a traditional pixelized detector, it leads to the particle detection event 
at any non-zero polar angle $\theta$.
However, the fact that, in this particular event, the particle hits the detector at angle $\theta$, 
does not mean that the quantization axis changes its direction.
It always points along axis $z$, not in the direction of $\theta$.
In order to detect the presence of the phase vortex, one would need to focus on the very forward or very backward
regions of small $\sin\theta$ and employ an OAM sorter, see, for example, a similar discussion of photons
emitted by atoms \cite{Zaytsev:2018}.

In that sense, detecting the OAM state is very different from detecting helicity or the spin state of a particle.
When measuring spin, we can use any convenient quantization axis,
for example, the direction in which the particle hit a local detector.
However we do not have this freedom when measuring the OAM: if a wave, prior to detection, contained a phase vortex, 
it was defined with respect to the quantization axis defined by the emission process itself.
If one still insists on detecting the vortex state and inserts an OAM sorter in a detector of finite size,
then a certain spectrum of the OAM states can indeed be observed. But if the phase vortex does not go through
the aperture of the detector, the resulting OAM spectrum will not be indicative of the OAM 
in the incident wave. It is a manifestation of the mere fact that even a plane wave 
with oblique incidence can be expanded in the basis of Bessel states as in Eq.~\eqref{PW-expansion}.

These remarks are especially relevant for vortex photons which can be emitted in a process that we have no control of.
For example, it was theoretically predicted that a Kerr black hole
twists the wavefront of a photon emitted nearby \cite{Tamburini:2011tk}.
Every vortex photon leaving the Kerr black hole vicinity will posses a phase vortex
with respect to the phase singularity axis. 
However there is no chance of detecting its OAM with telescopes, 
as it would only be feasible under the unrealistic condition that 
the phase singularity axis passes straight through the telescope aperture.
Equipping a telescope with OAM diagnostic instrumentation such as vortex masks 
will simply impart an instrument-induced OAM on the wave front,
not helping to detect the original OAM of the photon.
Multipoint interferometry suggested in \cite{Berkhout:2008zz}, too, is of use only 
in the vicinity of an optical vortex.
If the phase singularity axis is light years away, its presence cannot be detected locally.

It must be mentioned that the recent works \cite{Tamburini:2019vrf,Tamburini:2021lyi} claim to have detected the 
OAM structure of the radio emission from the black hole in M87 collected by the Event Horizon Telescope.
This report is hard to reconcile with the above discussion. 

Another example where similar concerns apply is the recent suggestion of \cite{Zou:2021pnu} 
that photons and charged particles produced in non-central heavy ion collisions are in fact twisted
by the strong local magnetic fields. 
They may well be twisted, but only with respect to a specific axis
defined by each non-central collision event. The photons and charged particles
are emitted at any angle with respect to this axis, and when they hit a local detector,
they do not reveal their twisted state.
Thus, it remains a major challenge to devise an experiment which could detect the OAM of final particles
in such collisions.


\subsection{The vortex photon}

Historically, the physics of vortex states began with vortex laser beams in specific LG modes \cite{Allen:1992zz}.
However, to illustrate the principles of vortex photon construction in potential application to high-energy collisions, 
we limit the discussion below to the Bessel photon states, which are exact monochromatic 
solutions of the Maxwell equations applicable within and beyond the paraxial approximation. 
Bessel photons were introduced in \cite{durnin1987exact,durnin1988comparison}.
Although it is possible to quantize the electromagnetic field directly in the space 
of Bessel photon states \cite{jauregui2005quantum}, this is not needed,
as one can define Bessel photons as superposition of plane wave photon states.

The first high-energy process with vortex photons studied theoretically was Compton backscattering 
in which a vortex optical photon collides with a plane wave ultrarelativistic electron \cite{Jentschura:2010ap,Jentschura:2011ih};
that work set off exploration of vortex states in particle physics.
A review of properties of vortex photons convenient for particle physics calculations can be found, for example, in \cite{Knyazev-Serbo:2018}.
The discussion below follows these references.

It is well known that, in quantum field theory, the operator of the total angular momentum (AM) $\hat{\bj}$ 
commutes with the Hamiltonian, while the spin and OAM operators do not.
Therefore, it is possible to construct one-particle states which are eigenstates 
of $\hat{j}_z = \hat{s}_z + \hat{L}_z$ but not of $\hat{s}_z$ and $\hat{L}_z$ individually.
Put simply, a vortex photon cannot be, strictly speaking, in a state of definite OAM.
This feature can be recast in the form of spin-orbital interaction in free propagating light 
and it was extensively explored by the optics community \cite{Bliokh:2015yhi}.
Thus, when we say ``vortex photon'', we do not mean that it is an eigenstate of $\hat{L}_z$.
We only mean that the OAM values involved in its construction are nonzero and potentially large.
However, as explained already in 1992 \cite{Allen:1992zz}, within the paraxial approximation when the cone opening angle
$\theta_k \ll 1$, one can talk about approximately conserved $s_z$ and $L_z$,
in the sense that the photon is mostly in the state of a single $L_z$ and that its $L_z$ evolution along the $z$ direction
can be neglected within typical longitudinal scales of the problem.

To construct the exact Bessel one-photon state, we begin with a monochromatic plane-wave electromagnetic field with helicity
$\lambda = \pm 1$, which is described in the Coulomb gauge by
\begin{equation}
\bA_{\bk \lambda}(\br) = \frac{1}{\sqrt{2\omega_{\bk}}}N_{\rm PW}\, \bbe_{\bk \lambda}\,e^{i\bk \cdot\br}\,,\label{PW1}
\end{equation}
with the familiar plane-wave normalization factor.
The polarization vector is orthogonal to the wave vector: $\bbe_{\bk\lambda}\cdot \bk = 0$.
Quantization of this field produces plane wave photons with momentum $\bk$.
This state is an eigenstate of the helicity operator $\hat{\Lambda} = \hat{\bs}\cdot \bn_k$, 
where $\bn_k = \bk/|\bk|$.

Let us now fix a reference frame, select axis $z$, and construct the Bessel photon 
with helicity $\lambda = \pm 1$ and the total angular momentum $j_z = m$ with respect to axis $z$
as a monochromatic superposition of plane waves with fixed longitudinal momentum
$k_z = |\bk|\cos\theta_k$, fixed modulus of the transverse momentum $\varkappa = |\bk_\perp| = |\bk|\sin\theta_k$,
but arriving from different azimuthal angles $\varphi_k$.
The usual dispersion relation holds for every plane wave component: $k_z^2 + \varkappa^2 = \omega_{\bk}^2$.
Using the Coulomb gauge for all plane wave components, we get
\begin{equation}
\bA_{\varkappa m \lambda}(\br) = \frac{1}{\sqrt{2\omega_{\bk}}}N_{\rm Bes}\,
\int {d^2\bk_\perp \over (2\pi)^2}\, a_{\varkappa m}(\bk_\perp)\, \bbe_{\bk \lambda}\, e^{i\bk\cdot \br} \,,
\quad 
a_{\varkappa m}(\bk_\perp) = i^{-m} e^{im\varphi_k} {2\pi \over \varkappa}\delta(|\bk_\perp| - \varkappa)\,.
\label{Bessel-photon-1}
\end{equation}
Notice that the Fourier amplitude $a_{\varkappa m}(\bk_\perp)$ is the same as in the scalar case \eqref{twisted-scalar-1}
but it now depends on the conserved total angular momentum $z$-projection $m$, not on the OAM.
However, being a scalar quantity, $a_{\varkappa m}(\bk_\perp)$ is an eigenfunction of both $\hat{j}_z$ 
and of $\hat{L}_z = -i \partial/\partial \varphi_k$ with the same eigenvalue $m$.

The polarization vector $\bbe_{\bk \lambda}$ inside the integral \eqref{Bessel-photon-1}
depends on $\bk$ and cannot be taken out of the integral. 
It means that the polarization state of a vortex photon must be described, in coordinate space, 
with a polarization field rather than a fixed polarization vector.
Since each plane wave component of the vortex photon comes with its own polarization vector,
it is convenient to describe it in a uniform fashion with respect to axis $z$.
Namely, define spin $z$-projection operator $\hat{s}_z$ and construct its eigenvectors ${\bm \chi}_\sigma$:
$\hat{s}_z {\bm \chi}_\sigma = \sigma {\bm \chi}_\sigma$, where $\sigma= \pm 1, 0$.
There explicit form is
\begin{equation}
{\bm \chi}_{0}=
(0, 0, 1)^T,\quad
{\bm \chi}_{\pm 1}= \frac{\mp 1}{\sqrt{2}}(1, \pm i, 0)^T\,, 
\quad {\bm \chi}^*_\sigma \cdot {\bm \chi}_{\sigma^\prime} = \delta_{\sigma\sigma^\prime}\,.
\label{chi}
\end{equation}
Now, the polarization vector of any plane wave component can be expanded in the basis of ${\bm \chi}_\sigma$
with the aid of Wigner's $d$-functions~\cite{Varshalovich}:
\begin{equation}
\bbe_{\bk \lambda}=
\sum_{\sigma=0,\pm 1} e^{-i\sigma \varphi_k}\,
d^{1}_{\sigma \lambda}(\theta)  \,\bm \chi_{\sigma}\,, \qquad d^{1}_{\sigma\lambda}(\theta) = \mmmatrix{\cos^2\frac{\theta}{2}}{-\frac{1}{\sqrt{2}}\sin\theta}{\sin^2\frac{\theta}{2}}%
{\frac{1}{\sqrt{2}}\sin\theta}{\cos\theta}{-\frac{1}{\sqrt{2}}\sin\theta}%
{\sin^2\frac{\theta}{2}}{\frac{1}{\sqrt{2}}\sin\theta}{\cos^2\frac{\theta}{2}}\,.
\label{bbe-chi}
\end{equation}
The first, second, and third rows and columns of this matrix correspond to the indices $+1,\, 0,\, -1$.
Performing the summation in Eq.~\eqref{bbe-chi}, one gets explicit expressions for the polarization vectors:
\begin{equation}
\bbe_{\bk \lambda}= \frac{\lambda }{\sqrt{2}}\triplet{-\cos\theta\cos\varphi_k + i \lambda \sin\varphi_k}%
{-\cos\theta\sin\varphi_k - i \lambda \cos\varphi_k}{\sin\theta}\,,\quad \lambda= \pm 1\,.\label{bbe-explicit}
\end{equation}
Although the vectors $\bbe_{\bk \lambda}$ are not eigenstates of $\hat{s}_z$ and $\hat{L}_z$ separately,
they are eigenvectors of $\hat{j}_z$ with zero eigenvalue.
Thus, when the operator $\hat{j}_z$ acts on \eqref{Bessel-photon-1}, it extracts the value $m$ 
from $a_{\varkappa m}(\bk_\perp)$ alone.
Finally, performing the angular integration in Eq.~\eqref{Bessel-photon-1}, one can compactly write 
the Bessel photon in cylindrical coordinates:
\begin{equation}
\bA_{\varkappa m \lambda}(\br) =  \frac{1}{\sqrt{2\omega_{\bk}}}N_{\rm Bes}\,e^{ik_z z}  \sum_{\sigma = \pm 1, 0} i^{-\sigma} \, d^1_{\sigma \lambda}(\theta)\,
J_{m-\sigma}(\varkappa \rho) \, e^{i(m-\sigma)\varphi_r} {\bm \chi}_\sigma\,.
\end{equation}

The description of the Bessel photon simplifies in the paraxial approximation $\theta\ll 1$. 
In this case, $\bbe_{\bk \lambda} \approx e^{-i\lambda \varphi_k} \,\bm \chi_{\lambda}$,
so that the Bessel photon now possesses not only a definite value of $j_z = m$ but also (approximately) conserved
values of $s_z=\lambda = \pm 1$ and $L_z = m - \lambda$. Notice however that this expression cannot be 
extended to the exact forward limit, as the factor $e^{-i\lambda \varphi_k}$ becomes undefined.

Anticipating vortex particle collision kinematics, 
we remark that a counter-propagating vortex photon defined with respect to the same axis $z$, 
can be described by the above expressions assuming that $k_z < 0$
and replacing $m \to - m$ in the expression for the Fourier amplitude $a_{\varkappa m}$.
The expression for the polarization vector \eqref{bbe-explicit} stays unchanged,
but $\cos\theta < 0$. The paraxial limit is now given by $\theta \to \pi$, in which case
$\bbe_{\bk \lambda} \approx e^{+i\lambda \varphi_k} \,\bm \chi_{-\lambda}$.

The above formalism can be immediately extended to a massive spin-1 particle of mass $M$
described with the polarization vector $V_\mu(\lambda_V)$.
The only modifications are that the dispersion relation changes
to $k_z^2+\varkappa^2+M^2 = E^2$,
and that the third polarization state with $\lambda_V = 0$ is now available.
The orthogonality condition holds for each plane-wave component with four-momentum $k_\mu$ inside the vortex state: $k^\mu V_\mu(\lambda_V) = 0$.
For the transverse polarization states with $\lambda_V = \pm 1$, the expression \eqref{bbe-explicit} applies as it stands.
The longitudinal polarization now includes the time-like component and is described by the four-vector
$V_\mu (\lambda_V = 0) = \gamma (\beta, \bn_k)$,
where $\gamma$ and $\beta$ are the standard relativistic quantities.
One can express the space-like part of this four-vector in the same basis ${\bm \chi}_\sigma$:
\begin{equation}
\bn_k =
\sum_{\sigma=0,\pm 1} e^{-i\sigma \varphi_k}\,
d^{1}_{\sigma 0}(\theta)  \,\bm \chi_{\sigma}\,.
\label{bbe-chi-0}
\end{equation}
In this way one constructs a Bessel vortex state of a massive vector particle with arbitrary helicity $\lambda_V$
and total angular momentum projection $m$.

\subsection{The vortex electron}\label{subsection-vortex-e}

Although cylindrical solutions of the Dirac equation exhibiting a phase vortex were described as early as 1980 \cite{Bagrov:1980ct,Bagrov:1990},
the burst of experimental and theoretical activity on vortex electrons was triggered by \cite{Bliokh:2007ec}.
In this paper, vortex electrons were proposed as a novel physical system with rich dynamics, both in free space and in external electromagnetic fields.
Several methods to produce vortex electrons were also proposed in \cite{Bliokh:2007ec},
and a few years later, they were experimentally demonstrated \cite{Uchida:2010,Verbeeck:2010,McMorran:2011,Guzzinati:2012mb,Beche:2013wua,beche2017efficient}.
During the past decade, the field matured and found various applications,
which are covered in reviews \cite{harris2015structured,Bliokh:2017uvr,Lloyd:2017,larocque2018twisted}.

The exact Bessel solutions of the Dirac equation were constructed in \cite{Bliokh:2011fi}
and later, within other approaches, in \cite{Karlovets:2012eu,Serbo:2015kia,Ivanov:2016jzt,Barnett:2017wrr,Bialynicki-Birula:2016unl,Campos:2020cza}.
Our construction of Bessel vortex electrons given below follows \cite{Serbo:2015kia,Ivanov:2016jzt}; 
it resembles the construction of Bessel photons and is convenient for computation of vortex electron scattering. 
It goes without saying that it also applies to other spin-1/2 particles.

We begin with the plane-wave electron with the four-momentum $k^\mu = (E, \bk)$
and the definite helicity $\lambda = \pm 1/2$:
\begin{equation}
\Psi_{k \lambda}(\br)= {1 \over \sqrt{2E}}\, N_{\rm PW}\, u_{k \lambda}\,e^{i\bk\cdot\br}\,,
\label{electron-PW}
\end{equation}
We define the unit vector $\bn_k = \bk/|\bk| = (\sin\theta_k \cos\varphi_k, \sin\theta_k \sin\varphi_k, \cos\theta_k)$,
and write the bispinor $u_{k\lambda}$ as in \cite{Landau4}
\begin{equation}
u_{k\lambda} = \doublet{\sqrt{E+m_e}\,w^{(\lambda)}}{2 \lambda \sqrt{E-m_e}\,w^{(\lambda)}}\,, \quad 
w^{(+1/2)} = \doublet{\cos(\theta_k/2)\, e^{-i\varphi_k/2}}{\sin(\theta_k/2)\, e^{i\varphi_k/2}}\,,
\quad w^{(-1/2)} = \doublet{-\sin(\theta_k/2)\, e^{-i\varphi_k/2}}{\cos(\theta_k/2)\,e^{i\varphi_k/2}}\,.\label{PWspinors}
\end{equation}
The bispinors are normalized as 
$\bar u_{k\lambda_1} u_{k \lambda_2}= 2m_e\, \delta_{\lambda_1, \lambda_2}$.
The spinors $w^{(\lambda)} = w^{(\lambda)}({\bm n})$ are eigenvectors of the helicity operator $\Lambda({\bm n}) = \hat{{\bm \sigma}} \cdot {\bm n} /2$:
$\Lambda({\bm n}) w^{(\lambda)}({\bm n}) = \lambda w^{(\lambda)}({\bm n})$,
with $\hat{\bm \sigma}$ being the vector of Pauli matrices. 
The helicity $\lambda = \pm 1/2$ is the projection of the electron spin onto its propagation direction.

With the choice of spinors $w^{(\lambda)}$ which depend on the azimuthal angle as $\exp(\pm i\varphi_k/2)$, 
the bispinor is defined up to an overall sign and also does not have a well-defined forward limit, 
see discussion on this issue in \cite{Karlovets:2022mhb}.
For an alternative, equally possible convention, see \cite{Karlovets:2012eu}.
This is a minor inconvenience and it will disappear once we combine these states into the Bessel vortex electron.
For now, we only mention that, with this definition, the bispinor $u_{k\lambda}$ becomes also an eigenstate of the 
total angular momentum operator $\hat{j}_z = \hat{s}_z + \hat{L}_z$ with the zero eigenvalue.

Next, we use this basis of plane wave solutions to construct the Bessel vortex electron state:
\begin{equation}
\label{electron-Bessel}
\Psi_{\varkappa m k_z \lambda}(\br)= {1 \over \sqrt{2E}}\,  N_{\rm Bes}\, \int \frac{d^2 \bk_\perp}{(2\pi)^2}\,
a_{\varkappa m}(\bk_\perp)\, u_{k \lambda}\,e^{i\bk\cdot\br},\quad a_{\varkappa m}(\bk_\perp)=(-i)^m \,e^{im\varphi_k}\,
\sqrt{\frac{2\pi}{\varkappa}}\,\delta(|\bk_\perp|-\varkappa)\,.
\end{equation}
This state possesses the definite longitudinal momentum $k_z$, the definite conical transverse momentum $|\bk_\perp|=\varkappa$,
the well-defined helicity $\lambda$ as well as the definite value of the total angular momentum $j_z = m$,
which must be half-integer.
The Fourier amplitude $a_{\varkappa m}(\bk_\perp)$ has exactly the same form
as for the scalar case \eqref{twisted-scalar-1} but with the half-integer $m$ replacing the OAM value $\ell$.
Since $m\pm \lambda$ is an integer number, all components of the bispinor $a_{\varkappa m}(\bk_\perp)\, u_{k \lambda}$
depend on $\varphi_k$ as $\exp[i(m\pm \lambda)\varphi_k]$ and are well defined.

It must be stressed that, although the Bessel vortex electron \eqref{electron-Bessel} possesses a well-defined helicity,
it cannot be interpreted as a state with a well-defined spin projection on any fixed axis.
The helicity operator involves the direction of $\bk$ and it does not correspond to any component of the spin operator.

The OAM and spin are not separately conserved due to the intrinsic spin-orbital interaction 
within the vortex electron \cite{Bliokh:2011fi}.
However, in the paraxial approximation, when $\theta_k \ll 1$, this spin-orbital coupling is suppressed,
and one deals with two approximately conserved quantum numbers, $s_z \approx \lambda$
and $\ell = m - \lambda$.

Just as for scalar fields and photon, one can also construct Laguerre-Gaussian electron states \cite{Karlovets:2018iww}.
Properties of these vortex electrons including multipole moments were further investigated in \cite{Karlovets:2018zdk,Karlovets:2019qyv}.

The exact vortex solutions of the Dirac equation can also be found
for an electron moving in the field of an electromagnetic (EM) wave.
These solutions can be named Volkov--Bessel states as they extend 
the well-known Volkov solutions \cite{Landau4,Wolkow:1935zz} to the Bessel vortex electron.
Constructed and investigated first in \cite{Karlovets:2012eu} and later in \cite{Hayrapetyan:2014faa,Bandyopadhyay:2015eri}, 
they offer insights into modifications of the vortex electron properties inside a strong laser field,
which is either \cite{Karlovets:2012eu} or linearly \cite{Hayrapetyan:2014faa} polarized.
Dynamical properties such the expectation values of the spin and OAM were explored, see also \cite{Bliokh:2017uvr}.
The numerical analysis of the Bessel electron subject to a strong few-cycle linearly polarized laser pulse reported in \cite{Hayrapetyan:2014faa} 
demonstrated transverse oscillations of the electron probability density distribution. 
We also mention in passing that Volkov states of the electron in the field of a vortex 
electromagnetic wave in plasma were explored in \cite{Mendonca:2019vdp}.

	Vortex solutions of the Dirac equation in Coulomb field 
represent another class of twisted electron states, 
which are especially important for computation of twisted electron scattering on heavy atoms 
\cite{Zaytsev:2016gqp,kosheleva2018elastic,zaytsev2020atomic}.
For large nucleus charge $Z$, with the parameter $\alpha_{em}Z \sim 1$, 
the Born approximation of the scattering amplitude can be unreliable, 
and one should use the full non-perturbative vortex electron solution 
valid to all orders of $\alpha_{em}Z$.
These states were constructed and discussed in \cite{Zaytsev:2016gqp,kosheleva2018elastic}; 
below we briefly summarize the construction.

In similarity with the free-electron case,
one begins with the exact solution of the Dirac equation in the Coulomb potential $\Psi^{(+)}_{km}(\br)$,
which is composed, asymptotically, by the incident plane wave $\Psi_{k\lambda}(\br)$ given by \eqref{electron-PW}
and a scattered outgoing wave:
\begin{equation}
\Psi^{(+)}_{k\lambda}(\br) \xrightarrow[r\to \infty]{} \Psi_{k\lambda}(\br) + 
G^{(+)}_{\lambda}(\bn_k,\bn_r) \frac{e^{ikr}}{r}\,.\label{non-PT-1}
\end{equation}
$G^{(+)}_{\lambda}$ is the bispinor scattering amplitude, 
$\bn_k$ and $\bn_r$ are the unit vectors in the directions of $\bk$ and $\br$, respectively.
The exact solution with this asymptotic behavior can be constructed as \cite{Pratt:1973na,Eichler:1995}:
\begin{equation}
\Psi^{(+)}_{k\lambda}(\br) = \frac{1}{\sqrt{4\pi E k}}\sum_{\ell m_j} C^{j\lambda}_{\ell 0 \, 1/2 \lambda}\, i^\ell
\sqrt{2\ell+1}\, e^{i \delta_q}\, D^{j}_{m_j \lambda}(\varphi_k, \theta_k, 0)\, \Psi_{E q m_j}(\br)\,.
\label{non-PT-2}
\end{equation}
Here, $q = (-1)^{\ell+j+1/2}\times (j + 1/2)$ is the Dirac quantum number
with $j$ and $\ell$ being the total and orbital angular momenta,
respectively, $C^{JM}_{j_1m_1\, j_2m_2}$ is the Clebsch–Gordan coefficient, 
$\delta_q$ is the phase shift induced by the potential, 
$D^J_{MM'}$ is the Wigner rotation matrix \cite{Varshalovich},
and finally $\Psi_{E q m_j}(\br)$ is the partial-wave solution of the Dirac equation 
in the nucleus field which can be found, for example, in \cite{Landau4}. 

The non-perturbative vortex electron solution in the nucleus Coulomb field 
is constructed in a similar manner.
One looks for the solution which exhibits not the plane wave but the Bessel vortex state asymptotics given by \eqref{electron-Bessel}: 
\begin{equation}
\Psi^{(+)}_{\varkappa m k_z \lambda}(\br+\bb) \xrightarrow[r\to \infty]{} \Psi_{\varkappa m k_z \lambda}(\br+\bb) + 
G^{\rm (tw)}_{m\lambda, \bb}(\theta_k,\bn_r) \frac{e^{ikr}}{r}\,.\label{non-PT-3}
\end{equation}
In this expression, a non-zero shift $\bb$ is introduced, which also affects the scattering amplitude.
To find the exact solution with this asymptotics, 
one just uses the same momentum space integral as in \eqref{electron-Bessel} (with an appripriate $\bb$-shift factor)
and applies it to the exact solution \eqref{non-PT-2}.
The resulting exact expressions can be found in \cite{Zaytsev:2016gqp,kosheleva2018elastic}.

\subsection{Polarization of vortex states}\label{section-vortex-polarization}

\begin{figure}[h!]
	\centering
	\includegraphics[width=0.9\textwidth]{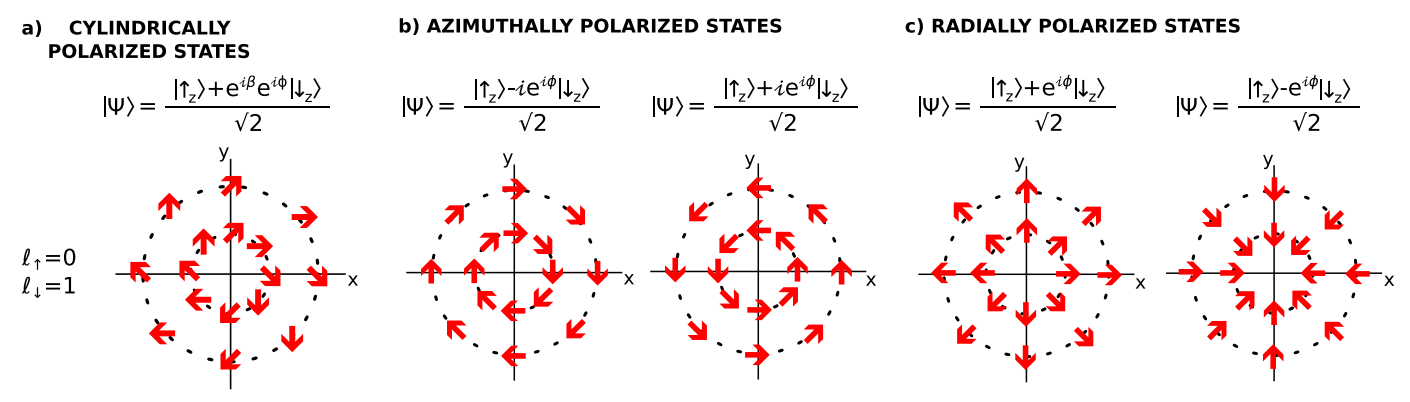}
	{\caption{\label{fig-spin-orbit} 
			Examples of spin-orbit coupled polarization states of the paraxial vortex fermion.
			The arrows show the local spin orientation in the transverse plane. 
			Reproduced from \cite{sarenac2018methods} with permission.
	}}
\end{figure}

Not only do the vortex photons and fermions carry a non-zero OAM, but they also allow for construction 
of exotic polarization states, which are impossible for plane wave.
A plane wave photon is described by a fixed polarization vector $e^\mu$, which is orthogonal to the photon momentum $k^\mu$
and is the same everywhere in the transverse plane.
Multiplying this polarization vector by a constant phase factor does not lead to any physical consequences.
The vortex photon, or in general any non--plane-wave light field, can be decomposed in various plane waves
each carrying its own polarization vector $e^\mu(k)$ as in \eqref{Bessel-photon-1}. 
As a result, the polarization state of the vortex photon is described not with a polarization vector
but with a polarization field, in which the Stokes parameters continuously change in space
and even exhibit singularity lines. Investigation of the three-dimensional structure of the optical polarization field
is the subject of singular optics \cite{soskin2001singular,dennis2009singular,rosales2018review}.
Beyond optical range, a mechanism of generation of vortex X rays with exotic polarization states,
including the so-called ``star'' Poincar\'e beams, was recently proposed for the free-electron laser facilities 
in \cite{Morgan:2020owk}.

A similar freedom exists for vortex fermions. When constructing the vortex electron,
one accompanies each plane wave with the corresponding bispinor $u_{s}(k)$ as in Eq.~\eqref{electron-Bessel}.
Again, the resulting fermion field is described with a polarization field, which can be defined as the local spin vector
smoothly depending on coordinates.
Fig.~\ref{fig-spin-orbit} shows a few examples of spin-orbit coupled states of a vortex fermion
constructed as superpositions of $s_z=+1/2$, $\ell = 0$ and $s_z=-1/2$, $\ell = +1$ states with various relative coefficients.
Since the weights of the $s_z=+1/2$ and $s_z=-1/2$ states are equal in this example, the state is polarized
in the transverse plane, with the spin orientation shown by the arrows depending on the azimuthal angle $\phi$.
Other polarization states of vortex fermions can also be constructed.

This freedom in preparing customized polarization states of vortex photons and fermions
is unthinkable for plane wave states. It represents a new degree of freedom, in addition to the OAM,
which can be imposed on the initial particles in vortex state collisions.
However, not much has been explored so far, at least with possible particle physics applications in mind.
In \cite{sarenac2018methods,sarenac2019generation}, experimental progress on generation of slow vortex neutrons 
was reported based on a scheme which relies on various spin-orbit coupled states of vortex neutrons. 
In theoretical analyses of vortex scattering processes, this opportunity is sometimes mentioned,
see e.g. \cite{Ivanov:2011kk}, but no systematic analysis exists of the effects to be expected.
In fact, in most cases, the vortex photons or electrons are considered as unpolarized.
A rare exception is the vortex muon decay calculations \cite{Zhao:2021joa,Zhao:2022} where 
the electron spectrum and azimuthal distribution
show significant dependence on the polarization state of the muon.
The full potential of this degree of freedom still awaits exploration.

Even defining the unpolarized vortex photon or electron, or the unpolarized cross section
with vortex particles, is not a trivial issue. 
When dealing with an unpolarized plane wave cross section,
we just sum over final helicities and average over the initial ones. 
Since the spin and coordinate dependences are factorized for plane waves,
no ambiguity arises.
In contrast, the exact expression for the vortex photon or fermion 
contains two relevant quantum numbers: the helicity $\lambda$ and the total AM $m$.
When averaging over $\lambda = \pm 1$, 
what should we keep fixed: the total angular momentum $m$ or the combination $m-\lambda$, 
which would correspond in the paraxial limit to the OAM?
There is no unique answer to this question.
It will eventually depend on the vortex photon or fermion preparation scheme
and, in particular, how well the paraxial approximation holds.
If one creates twisted photons or electrons by letting an unpolarized plane wave pass through
a fork diffraction grating which imposes a given OAM $\ell$ immediately behind the aperture, 
the intrinsic spin-orbital interaction will lead to $\ell$ evolution downstream the beam,
and the polarization state becomes non-trivial at the focal spot.
This spin-to-orbital conversion was already experimentally verified in optical vortex beams \cite{Zhao:2007}.
Also, an electron microscope can act as an efficient spin polarizer if one prepared a Bessel vortex electron 
and postselects it near the axis \cite{schattschneider2017spin}.
In any case, as was recently found in \cite{Ivanov:2019vxe,Ivanov:2020kcy},
different definitions of unpolarized vortex electron and photon cross sections
will lead to significantly different predictions for the polarization state of a produced resonance,
once high-energy vortex $e^+e^-$ or $\gamma\gamma$ collisions become possible.

\subsection{Beyond simple vortices}\label{section-beyond-simple}

We close this section with the remark that vortex states with a definite OAM or total AM value
represent just one example of a broader class of beams or wave packets with custom-tailored wave functions.
Let us briefly mention other options considered in literature and demonstrated experimentally.

First, one can take a superposition of vortex states with the same intensity distribution but with the opposite
OAM $z$-projection values. The coordinate wave function $|+\ell\rangle + |-\ell\rangle \propto \cos(\ell\varphi_r)$
leads to the intensity distribution with multi-petal azimuthal angular dependence. 
When such states scatter head-on, they lead to an azimuthally dependent signal which
was absent in the single-OAM case \cite{Ivanov:2011kk,Ivanov:2016xab}.
Behavior of such states of electrons in external magnetic fields was explored theoretically 
in \cite{bliokh2012electron,greenshields2012vacuum} and verified in experiment in \cite{Schattschneider:2014}.
The spin-orbit coupled states of vortex photon or fermion with an azimuthally-dependent polarization field, 
which were discussed above and shown in Fig.~\ref{fig-spin-orbit}, represent yet another example 
of superposition of different OAM.

In the simplest case, the non-trivial intensity distribution of an OAM superposition state 
does not evolve downstream the beam.
However it is also possible to prepare fine-tuned OAM superpositions which exhibit such evolution
in free space, without any external fields.
Very recently, the so-called coiling free electron waves were theoretically explored 
and experimentally produced with the aid of nanofabricated holographic masks in \cite{Pierce:2019gvo}. 
In these states, the electron density in the transverse plane contains two compact lobes
which rotate around each other as the beam propagates, and their angular velocity can be 
either constant or variable.

One can also consider an electron state with a transverse wave function in the form of a superposition
of two compact Gaussian states separated by a controlled distance, which may be larger than the size of 
the individual Gaussian envelop. Such states were called in \cite{Karlovets:2017gzk} Schr\"{o}dinger's cat
states following the quantum optics tradition. 
With suitable parameters, scattering of cat states on atomic targets proceeds in a regime
which significantly differs from approximate plane wave scattering. 
In particular, as shown in \cite{Karlovets:2017gzk}, this process allows one to probe
negative values of the Wigner function of the incident wave packet.

Other examples of states with highly non-trivial phase front include Airy beams \cite{berry1979nonspreading,efremidis2019airy}, 
which have already been demonstrated experimentally for electrons \cite{voloch2013generation},
and self-accelerated Dirac particles which are predicted to exhibit remarkable behavior \cite{kaminer2015self}.
One can also construct a more exotic form of vortex states called spatiotemporal vortices \cite{sukhorukov2005spatio,bliokh2012spatiotemporal}.
In momentum space, these states can also be approximated by a thin ring as for the Bessel state shown 
in Fig.~\ref{fig-vortex-2}. However, now this ring is not orthogonal to $\lr{\bp}$ and can even lie in a plane
that includes axis $z$. Such states are unavoidably non-monochromatic, thus they exhibit space and time evolution. 
Experimentally observed for photons in \cite{jhajj2016spatiotemporal,chong2020generation},
these spatiotemporal vortices are now a vibrant topic on their own.

In broader terms, the concept of wave function engineering for electrons and ions is now being proposed
as a novel technique which can expand horizons of traditional microscopy \cite{madan2020quantum,vanacore2020spatio}.
These states may eventually find applications in particle physics.
For example, freely propagating states demonstrating self-accelerating behavior 
are predicted to exhibit slowing down of an unstable particle decay \cite{kaminer2015self}.
Airy beams and other states with non-trivial phase structure may represent 
an additional probe of the overall phase of the scattering amplitude \cite{Karlovets:2014wva,Karlovets:2016dva,Karlovets:2016jrd}.
Scattering of precisely shaped electron wave packets offers an unprecedented control over 
the rate, the spatial and the spectral distributions of bremsstrahlung and, potentially, other QED processes \cite{wong2021control}.
Discussion of these states goes beyond the scope of the present review, 
but we strongly suggest that particle physics community keeps track of this topic and draws inspiration from the results.

	\newpage
	\section{Vortex state scattering: generic features}\label{section-scattering-general}

\subsection{Types of scattering processes with vortex states}

Several schemes for vortex states scattering are possible \cite{Bliokh:2017uvr}. 
They depend on whether the vortex states appear in the initial or final state 
and on how many particles involved are described as vortex states.
Each setting brings its own opportunities and challenges, which we now briefly describe.

Vortex states can appear in the initial or in the final state.
From the experimental perspective, these two options are fundamentally different.
Experiments with initial state particles in vortex states 
require new state preparation instrumentation,
but the outcome of the collision can analyzed with traditional detectors.
However if one aims at producing at least one final particle in a vortex state,
one must introduce new instrumentation capable of detecting that a final particle 
indeed possesses a phase vortex and of measuring its vortex state parameters.
This is a major additional experimental challenge. 

Indeed, an initial vortex state can be represented as a well-collimated, often paraxial wave packet or a beam.
Its vortex state parameters can be analyzed with diffraction gratings or other OAM sorters of small transverse size.
In contrast, an outgoing wave function emerging from a typical scattering process 
has a wide polar angle distribution. If one wishes to detect this outgoing state 
as a vortex state defined with respect to the collision axis,
one would need to coherently detect it in the entire $2\pi$ range of azimuthal angles $\varphi$. 
At present, no such detectors exist, at least for high-energy particles.
As we discussed in Section~\ref{section-qualitative}, a standard detector instrumentation will be of little use:
as long as one detects an outgoing particle in a certain direction,
one loses information about its $\varphi$-dependent phase \cite{Karlovets:2022evc,Karlovets:2022mhb}.
These challenge may be less severe for processes featuring very narrow
angular distributions of one of the final particles, such as Compton backscattering from ultrarelativistic
electrons \cite{Jentschura:2010ap,Jentschura:2011ih}.

Next, scattering of a vortex state can proceed in a fixed target or in a collider-like regime.
In fixed target scattering, a heavy scattering center can absorb any momentum transfer.
Since the vortex initial state is a coherent superposition of plane waves, 
its scattering amplitude to any final plane wave state 
can be written is a superposition of individual plane wave scattering amplitudes
corresponding to different momentum transfers.
This leads to modifications of the differential cross section 
and the polarization properties of final photons or electrons with respect to the plane wave case.
For a pedagogical exposition of wave packet scattering on atoms, see, for example, \cite{Karlovets:2015nva}.

These modifications depend not only on the vortex state parameters but also 
on the location of the scattering center $\bb = (\bb_\perp, b_z)$ with respect to the vortex axis.
For a single scattering center, such as a single atom or an ion in an trap which absorbs a twisted photon, 
the $\bb$-dependent modifications can be very profound.
However, if one considers scattering of vortex states on a macroscopic target 
with uniformly distributed scattering centers,
averaging of the scattering cross section over a range of $\bb$
washes out the azimuthal angle features and renders the results
insensitive to the OAM value $\ell$ \cite{matula2013atomic,Serbo:2015kia}.
However, structures in the polar angle distributions are preserved and can be used to investigate vortex 
state scattering on atoms. 
Uniform averaging does not apply to crystals and, in particular, to chiral crystals; 
as a result, elastic scattering of vortex electron beams can be used to investigate 
chirality in crystalline materials \cite{Juchtman2015-II}.
Finally, a mesoscopic target of a spatial extent comparable with the waist of the focused vortex beam
interpolates between the two cases and allows one to observe residual dependences 
which were pronounced for the single scattering center but disappeared for a macroscopic target.

In collider-type scattering, where the two initial particles 
are described as freely propagating wave packets or monochromatic beams,
energy and momentum conservation imposes additional constraints 
on interference of individual plane wave components from the initial state scattering to a given final state.
In single-vortex scattering (or single-Bessel scattering if Bessel states are used), 
a vortex state collides with a counterpropagating plane wave.
If one studies the angular distribution of the final particles, the interference disappears, see details below.
Decay of a vortex state particle into plane wave final states also follows this scheme.

In contrast, double-vortex scattering, in which two counterpropagating vortex states collide,
will exhibit a remarkable interference pattern between different pairs of initial plane wave components
scattering into the same final state.
Thus, double-vortex scattering gives access to new observables, which cannot be studied with the single-vortex setting.
Below we will consider in detail the main kinematic features in these two scattering regimes.

\subsection{Single-vortex scattering}\label{subsection-single-Bessel}

\subsubsection{General expressions}

In order to illustrate how one calculates cross sections of processes involving vortex states,
let us review the standard plane-wave scattering computation with all the factors taken into account.
For definiteness, we focus on a $2 \to 2$ scattering, not necessarily elastic, 
with the initial four-momenta $k_1$, $k_2$ and the final four-momenta $k'_1$ and $k'_2$. 
The total four-momentum is denoted as $K = k_1 + k_2 = k'_1 + k'_2$. 
With the normalization conditions adopted in Section~\ref{section-describing} (one particle per large volume), the scattering matrix element is
\begin{equation}
S_{\rm PW}\left(k_1,k_2; k'_1,k'_2\right) = \frac{i (2\pi)^4\, \delta^{(4)}(k_1+k_2 - k'_1 - k'_2)\, N_{\rm PW}^4}{\sqrt{16E_1E_2E'_1E'_2}}\,
{\cal M}\!\left(k_1,k_2; k'_1,k'_2\right)\,,
\label{S-PW}
\end{equation}
where the invariant amplitude ${\cal M}$ is calculated according to the standard Feynman rules
and $N_{\rm PW} = 1/\sqrt{V}$.
Squaring the scattering matrix element, we obtain the transition probability to a particular final state.
The square of the delta function is regularized as \cite{Landau4,Peskin:1995ev}
\begin{equation}
[\delta^{(4)}(k_1+k_2 - k'_1 - k'_2)]^2 = \delta^{(4)}(k_1+k_2 - k'_1 - k'_2) \frac{V T}{(2\pi)^4}\,.
\end{equation}
Dividing this probability by the observation time $T$ and summing over the final phase space volume, 
we obtain the differential transition rate
\begin{eqnarray}
d\nu_{\rm PW} &=& \frac{(2\pi)^4\, \delta^{(4)}(k_1+k_2 - k'_1 - k'_2)\, V\, }{16 E_1 E_2 E_1' E_2' V^4}\, |{\cal M}|^2
\frac{Vd^3k'_1}{(2\pi)^3} \frac{Vd^3k'_2}{(2\pi)^3} = \frac{\delta^{(4)}(k_1+k_2 - k'_1 - k'_2)}{(2\pi)^2 V \cdot 16 E_1 E_2 E_1' E_2'}\,
|{\cal M}|^2 \, d^3k'_1d^3k'_2\,.
\end{eqnarray}
To obtain the differential cross section $d\sigma$, we divide the event rate by the collision luminosity $L$, 
which, with the normalization of one particle per volume $V$, has the form
\begin{equation}
L = v_{\rm rel}\int d^3 \br |\psi_1(\br)|^2 |\psi_2(\br)|^2 = v_{\rm rel}\frac{1}{V}\,,\label{lumi}
\end{equation}
where $v_{\rm rel} = |\bv_1 - \bv_2|$ is the relative velocity.
The cross section is then written as
\begin{equation}
d\sigma = \frac{\delta(E_1+E_2 - E'_1 - E'_2)}{(2\pi)^2 \cdot 4 E_1 E_2 v_{rel} \cdot 4 E_1' E_2'}\,
|{\cal M}|^2 \, d^3k'_1\,.\label{dsigma-PW-3}
\end{equation}
The remaining energy delta function is usually removed in the center of motion frame, leading to the familiar expression
\begin{equation}
d\sigma = \frac{|{\cal M}|^2}{64 \pi^2 s} \frac{|\bk'_1|}{|\bk_1|} d\Omega_1\,.\label{dsigma-PW-4}
\end{equation}

Let us now consider the single-Bessel scattering process, in which the first particle is prepared in the Bessel state 
with parameter $\varkappa$ and the total angular momentum $m$.
The $S$-matrix element is now written as
\begin{eqnarray}
S_{\rm Bes} &=& \frac{N_{\rm Bes}}{N_{\rm PW}}\int \frac{d^2\bk_{1\perp}}{(2\pi)^2} a_{\varkappa,m}(\bk_{1\perp}) S_{\rm PW}\left(k_1,k_2; k'_1,k'_2\right) \nonumber\\
&=& \frac{i (2\pi)^4\delta(E)\delta(k_z)}{\sqrt{16E_1E_2E'_1E'_2}}\, N_{\rm PW}^3 N_{\rm Bes}\,
\int \frac{d^2\bk_{1\perp}}{(2\pi)^2} a_{\varkappa,m}(\bk_{1\perp})\, 
\delta^{(2)}(\bk_{1\perp} + \bk_{2\perp} - \bK_\perp)\,{\cal M}(k_1)\,.
\label{S-1Bessel-1}
\end{eqnarray}
Here, we use the shorthand notation $\delta(E) = \delta(E_1+E_2-E_1'-E_2')$, 
$\delta(k_z) = \delta(p_{1z}+p_{2z} - k'_{1z} - k'_{2z})$, ${\cal M}(k_1) = {\cal M}\left(k_1,k_2; k'_1,k'_2\right)$,
and the total final transverse momentum $\bK_\perp$.
When squaring the transverse integral, we get
\begin{eqnarray}
\left|\int \frac{d^2\bk_{1\perp}}{(2\pi)^2} \dots\right|^2 &= &
\int \frac{d^2\bk_{1\perp}}{(2\pi)^2}\, \frac{d^2\tilde \bk_{1\perp}}{(2\pi)^2}\, a_{\varkappa,m}(\bk_{1\perp}) 
a^*_{\varkappa,m}(\tilde \bk_{1\perp})\, \delta^{(2)}(\bk_{1\perp} + \bk_{2\perp}-\bK_\perp)\, \delta^{(2)}(\tilde \bk_{1\perp} + \bk_{2\perp}-\bK_\perp)\, 
{\cal M}(k_1) {\cal M}^*(\tilde k_1)
\nonumber\\
&=& \int \frac{d^2\bk_{1\perp}}{(2\pi)^4}\, |a_{\varkappa,m}(\bk_{1\perp})|^2 
\delta^{(2)}(\bk_{1\perp} + \bk_{2\perp}-\bK_\perp)\, |{\cal M}(k_1)|^2\nonumber\\
&=& \frac{R}{\pi} \frac{1}{(2\pi)^2} \int \frac{d\varphi_k}{2\pi} \delta^{(2)}(\bk_{1\perp} + \bk_{2\perp}-\bK_\perp)\, |{\cal M}(k_1)|^2\,.\label{single-Bes-2}
\end{eqnarray}
In the last line, we used the radial regularization prescription \eqref{radial-reg}. 
Notice that the factor $R$ will be compensated by $1/R$ in $N_{\rm Bes}^2$.
Bringing all the coefficients together, we obtain the event rate for single-Bessel scattering:
\begin{equation}
d\nu_{\rm Bes} = \int \frac{d\varphi_k}{2\pi}\, d\nu_{\rm PW}\,.\label{single-Bes-3}
\end{equation}
In general wave packet scattering, the luminosity function should be defined
with the aid of the Wigner function, see Section~\ref{subsection-Wigner-formalism}.
However, for the single-Bessel scattering, in which the plane wave propagates along the same axis
with respect to which the Bessel state is defined, the relative velocity
$v_{\rm rel,Bes} = |\bv_1 - \bv_2| = v_1\cos\theta_k + v_2$ is the same for all plane-wave components. 
As a result,
\begin{equation}
L = v_{\rm rel,Bes}\int d^3 r |\psi_1(\br)|^2 |\psi_2(\br)|^2 = v_{\rm rel,Bes}\frac{1}{V}\,.\label{lumi-Bes}
\end{equation}
Thus, the cross section for the single-Bessel scattering is represented as the azimuthally averaged plane wave cross section
corrected by the relative velocity ratio:
\begin{equation}
d\sigma_{\rm Bes} = \frac{v_{\rm rel,PW}}{v_{\rm rel,Bes}} \int \frac{d\varphi_k}{2\pi}\, d\sigma_{\rm PW}\,.\label{single-Bes-4}
\end{equation}
The velocity ratio can be neglected in the paraxial approximation $\theta_k \ll 1$.

\subsubsection{Physical consequences}

Most publications reporting calculations of various processes with vortex states 
adopt the single-vortex (or, specifically, single-Bessel) scheme.
The findings reported share much in common,
as many of them are driven by the single-vortex kinematics, not by concrete physical process. 
Here, we will illustrate the main features which arise when passing from the fully plane wave to the single-vortex setting
with an example which may at first look peculiar: Vavilov-Cherenkov radiation 
emitted by a sufficiently fast electron moving through medium with refraction index $n$.
Vavilov-Cherenkov radiation from vortex electrons was first studied in \cite{Kaminer:2014iia}
and then addressed in more detail in \cite{Ivanov:2016xab}; below, we follow the exposition of \cite{Ivanov:2016xab}.

It is well known that Vavilov-Cherenkov radiation can be described in a fully quantum approach
as a decay process $e(p) \to e(p')\gamma(k)$, where the photon propagates inside the medium
and obeys $|\bk| = n\omega$, $k^2 = -\omega^2(n^2-1)$ \cite{Ginzburg:1979,Afanasev:2004yh}. 
Energy and momentum conservation laws uniquely determine the angle $\theta_{pk}$ between the photon emission direction 
and the initial electron momentum: $\theta_{pk} = \theta_0$, where the Cherenkov cone opening angle is given by
\begin{equation}
\cos{\theta_0}=\frac{1}{v n}+\frac{\omega}{2E_e}\,\frac{n^2-1}{vn}\,,
\label{VCh-angle}
\end{equation}
with $E_e$ and $v$ denoting the electron energy and its velocity.
In most cases, the second term here is suppressed and can be dropped.
The differential ``decay rate'' defines the spectral-angular distribution of the emitted photons:
\begin{equation}
\frac{d\Gamma_{\rm PW}}{d\omega d\Omega_k} = 
\frac{\alpha_{\rm em}}{2\pi}\,\left[\frac{\left|\bp \cdot \bbe \right|^2}{vE_e^2} + \frac{\omega^2}{4vE_e^2}\,(n^2-1)\right]\,
\delta(\cos\theta_{kp}-\theta_0)\,,
\label{VCh-2}
\end{equation}
where $\bbe$ is the photon polarization vector, which is purely spatial in the Coulomb gauge
and which is orthogonal to the photon's direction.
The angular delta function is the hallmark of the Cherenkov cone.
Since the second term in the brackets of Eq.~\eqref{VCh-2} is small, one also concludes that 
the emitted photons are almost 100\% polarized in the scattering plane.
The calculations can also be conducted for a polarized electron 
(appendix A of \cite{Ivanov:2016xab}).

Suppose now that the initial electron is in a Bessel vortex state propagating on average along axis $z$.
It carries the total AM $m$ and has the vortex cone opening angle $\theta_p$, 
or the corresponding conical transverse momentum $\varkappa$. 
The scattering amplitude can be written as
\begin{equation}
S_{\rm Bes} = \frac{N_{\rm Bes}}{N_{\rm PW}} \int \frac{d^2\bp_{\perp}}{(2\pi)^2}\, a_{\varkappa,m}(\bp_{\perp}) S_{\rm PW}\left(p, p',k\right)\,.
\label{single-1}
\end{equation}
Applying the same formalism as above, we obtain the spectral-angular distribution of the Vavilov-Cherenkov radiation 
emitted by the vortex electron:
\begin{equation}
\frac{d\Gamma_{\rm Bes}}{d\omega d\Omega_k} = 
\int \frac{d\varphi_p}{2\pi} \, \frac{d\Gamma_{\rm PW}(p)}{d\omega d\Omega_k}\,.\label{VCh-3}
\end{equation}
The spectral-angular distribution for the Bessel vortex state is given by incoherent averaging 
over all plane wave components residing inside the Bessel electron.
It is crucial to realize that processes induced by all these plane wave components do not interfere
when we study the angular distribution of the emitted photons.
Indeed, if the photon momentum $\bk$ is fixed, then different initial electron momenta $\bp$ lead to different final electron momenta $\bp'$,
which precludes any interference. 
In order to detect that outgoing photon wave function carries a non-trivial phase profile, 
one would need to employ a special post-selection protocol: employ a hypothetical macroscopic detector 
which would coherently analyze the wave function in the entire $2\pi$ domain
of azimuthal angles \cite{Karlovets:2022evc,Karlovets:2022mhb}, see also Section~\ref{subsection-schemes-scattering}.

\begin{figure}[h!]
	\centering
	\includegraphics[width=0.25\textwidth]{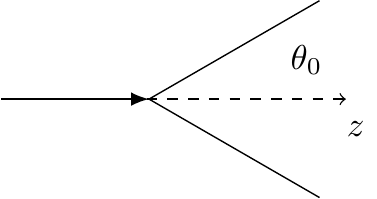}\hspace{1cm}
	\includegraphics[width=0.3\textwidth]{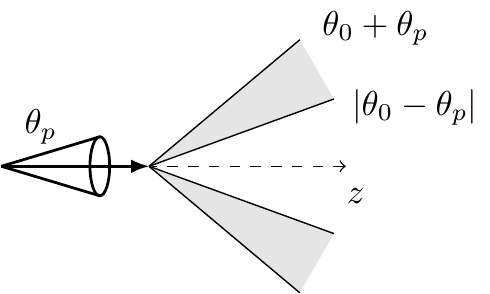}\hspace{1cm}
	\includegraphics[width=0.2\textwidth]{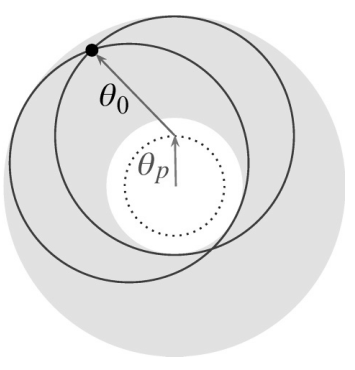}
	{\caption{\label{fig-VCh-ring} 
			Left: Side view of the Vavilov-Cherenkov cone from a plane wave (or pointlike classical)
			charged particle.
			Center: Side view of the broad Vavilov-Cherenkov cone from a Bessel electron with cone opening angle $\theta_p$.
			Right: Forward view of the angular distribution of the Vavilov-Cherenkov radiation by a Bessel
			vortex electron. The dotted circle shows the opening angle of the vortex
			electron; the solid circles correspond to the Vavilov-Cherenkov cones from selected
			plane-wave configurations inside the vortex electron. 
	}}
\end{figure}

The plane-wave distribution $d\Gamma_{\rm PW}(p)$ under the integral in \eqref{VCh-3}
depends on the angle $\varphi_p$ through $\cos\theta_{pk} = (\bp \cdot \bk)/|\bp| |\bk|$ and
through $(\bp \cdot \bbe)$, which leads to modifications of the angular distribution 
and the polarization properties of the emitted radiation.
The angular delta function now depends on 
$\cos\theta_p\cos\theta_k + \sin\theta_p \sin\theta_k \cos(\varphi_p-\varphi_k) - \cos\theta_0$.
As $\varphi_p$ sweeps its entire range, the polar angle of the photon emission $\theta_k$ varies within the interval
\begin{equation}
|\theta_0 - \theta_p| \le \theta_k \le \theta_0 + \theta_p\,.
\end{equation}
Instead of an infinitely narrow Cherenkov ring, we obtain an annular region of the finite width $2\theta_p$, 
see Fig.~\ref{fig-VCh-ring}.
Emission of the photon in any particular direction inside this ring (the black dot in Fig.~\ref{fig-VCh-ring}, right) comes from
two plane wave components of the vortex electron.
The spectral-angular distribution for the vortex electron Vavilov-Cherenkov radiation can then be compactly written as
\begin{equation}
\frac{d\Gamma_{\rm Bes}}{d\omega\,d\Omega}=
\frac{\alpha_{\rm em}}{2\pi v }\,\left[\frac{\langle\left|\bp \cdot \bbe \right|^2\rangle}{E^2}
+\frac{\omega^2}{4E^2}\,(n^2-1)
\right]\, F(\theta_k, \theta_p, \theta_0),
\label{satw}
\end{equation}
where $\langle\left|\bp \cdot \bbe \right|^2\rangle$ is the average of $|\bp \cdot \bbe|^2$ 
computed for the two contributing plane wave configurations, and the function $F(\theta_k, \theta_p, \theta_0)$ 
effectively replaces the angular delta function for the plane wave case
\begin{equation}
F(\theta_k, \theta_p, \theta_0) = \frac{1}{\pi}\frac{1}{\sqrt{[\cos\theta_k - \cos(\theta_0+\theta_p)][\cos(\theta_0-\theta_p) - \cos\theta_k]}}\,,
\quad
\int_{|\theta_0-\theta_p|}^{\theta_0+\theta_p} F(\theta_k, \theta_p, \theta_0)\, \sin\theta_k d\theta_k = 1\,.
\label{F-single}
\end{equation}
In the above analysis as well as in Fig.~\ref{fig-VCh-ring} we implicitly assume that the cone opening angle
of the vortex electron $\theta_p$ is smaller (or even much smaller) than the standard Cherenkov emission angle $\theta_0$.
In the opposite case, $\theta_p > \theta_0$, the above analytical expressions
remain valid, the angular width of the ring is $2\theta_0$, and the plots in Fig.~\ref{fig-VCh-ring} 
will be modified \cite{Ivanov:2016xab}.

As for the average polarization of the Vavilov-Cherenkov photons, it now acquires a component
orthogonal to the scattering plane, a situation that was impossible for the plane wave electron.
This peculiar feature is of purely kinematical origin as it arises from the mismatch 
of the true scattering plane $(\bp, \bk)$ and the overall scattering plane $(\lr{\bp}, \bk)$.
For $\theta_0 \sim \theta_p$, the average degree of linear polarization can dramatically depart from one \cite{Ivanov:2016xab}.

Next, notice that the value of the total AM or the OAM of the vortex electron disappeared in the second line of Eq.~\eqref{single-Bes-2}. 
As a result, the photon angular distribution 
is azimuthally symmetric and is independent of the OAM parameter.
This OAM blindness is a drawback of the single-vortex scattering setting, as it makes it difficult
to detect whether the incident particle indeed possesses a phase vortex.
However, if one considers the vortex state in a superposition of two distinct OAM or total AM states offset by $\Delta m$, 
then the azimuthal symmetry is broken, and the angular distribution exhibits multi-petal structure, 
repeating itself $|\Delta m|$ times around the circle \cite{Ivanov:2011kk,Ivanov:2016xab}.

Let us summarize the key features of this example which can be expected in other single-vortex scattering processes,
at least with Bessel vortex state.
\begin{itemize}
	\item The single-Bessel cross section is an incoherent average of the plane-wave cross section over all 
	plane wave components of the vortex state.
	\item
	For a pure OAM state, the single-vortex cross section is insensitive to the value of the OAM.
	However, this sensitivity arises if a superposition of two distinct OAM states is used.
	\item
	The polar angle distribution of the plane wave cross section becomes smeared by the angle $\pm \theta_p$
	for the single-vortex case. If the angular distribution in a plane wave process is itself very narrow 
	while the cone opening angle is large, one expects to see a novel ring-like structure of the angular distribution for the single-vortex scattering.
	\item
	The average polarization properties of the final particles may differ dramatically from the plane wave case.
	This modification is driven not by new physics effects but by a mismatch of the scattering planes.
\end{itemize}

\subsection{Double-vortex scattering}\label{subsection-double-vortex}

\subsubsection{Novel kinematical features}\label{subsubsection-novel-kinematical}

In the double-vortex scattering regime, we assume that the two initial particles are prepared
in vortex states. The final particles in the $2\to 2$ scattering are described as plane waves with momenta 
$k_1^{\prime}$ and $k_2^{\prime}$.
In order to illustrate the novel kinematical effects, 
let us assume the initial particles to be spinless and describe them as pure Bessel states
with the longitudinal momenta $k_{1z} > 0$ and $k_{2z}<0$, the conical opening angles $\varkappa_1$ and $\varkappa_2$,
and the orbital helicities $\ell_1$ and $\ell_2$.
Since the second particle propagates in the $-z$ direction, its OAM projection on the common $z$ axis is $-\ell_2$.
The $S$-matrix element of the double-Bessel scattering can be written as
\begin{eqnarray}
S_{\rm 2Bes} &=& \frac{N_{\rm Bes}^2}{N_{\rm PW}^2}\int \frac{d^2\bk_{1\perp}}{(2\pi)^2} \frac{d^2\bk_{2\perp}}{(2\pi)^2}\,
a_{\varkappa_1,\ell_1}(\bk_{1\perp}) a_{\varkappa_2,-\ell_2}(\bk_{2\perp}) S_{\rm PW}\left(k_1,k_2; k'_1,k'_2\right) \nonumber\\
&=&\frac{i (2\pi)^4\delta(E)\delta(k_z)}{\sqrt{16E_1E_2E'_1E'_2}} N_{\rm Bes}^2 N_{\rm PW}^2 \cdot  \frac{(-i)^{\ell_1-\ell_2}}{(2\pi)^3\sqrt{\varkappa_1\varkappa_2}} \cdot {\cal J}\,,
\label{S-2Bessel-1}
\end{eqnarray}
where $\delta(E) = \delta(E_1+E_2-E_1'-E_2')$, $\delta(k_z) = \delta(k_{1z}+k_{2z} - k'_{1z} - k'_{2z})$, and the two-Bessel
amplitude is defined as   
\begin{eqnarray}
{\cal J} = \int d^2\bk_{1\perp} d^2\bk_{2\perp} \, e^{i\ell_1\varphi_1 - i\ell_2\varphi_2}\,
\delta(|\bk_{1\perp}|-\varkappa_1) \delta(|\bk_{2\perp}|-\varkappa_2)\, \delta^{(2)}(\bk_{1\perp} + \bk_{2\perp} - \bK_\perp)
\cdot {\cal M}(k_1,k_2; k'_1,k'_2)\,.\label{J}
\end{eqnarray}
Squaring \eqref{S-2Bessel-1}, regularizing the squares of the delta functions as $[\delta(E)\delta(k_z)]^2 = \delta(E)\delta(k_z)\cdot T d_z/(2\pi)^2$,
we obtain the differential event rate $d\nu_{\rm 2Bes}$. Dividing by the appropriately defined flux 
and using $dk'_{1z} dk'_{2z}$ to eliminate $\delta(E)\delta(k_z)$, we can obtain the differential cross section
\cite{Ivanov:2012na,Ivanov:2016oue}. 

We will not, however, give these expressions in detail due to several reasons. First, 
our main focus now is on the kinematical distributions, not on the absolute values of the cross section.
Second, the definition of the flux is not as straightforward as for the single-Bessel case
and should be expressed via the Wigner functions.
Third, as we will find shortly, scattering of two exact Bessel states leads to a divergence
which stems from their non-normalizable nature. 
Thus, for a physically relevant situation, we need to replace the Bessel beams
with transversely localized vortex states, which will lead to a different expression for the flux.
Therefore, returning to the double-Bessel scattering, we represent the cross section simply as
\begin{equation}
d\sigma \propto |{\cal J}|^2 d^2\bk'_{1\perp} d^2\bk'_{2\perp}
\label{S-2Bessel-2}
\end{equation}
and focus now on the transverse momentum distributions.

\begin{figure}[h]
	\centering
	\includegraphics[width=0.6\textwidth]{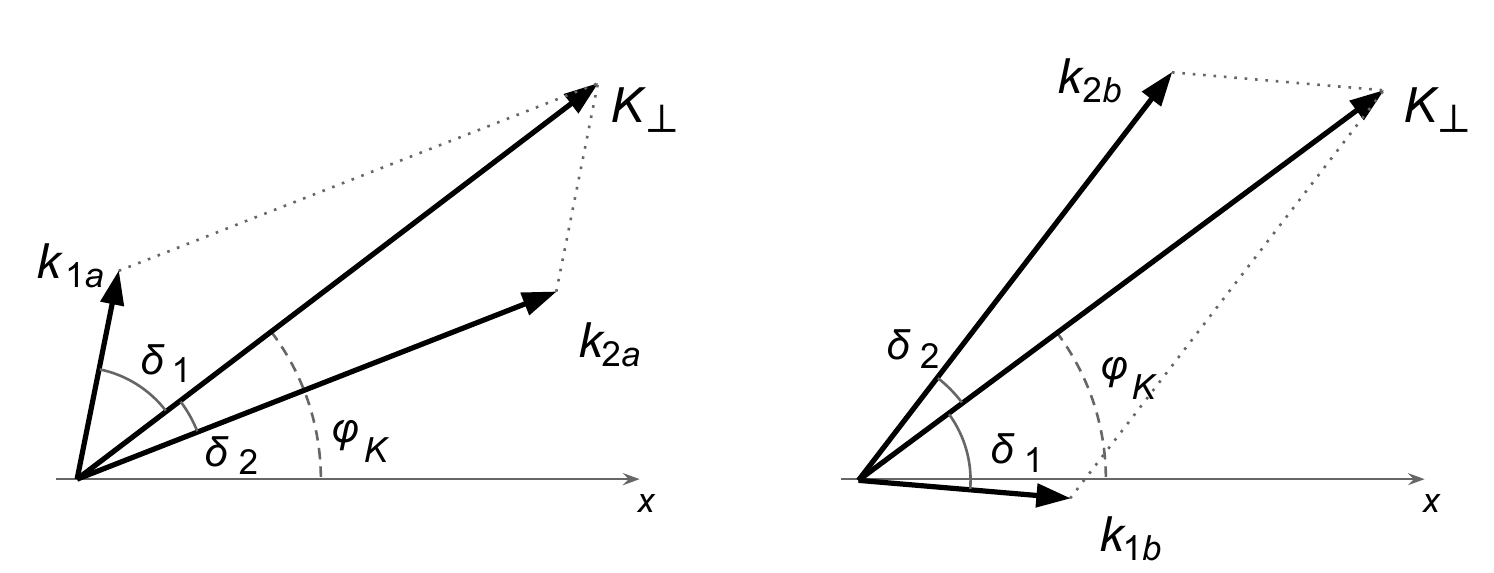}\hspace{5mm}
	\includegraphics[width=0.3\textwidth]{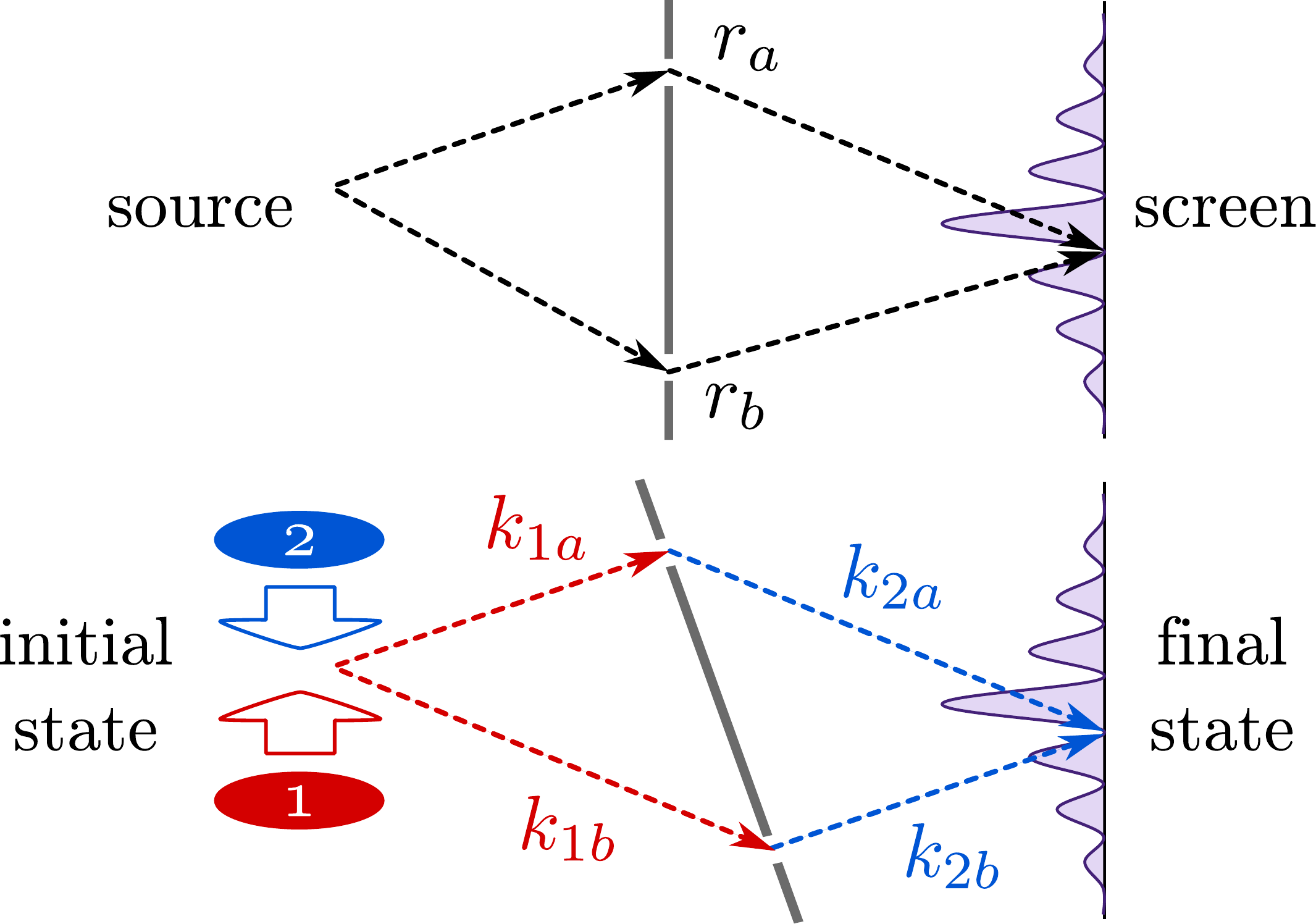}
	{\caption{\label{fig-2configurations} Left: the two plane wave configurations 
			in the transverse momentum plane that can lead to the final transverse momentum 
			$\bK_\perp$ in double-Bessel scattering. Right: schematic illustration of Young’s experiment in the coordinate space (upper plot) 
			and of the double-Bessel collision in momentum space (lower	plot). 
			Reproduced from \cite{Ivanov:2016oue} with permission.}}
\end{figure}

The dynamics of the scattering process is determined by the vortex amplitude ${\cal J}$ defined in (\ref{J}).
It contains four integrations and four delta functions, so that the integral can be done exactly for any non-singular ${\cal M}$. 
It is nonzero only when the total transverse momentum $\bK_{\perp} = \bk'_{1\perp} + \bk'_{2\perp}$ lies within the annular region
defined by the initial conical momenta:
\begin{equation}
|\varkappa_1 - \varkappa_2| \le |\bK_\perp| \le \varkappa_1 + \varkappa_2\,.\label{ring}
\end{equation}
For any value of $\bK_{\perp}$ strictly inside this region,
the integral comes only from two points in the entire $(\bk_{1\perp},\,\bk_{2\perp})$ space,
at which the momenta $\bk_{1\perp} = \varkappa_1 (\cos\varphi_1, \sin\varphi_1)$ 
and $\bk_{2\perp} = \varkappa_2 (\cos\varphi_2, \sin\varphi_2)$ sum up to $\bK_\perp$, see Fig.~\ref{fig-2configurations}.
This happens at the following azimuthal angles:
\begin{eqnarray}
\mbox{configuration a:} &&\varphi_1 = \varphi_{K} + \delta_1\,,\quad \varphi_2 = \varphi_{K} - \delta_2\,,\nonumber\\
\mbox{configuration b:} &&\varphi_1 = \varphi_{K} - \delta_1\,,\quad \varphi_2 = \varphi_{K} + \delta_2\,.\label{phi12}
\end{eqnarray}
Here $\delta_1$ and $\delta_2$ are the internal angles of the triangle with the sides $\varkappa_1$, $\varkappa_2$, $K \equiv |\bK_\perp|$:
\begin{equation}
\delta_1 = \arccos\left({\varkappa_1^2 + K^{2} - \varkappa_2^2 \over 2\varkappa_1 K}\right)\,,
\quad
\delta_2 = \arccos\left({\varkappa_2^2 + K^{2} - \varkappa_1^2 \over 2\varkappa_2 K}\right)\,.
\end{equation}
With these definitions,
the result for the vortex amplitude ${\cal J}$ can be compactly written as \cite{Ivanov:2011kk}
\begin{equation}
{\cal J} = \frac{e^{i(\ell_1 - \ell_2)\varphi_{K}}}{\sin(\delta_1 + \delta_2)}
\left[{\cal M}_{a}\, e^{i (\ell_1 \delta_1 + \ell_2 \delta_2)} + {\cal M}_{b}\, e^{-i (\ell_1 \delta_1 + \ell_2 \delta_2)}\right]\,.\label{J2}
\end{equation}
Notice that the plane-wave amplitudes ${\cal M}_{a}$ and ${\cal M}_{b}$ entering this expression are calculated 
for the two distinct initial momentum configurations shown in Fig.~\ref{fig-2configurations} 
but the same final momenta $k_1'$ and $k_2'$. 
They interfere, and this is a crucial change with respect to the single-Bessel case.
One can view these two configurations as two distinct paths in momentum space to
arrive at the same final state from the initial vortex states.
In a sense, double-Bessel scattering represents the momentum space counterpart
of the Young double-slit experiment \cite{Ivanov:2016jzt,wong2021control}.
Squaring ${\cal J}$, we obtain
\begin{equation}
|{\cal J}|^2 = \frac{1}{\sin^2(\delta_1 + \delta_2)}\left\{|{\cal M}_{a}|^2 + |{\cal M}_{b}|^2
+ 2\Re \left[{\cal M}_{a}{\cal M}_{b}^*\, e^{2i(\ell_1 \delta_1 + \ell_2 \delta_2)}\right]\right\}\,.\label{J3}
\end{equation}
With this expression, the differential cross section \eqref{S-2Bessel-2} demonstrates several features.
First, double-vortex collision offers a new dimension in the final state kinematical analysis. 
Writing $d^2\bk'_{1\perp} d^2\bk'_{2\perp}$ as $d^2\bk'_{1\perp} d^2\bK_{\perp}$, one can fix $\bk'_{1\perp}$
and explore the distribution in $d^2\bK_{\perp}$. Unlike the usual plane wave case, where the total final momentum is fixed 
by the initial state kinematics, $d\sigma_{\rm PW} \propto |{\cal M}|^2 \delta^{(2)}(\bk_{1\perp} + \bk_{2\perp} - \bK_\perp) d^2\bk'_{1\perp} d^2\bK_{\perp}$,
here we observe a non-trivial distribution in $\bK_{\perp}$ within the ring defined by \eqref{ring}.

Second, the behavior of the cross section within this ring reveals remarkable interference fringes,
which arise from the interference term in \eqref{J3}.
As one varies $K$ from $|\varkappa_1 - \varkappa_2|$ to $\varkappa_1 + \varkappa_2$,
the angles $\delta_1$ and $\delta_2$ change, which leads to oscillations of the cross section.
The number, the exact location, and the contrast of these interference fringes depend on the initial Bessel state parameters
as well as on the resolution in the final particles momentum measurement. 

\begin{figure}[h!]
	\centering
	\includegraphics[width=0.45\textwidth]{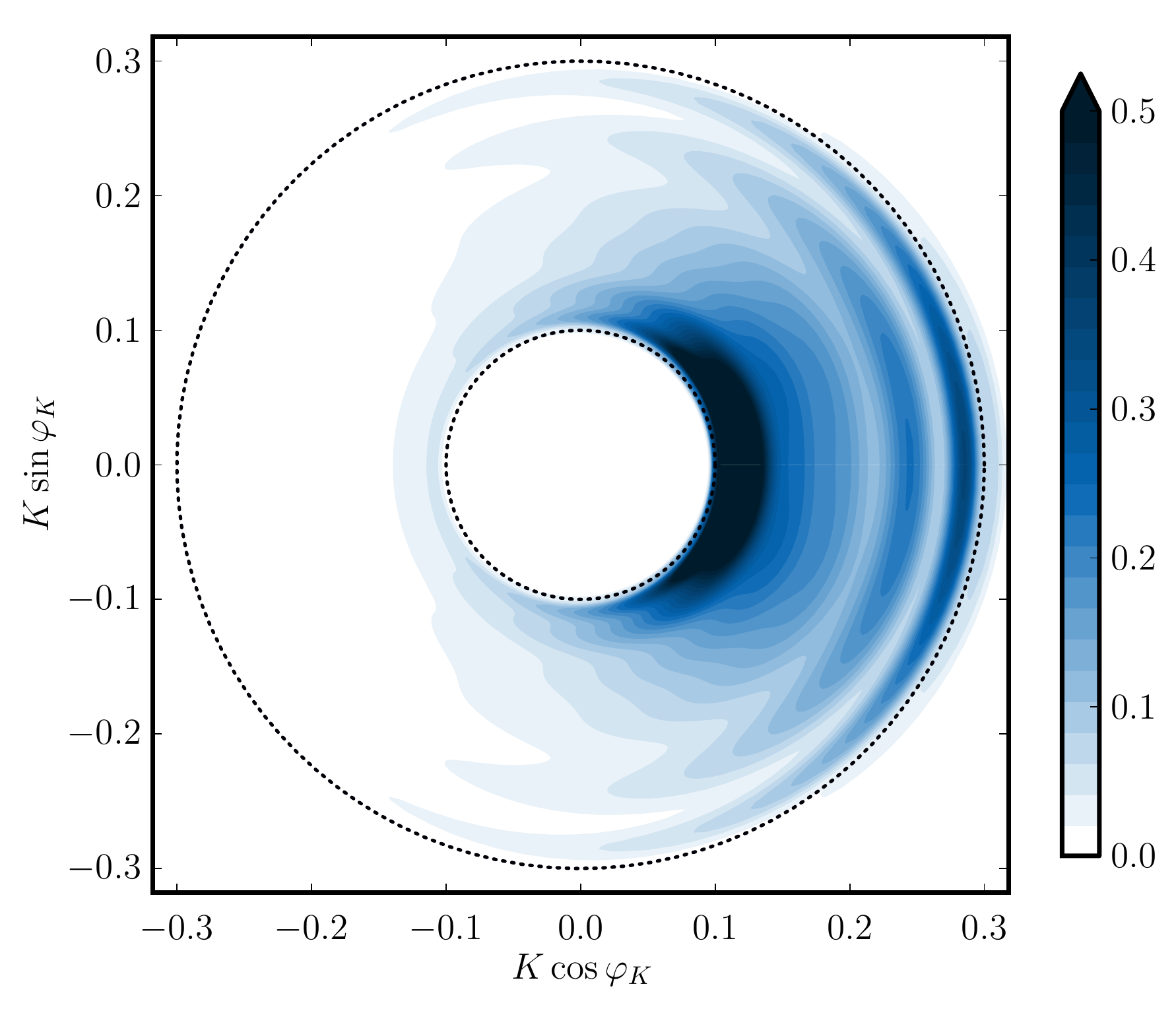}
	\includegraphics[width=0.45\textwidth]{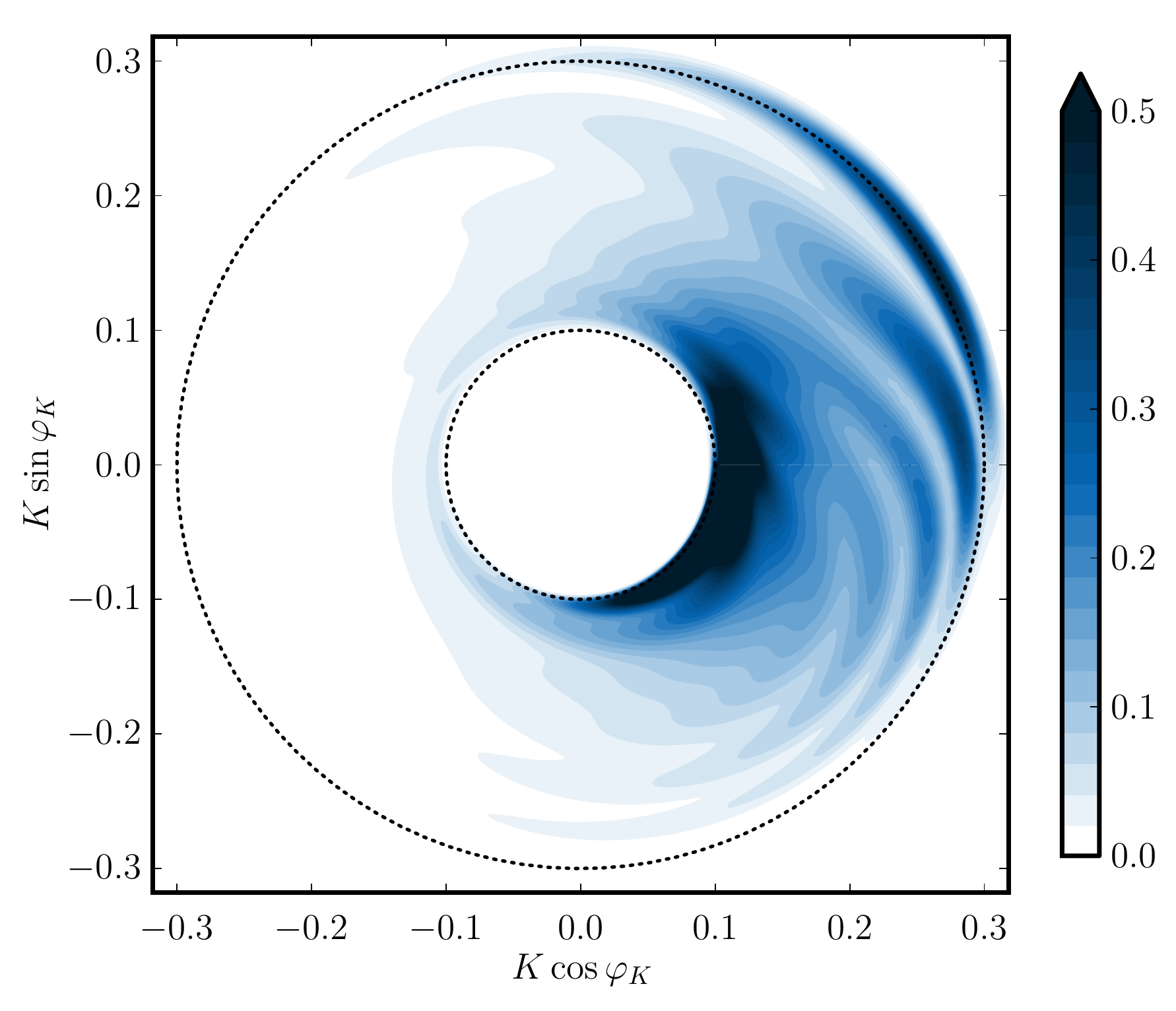}
	{\caption{\label{fig-Moller} The cross section of the elastic electron-electron scattering (arbitrary units, shown in shades)
			in the $\bK_\perp$ plane (shown in MeV) for a fixed $\bk'_{1\perp}$ pointing to the right. 
			Left: the plane wave amplitude is assumed to be real. 
			Right: the plane wave amplitude includes the Coulomb phase which, for visibility
			is enhanced by the factor of 1000. Reproduced from \cite{Ivanov:2016oue} with permission.}}
\end{figure}

Third, the expression \eqref{J3} displays $1/(\sin^2(\delta_1 + \delta_2))$ divergence at the edges of the annular region.
Unlike the single-Bessel counterpart $F(\theta_k, \theta_p, \theta_0)$ in \eqref{F-single}, 
this divergence is not integrable. One can trace this divergence to the non-normalizable nature of the exact Bessel beams 
\cite{Ivanov:2011kk,Ivanov:2016oue}.
With a large but finite radius $R$ of the normalization volume, one can regularize this divergence, 
with a similarly behaving luminosity factor in the denominator. 
However, any realistic beam is certainly localized in the transverse plane.
Therefore, a more sensible approach is to replace the exact Bessel beams with less singular vortex states.
For instance, the computation of the double-twisted M\o{}ller scattering $e^-e^- \to e^-e^- $
reported in \cite{Ivanov:2016oue} adopted the $\varkappa$-smearing procedure \eqref{smearing}.
Fig.~\ref{fig-Moller}, left, shows the numerical results of the $\bK_\perp$-distribution
for the following parameters: $E_1 = 2.1$ MeV, $m_1 = 1/2$, $m_2 = 13/2$, 
$\varkappa_1 = 200$ keV, $\varkappa_2 = 100$ keV with 5\% smearing in each $\varkappa$,
and for the final momentum $|\bk'_{1\perp}| = 500$ keV directed to the right. 
One clearly observes the fringes arising the interference between the two plane-wave amplitudes.
We repeat once again that in plane-wave scattering the entire $\bK_\perp$-distribution would collapse to a single point.

\subsubsection{Probing the overall phase of a scattering amplitude}\label{subsection-phase}

The interference fringes of the double-vortex scattering cross section represent a new observable which cannot be accessed
in the plane wave collisions, for which $\bK$ is fixed.
One may wonder if these fringes can provide any new physical insights.
The answer is in the affirmative: the double-vortex cross section provides access
to a quantity which is not measurable in plane wave scattering, namely, the overall phase of the scattering amplitude
\cite{Ivanov:2012na}.

Consider a plane wave invariant amplitude ${\cal M}$, which can be complex, ${\cal M} = |{\cal M}| e^{i\zeta}$,
and both $|{\cal M}|$ and $\zeta$ can depend on the collision energy and on the scattering angle $\theta$ 
(or the Mandelstam variables $s$ and $t$).
Such a dependence of the overall phase is ubiquitous in elastic hadron scattering \cite{Dremin:2012ke} 
and in hadron photoproduction processes such as $\gamma p \to K^+\Lambda$, 
in which several interfering partial waves involve different intermediate resonances
and, therefore, bear different phases \cite{Sandorfi:2010uv}. 
As a result, the total amplitude has a phase which depends on $\theta$ in a non-trivial way,
but in the usual plane-wave scattering its determination is deemed impossible \cite{Wunderlich:2019imi}.
Another example is the small angle $pp$ or $p\bar p$ scattering at high energies,
where the predominantly imaginary Pomeron exchange amplitude may have a non-trivial
phase evolution, which is usually quantified by the parameter $\rho(s,t) = \Re{\cal M}_{\rm P}/\Im{\cal M}_{\rm P}$.
In the usual scattering setting, this phase can only be accessed in the very narrow $t$-region
of the Coulomb-strong interference \cite{TOTEM:2016lxj} or, at larger $t$, indirectly, by invoking the optical theorem
\cite{LHCForwardPhysicsWorkingGroup:2016ote}.

In addition, multi-photon exchanges in elastic scattering of charged particles generate the so-called Coulomb phase of the amplitude. 
First calculated by Bethe for small-angle elastic proton-nucleus collision \cite{Bethe:1958zz},
it was reevaluated by other authors and sparked some debates in 1960's, 
until the controversy was settled by West and Yennie \cite{West:1968du} with a direct diagrammatic computation.
Later, refined calculations of the Coulomb phase \cite{Kopeliovich:2000ez}
were used to gain novel insights into the electron-nucleus deep inelastic scattering, 
the elastic small-angle $pp/p\bar p$ scattering \cite{Kopeliovich:2001dz,Petrov:2002nt}, and 
elastic scattering of other hadrons \cite{Dremin:2012ke}.

In the traditional plane wave scattering, the cross section $d\sigma/d\cos\theta \propto |{\cal M}|^2$
is blind to the phase evolution $\zeta(\theta)$. In the double-Bessel scattering, 
the two interfering plane-wave amplitudes correspond to the same final but different initial momenta. 
Therefore, they involve different scattering angles $\theta_a$ and $\theta_b$ and, as a consequence,
the two amplitude may have different phases $\zeta_a \not = \zeta_b$.
The interference term in the cross section
\begin{equation}
d\sigma_{int} \propto |{\cal M}_a||{\cal M}_b|\cos\left(2\ell_1\delta_1 + 2\ell_2\delta_2 + \zeta_a - \zeta_b\right)
\end{equation}
allows one to detect $\zeta_a - \zeta_b$ via the distortion of the interference pattern
with respect to the case where ${\cal M}$ is purely real.

The magnitude and the shape of this distortion depend on the process.
For the purpose of illustration, we show in Fig.~\ref{fig-Moller}, right, the same interference fringes for the 
double-Bessel M\o{ller} scattering but now with the Coulomb phase included and enhanced by factor of about 1000 \cite{Ivanov:2016oue}.
Although unrealistic, this example helps visualize the pattern of interference fringe distortion.
One clearly sees that the up-down symmetry which the $\bK_\perp$ distribution exhibited on this plot for the purely real amplitude
is broken by the phase evolution.

The origin of this phase-induced asymmetry can be traced to the fundamental feature of the vortex states: 
their built-in parity breaking, that is, non-invariance under a mirror reflection in the transverse plane
Indeed, upon such reflection, for example $y \to -y$, the topological charge flips the sign $\ell \to -\ell$. 
The intensity distribution $|\psi_{\varkappa,\ell}(\br)|^2$, however, remains invariant because the mirror reflection can be compensated by complex conjugation:
$\psi_{\varkappa,\ell}(\br) = \psi_{\varkappa,-\ell}^*(\br)$.
If one studies the differential cross section of a head-on collision of two such states, 
the mirror reflection can also be compensated by the complex conjugation provided the plane-wave invariant amplitude is real.
If it is not, the compensation does not apply, and one expects to observe a parity-asymmetric
differential distribution. In the above example, let us define by $\varphi'_{12} = \varphi_1' - \varphi_2'$ 
the relative azimuthal angle between the transverse momenta $\bk'_{1\perp}$ and $\bk'_{2\perp}$ of the two final particles.
Then, one can detect an asymmetry with respect to replacement $\varphi'_{12} \to - \varphi'_{12}$,
which is exactly what is seen in Fig.~\ref{fig-Moller}, right.
Quantitatively, this azimuthal asymmetry can be extracted via the quantity $A_\perp = \int d\sigma\sin(\varphi'_1 - \varphi_K) /\int d\sigma$.
For realistic parameters of the elastic electron scattering, $A_\perp \sim 10^{-4}$ was predicted in \cite{Ivanov:2016oue,Karlovets:2016dva,Karlovets:2016jrd}.
Measuring this delicate asymmetry is a challenging but not an impossible task.
Estimates for this asymmetry in high energy vortex $pp$ elastic scattering found in \cite{Karlovets:2016dva,Karlovets:2016jrd}
are much lower due to the high energy of the proton and the expected values of $\varkappa$ which could be realistically achieved for vortex protons.

It was further demonstrated in \cite{Karlovets:2016dva,Karlovets:2016jrd} that the overall phase of the scattering amplitude 
can be probed not only via vortex state scattering but also in the broader setting of collisions of wave packets
with non-trivial phase dependence of their momentum space wave functions $\varphi_i(\bk_i)$.
These include, apart from the already familiar Bessel or LG states, such exotic states as Airy beams \cite{Karlovets:2014wva}, 
which have also been demonstrated experimentally for electrons \cite{voloch2013generation}.
A general theory of how to extract the phase of the scattering amplitude in collision of such wave packets
was constructed and illustrated with examples.
Another possibility explored in \cite{Karlovets:2016dva,Karlovets:2016jrd} was collision
of wave packets with manifestly broken azimuthal symmetry and with a controlled impact parameter,
which can also probe the evolution of $\zeta(\theta)$.

\subsubsection{$s$-channel resonance production}\label{subsubsection-s-channel}

Another unusual kinematical feature of double-vortex scattering is that, even for monochromatic beams, 
the invariant mass of the colliding system is not fixed. 
Once again, this can be most clearly seen in the double-Bessel kinematics.
Pick up any pair of the plane wave components with the transverse momenta $\bk_{i\perp}$ in each of the two colliding Bessel states
and compute their invariant mass: 
\begin{equation}
M_{inv}^2 = (E_1+E_2)^2 - (p_{1z}+p_{2z})^2 - (\bk_{1\perp}+\bk_{2\perp})^2 = (E_1+E_2)^2 - K_z^2 - \bK_{\perp}^2\,.
\label{invmass}
\end{equation}
Since $|\bK_{\perp}|$ is not fixed but resides inside the annular region \eqref{ring},
we conclude that collision of two monochromatic Bessel states is capable of simultaneously producing systems
within a range of invariant masses:
\begin{equation}
(E_1+E_2)^2 - K_z^2 - (\varkappa_1+\varkappa_2)^2 \le M_{inv}^2 \le (E_1+E_2)^2 - K_z^2 - (\varkappa_1-\varkappa_2)^2\,.
\label{invmass2}
\end{equation}
Different circles on the $\bK_\perp$ distribution in Fig.~\ref{fig-Moller} correspond, in fact, to
different invariant masses within this range.

This phenomenon looks even more peculiar if, instead of elastic scattering, we consider $s$-channel production 
of several narrow resonances with close but different masses \cite{Ivanov:2019pdt}.
In the double-Bessel kinematics, the longitudinal momentum of the produced resonance is fixed,
but the transverse momentum varies. The polar angle of production of a resonance of mass $M$
is $\cos\theta = K_z/\sqrt{(E_1+E_2)^2-M^2}$. Thus, two resonances with masses $M$ and $M'$
produced in the same experiment running at the incident particle energies $E_1$ and $E_2$
will be emitted at different polar angles $\theta$ and $\theta'$, which can be reconstructed from the 
momenta of their decay products. Thus, the double-Bessel cross section
possesses a built-in self-analyzing spectrometric feature.
Notice that the cross section for each resonance production in the double-Bessel collision
depends not only on the intrinsic plane wave amplitude given by the Feynman diagrams 
but also on the interference between two plane-wave amplitudes discussed above.
Thus, the relative intensity of the two resonances can be easily varied by adjusting 
their position on the interference fringes.

As a result, one will be able to arrange for a process unthinkable in the plane wave case:
simultaneous $s$-channel production, in monochromatic beam collision, of resonances of different masses,
which are automatically separated in their polar angles. This scheme can become an intriguing alternative
to beam energy scans when studying narrow resonant structures in particle production.
In anticipation of future development of high-energy vortex state generation,
one could verify the atomic spectroscopy counterparts of these predictions,
where close atomic excited states replace the resonances
\cite{Ivanov:2021-delocalized}. 

In addition,
interesting polarization effects were predicted in \cite{Ivanov:2019vxe,Ivanov:2020kcy} 
for the $s$-channel resonance production in double-Bessel $e^+e^-$ or $\gamma\gamma$ collision.
As discussed in Section~\ref{section-vortex-polarization},
the notion of unpolarized cross section becomes ambiguous for vortex state collisions.
If one describes the colliding electron and positron as Bessel states with definite values of the total AM $m_1$ and $m_2$
and with definite helicities $\lambda_1$ and $\lambda_2$, then the unpolarized production cross section
may be defined as an average over all $\lambda_1, \lambda_2 = \pm 1/2$ at fixed $m_1$ and $m_2$.
This definition of the unpolarized double-vortex $e^+e^-$ cross section leads to production 
of vector mesons which are polarized. The reason is that, for fixed $m_1, m_2$, different helicities
lead to distinct interference patterns discussed above. Thus, by adjusting the initial state kinematics,
one can put different helicity combinations on a peak or in a dip of the interference fringes.
As a result, the degree of the final resonance polarization can be as high as 90\%,
with the dominant helicity being $+1$ or $-1$, even in the initial vortex particles are unpolarized \cite{Ivanov:2019vxe,Ivanov:2020kcy}.

It should be mentioned in passing that using vortex states instead of plane waves 
cannot induce processes which are altogether impossible in the plane wave collisions. 
For example, the famous Landau-Yang theorem \cite{Landau:1948kw,Yang:1950rg} forbids production of spin-1 particles in collisions
of two photon. Using vortex photons does not make this process allowed:
one can couple two vortex photons two total AM $J=1$ states, 
but not to the total spin $S=1$ states \cite{Ivanov:2019lgh}.
Two photons can exist in the state $J=1$, and so can a spin-1 particle, 
but the two systems cannot be coupled in a way which respects the Lorentz and gauge invariance and the Bose statistics. 

We summarize the above discussion with a list of key feature of the double-vortex scattering.
\begin{itemize}
	\item 
	Collisions of two vortex states leads to remarkable interference of different pairs of plane-wave components
	scattering into the same final state. For the double-Bessel case, this interference
	takes place between exactly two distinct plane-wave amplitudes.
	\item
	A new dimension for the angular analysis opens up: the total final transverse momentum distribution.
	\item
	The double-vortex scattering gives direct access to the total phase of the scattering amplitude $\zeta(\theta)$,
	which can be detected via azimuthal asymmetry of the total momentum distribution.
	This azimuthal asymmetry, in turn, relies on the fact that the initial vortex states explicitly break parity
	despite having azimuthally symmetric intensity distribution.
	\item
	A double-vortex collision experiment, even with monochromatic beams running at fixed energies, 
	can simultaneous produce $s$-channel resonances with close but different masses. These resonances are
	automatically separated in their polar angles. 
	\item 
	The resonances produced can be polarized in a controlled way even if the colliding vortex $e^+e^-$ or $\gamma\gamma$ 
	pairs are unpolarized.
\end{itemize}

\subsection{General frameworks for wave packet collisions}\label{subsection-Wigner-formalism}

If we deal with collision of two wave packets in pure states described with the momentum space wave functions $\varphi_1(\bk_1)$ and $\varphi_2(\bk_2)$,
we can calculate the scattering matrix element as \cite{Peskin:1995ev}
\begin{equation}
S = \int \frac{d^3\bk_1}{(2\pi)^3} \frac{d^3\bk_2}{(2\pi)^3}\,
\varphi_1(\bk_1) \varphi_2(\bk_2) S_{\rm PW}\left(k_1,k_2; k'_1,k'_2\right) \,. \label{S-packets}
\end{equation}
The normalization factors should be adjusted accordingly.
The final particle energies and angular distributions will depend on how singular $\varphi_i(\bk_i)$ are.
For example, if the initial states are localized wave packets, then $\varphi_i(\bk_i)$ are smooth functions.
The six integrations in \eqref{S-packets} eliminate the four delta functions of $S_{\rm PW}$ completely,
and we expect to see the full six-dimensional final distributions in $\bk'_1$ and $\bk'_2$.
If the initial states are considered to be non--plane-wave monochromatic beams,
then each $\varphi_i(\bk_i)$ contains an additional delta function of the form $\delta(\bk_i^2 + M_i^2 - E_i^2)$.
The energy delta function $\delta(E_1+E_2-E_1'-E_2') = \delta(E)$ can be taken out of the integral,
and it will reduce the dimensionality of the final phase space available.
For the double-Bessel scattering, each $\varphi_i(\bk_i)$ contains two delta functions, and
the cross section becomes differential only in $\bk'_{1\perp}$ and $\bk'_{2\perp}$.
An exposition of this treatment with plane wave and Bessel state limits can be found, 
for example, in Appendix A of \cite{Ivanov:2016oue}.

This treatment can be generalized beyond pure states. 
Scattering theory of arbitrarily shaped, partially coherent beams
was developed in the paraxial approximation back in 1992 in \cite{Kotkin:1992bj} 
and recently extended beyond the paraxial approximation in \cite{Karlovets:2016dva,Karlovets:2016jrd,Karlovets:2020odl}.
This formalism is based on the Wigner function approach to describing initial states,
which was formulated in \cite{Karlovets:2020odl} in relativistic form.
For a pedagogical discussion of the Wigner functions applied to particle collisions, see \cite{Karlovets:2016jrd}.
The Wigner function of a paraxial vortex electron can be found in \cite{Karlovets:2018fof,Karlovets:2020odl}. 

If one considers a pure state with the momentum space wave function $\psi(\bp)$
normalized according to \eqref{norm-lorentz} and accompanies it with the time dependence as
$\psi(\bp,t) = \psi(\bp) e^{-iE(\bp)t}$, then the relativistic Wigner function is defined as
\begin{equation}
n({\br}, {\bp}, t) = \int \frac{d^3\bk}{(2\pi)^3}\, e^{i {\bk} \cdot {\br}} \frac{\psi^*({\bp} - {\bk}/2, t)}{\sqrt{2E({\bp} - {\bk}/2)}} \,
\frac{\psi({\bp} + {\bk}/2, t)}{{\sqrt{2E({\bp} + {\bk}/2)}}} \label{wigner-1}
\end{equation}
and is normalized according to
\begin{equation}
\int \frac{d^3\bp\, d^3 \br}{(2\pi)^3}\, n({\br}, {\bp}, t) = 1\,.
\end{equation}
For a collision process of two initial states, one defines the two-particle correlator
\begin{eqnarray}
\mathcal L ({\bp}_i, {\bk}) = \upsilon ({\bp}_i) \int\, d^4 x\,d^3 \bR\, e^{i{\bk}\cdot {\bR}}\, n_1({\br}, {\bp}_1, t)\, n_2({\br} + {\bR}, {\bp}_2, t),
\label{wigner-2}
\end{eqnarray}
where
\begin{eqnarray}
\upsilon ({\bp}_i) = \frac{\sqrt {({p_1}_{\mu} p_2^{\mu})^2 - m_1^2 m_2^2}}{E_1({\bp}_1) E_2({\bp}_2)}\,,
\quad
E_i({\bp}_i) \equiv \sqrt{{\bp_i}^2 + m_i^2}.\label{wigner-3}
\end{eqnarray}
The amplitude of the scattering process $2\to n_f$ is described with
\begin{equation}
T_{fi}^{(\text{PW})} = \frac{M_{fi}^{(\text{PW})}}{\sqrt{2 E_1 2 E_2 \prod_{f}2E_f}}\,,
\end{equation}
where $M_{fi}^{(\text{PW})}$ is the plane wave invariant amplitude calculated according to the standard Feynman rules.
With these definitions, the generalized cross section can be defined as $d\sigma_{\rm gen} = dW/L$, with
the differential scattering probability
\begin{eqnarray}
dW&=& |S_{fi}|^2\, \prod\limits_{f} V \frac{d^3 \bp_f}{(2\pi)^3} =  
\int \frac{d^3 \bp_1}{(2\pi)^3}\frac{d^3 \bp_2}{(2\pi)^3}\frac{d^3 \bk}{(2\pi)^3}\,\, 
\mathcal L({\bp}_i, {\bk})\, d \sigma({\bp}_i, {\bk}),\nonumber\\
d \sigma({\bp}_i, {\bk}) &=& (2\pi)^4\, \delta \Big [E_1({\bp}_1 + {\bk}/2) + E_2({\bp}_2 - {\bk}/2) - \sum\limits_{f}E_f({\bp}_f) \Big]\, 
\delta^{(3)}\Big({\bp}_1 + {\bp}_2 - \sum\limits_{f}{\bp}_f\Big) \times\nonumber\\
&&\times\ 
T_{fi}^{(\text{PW})}({\bp}_1 + {\bk}/2, {\bp}_2 - {\bk}/2)\, {T_{fi}^{(\text{PW})}}^* ({\bp}_1 - {\bk}/2, {\bp}_2 + {\bk}/2)\, 
\frac{1}{\upsilon({\bp}_i)}\prod\limits_{f} \frac{d^3 \bp_f}{(2\pi)^3},
\label{wigner-4}
\end{eqnarray}
and the luminosity is 
\begin{equation}
L = \int \frac{d^3 \bp_1}{(2\pi)^3}\frac{d^3 \bp_2}{(2\pi)^3}\frac{d^3 \bk}{(2\pi)^3}\,\, \mathcal L ({\bp}_i, {\bk}) 
= \int \frac{d^3 \bp_1}{(2\pi)^3}\frac{d^3 \bp_2}{(2\pi)^3}\,\, d^4x\,\, \upsilon ({\bp}_i)\, n_1 ({\br}, {\bp}_1, t) n_2 ({\br}, {\bp}_2, t).
\label{wigner-5}
\end{equation}
Both $dW$ and $L$ are Lorentz invariant. Notice that, unlike \eqref{lumi-Bes}, the luminosity function is uniquely defined for 
any initial state configuration. For $\bk = 0$, the generalized cross section
turns into the usual plane wave cross section, in which the amplitudes of different momenta do not interfere.

This general representation can be applied to collision of any initial state wave packets. 
A detailed theoretical study of various properties of this expression, its simplification in the paraxial limit,
its decomposition into an incoherent part $d\sigma^{\text{incoh}}$ and the self-interference part $d\sigma^{\text{int}}$,
as well as numerical estimates for various $s$-channel and $t$-channel processes 
with Gaussian and LG vortex states can be found in \cite{Karlovets:2020odl}.

Although this formalism may seem heavy, it was used back in 1992 to successfully describe phenomena in $e^+e^-$ scattering
in which the transverse extent of the beam played crucial role in cross section evaluation \cite{Kotkin:1992bj}.
Its recent reincarnation \cite{Karlovets:2020odl} has not yet been widely adopted in analyses of specific processes.
However this is the method of choice if one wishes to put analysis of generic non--plane-wave states scattering 
and, in particular, vortex state collisions on a firm ground.

	\newpage
	\section{Specific processes with vortex states}\label{section-processes}

\subsection{Vortex states for atomic physics}\label{subsection-atomic}

Although our main focus is on high energy physics applications of vortex states,
it is instructive to briefly review the progress and highlight selected results on low-energy vortex photon and electron interaction with atoms.
The formalism used often relies on the exact relativistic dynamics, 
and the ideas and results in this field may become a source of inspiration for high energy vortex states collisions.

Since twisted light is studied since decades \cite{Allen:1992zz,Paggett:2017,Knyazev-Serbo:2018} and 
has found numerous applications \cite{Torres-applications,andrews2012angular}, 
there is vast literature on interaction of atoms with optical vortices, 
see review \cite{babiker2018atoms} and references therein.
A photon with a non-zero OAM absorbed by an atom can excite multipole transitions suppressed for plane wave photons.
Theoretical investigation of this process began almost two decades ago \cite{babiker2002orbital},
leading to its experimental demonstration in \cite{schmiegelow2016transfer,afanasev2018experimental}.
In these experiments, a single Ca$^{+}$ ion was trapped at the center of Paul trap. 
With its thermal spatial spread reduced to 60 nm, 
much smaller than the laser beam waist, 
it represented a localized probe of the optical vortex.
The laser beam was shaped to various LG modes, with adjustable circular polarization.
Quadrupole transitions between $4^2S_{1/2}$ and $3^2D_{5/2}$ were explored in external magnetic field,
which allowed to spectroscopically resolve 
transitions between all the initial $m_i = \pm 1/2$ and final $m_f = \pm 1/2, \pm 3/2, \pm 5/2$ sublevels. 
In particular, transitions were observed in which a single LG photon provided two units of angular momentum
$m_f - m_i = M = \pm 2$: one from spin, the other from the OAM.

Intensity of transitions strongly depends on the distance $b$ between the ion and the optical vortex axis.
Indeed, if one considers a vortex photon with a well-defined total AM projection $m_\gamma$ 
and shifts its phase singularity axis by distance $b$, one can represent the resulting optical field
as a superposition of vortex states with $z$ projections $M$ defined with respect to the new axis, on which the ion resides.
For $b=0$, $M = m_\gamma$ is the only choice, but for a non-zero $b$, one expects a distribution over $M \not = m_\gamma$, \cite{afanasev2018experimental,duan2019selection}.
For example, for Bessel beams, it is governed by the function $J_{m_\gamma - M}(\varkappa b)$, see Eq.~\eqref{shifted}. 
Thus, vortex light beam contains a rich position-dependent multipole composition, capable of exciting various transitions.

Position-dependent modifications of the atomic transition selection rules 
were further investigated in \cite{solyanik2019excitation,schulz2019modification,schulz2020generalized}.
Excitation \cite{scholz2014absorption,surzhykov2015interaction} 
and ionization of atoms \cite{matula2013atomic,muller2016photoionization,kaneyasu2017limitations} and molecules \cite{peshkov2015ionization} by twisted photons,
including two-photon ionization with a combination of twisted and plane-wave photons \cite{Kosheleva:2020},
as well as scattering of twisted light by atoms \cite{Peshkov:2018},
were also theoretically explored.

Apart from modifying internal atomic transitions, twisted light may affect linear motion of atoms
in a surprising way. An atom placed in the vicinity of an optical vortex axis was predicted in \cite{barnett2013superkick} 
to acquire, upon absorption of a single photon, a transverse momentum much larger than this photon carried.
This unexpectedly large momentum transfer was called in \cite{barnett2013superkick} a ``superkick''.
Semiclassical arguments behind this phenomenon were outlined in \cite{barnett2013superkick} and later
investigated with specific examples in \cite{Afanasev:2020nur}.
Consider a circular path with radius $b$ around the phase singularity with the topological charge $\ell$. 
Along this path, the phase factor $\exp(i\ell\varphi_r)$ generated the total phase 
change of $2\pi \ell$ over the circumference length $2\pi b$. 
The resulting phase gradient, which can be interpreted as the local momentum in the optical vortex \cite{berry2013five},
is equal to $\ell/b$. For small $b$, it can become arbitrarily large and exceed $\varkappa$.
Thus, we arrive at a seemingly paradoxical conclusion: 
photon absorption transfers to an atom a much larger transverse momentum 
than the photon actually carries.
To resolve it, one needs to consider the atom not as a classical entity but as a compact wave packet
of spatial extent $d$ \cite{barnett2013superkick}. Small $d$ implies that, in momentum space, the wavefunction 
extends up to large momenta $\sim 1/d$.
In the initial wave packet, all these plane-wave components balance each other leading to $\langle\bp_\perp\rangle = 0$.
Distorted by the vortex photon absorption, this set of plane wave components produces a nonzero average transverse momentum.
The photon does not supply the large transverse momentum to the atom, it only rearranges the plane-wave components
inside the wave packets which produce the superkick.

The superkick effect deserves mentioning here because, as it was recently suggested in \cite{Afanasev:2020nur,Afanasev:2021fda}, 
similar phenomena may be observable in nuclear and hadronic physics realm provided one can achieve the required extreme focusing
and alignment.
A quantum-field-theoretical description of this effect through a vortex beam collision with a compact Gaussian beam
can be found in \cite{Ivanov:2022sco}, although it relied on a somewhat artificial regularization of the exact Bessel vortex state.
Interestingly, the recent work \cite{Karlovets:2020odl} used the LG vortex state collision with a Gaussian state 
to illustrate the general formalism of arbitrary wave packet scattering,
but it did not explore it in the superkick-related regime. 
More work is needed to see which observable superkick-related effects can be realistically expected in experiment.

Scattering of vortex electrons on atoms and molecules has also received much attention in the past decade.
Interaction of focused vortex electrons with a sample placed in the focal plane of an electron microscope 
is proposed as a novel atomic scale probe of magnetism and other material properties;
experimental progress in this direction was reviewed, for example, in \cite{Bliokh:2017uvr,Lloyd:2017}.
On the theoretical side, one first needs to adapt the general theory
of electron-atom interaction to twisted electron wave packets or beams.
A generalized Born formula for elastic scattering of a generic non-relativistic electron wave packet on a fixed potential
was derived in \cite{Karlovets:2015nva} and was applied to the elastic scattering of twisted electrons
on atoms in \cite{Karlovets:2016uhb}. The cases of an individual scattering center and 
of a macroscopic target with a uniform impact parameter distribution were considered.
For the relativistic twisted electrons, the elastic electron-atom scattering theory was constructed in 
\cite{Serbo:2015kia} using the exact Bessel electron states.
For the polar angle distribution, the sharp forward peak characteristic for plane-wave electrons
turned into the already familiar ring for the Bessel electron as discussed in Section~\ref{subsection-single-Bessel}.
The scattered electron polarization as a function of the scattering angle was also modified
with respect to the plane-wave case. All these features survive after averaging over
impact parameters for a macroscopic target and showed only moderate dependence on the electron energy. 

Extension of this fully relativistic theory beyond the Born approximation, which allows for electron-atom interaction treatment
to all orders in $\alpha_{\rm em}Z$, as described in Section~\ref{subsection-vortex-e}, was done in \cite{Zaytsev:2016gqp}
and then applied to the elastic electron-atom scattering in \cite{kosheleva2018elastic}. A brief summary of several
atomic processes with twisted electrons calculated within this framework can be found in \cite{zaytsev2020atomic}.
Elastic scattering of twisted electrons on molecules were studied in \cite{Maiorova:2018inm}.
Further prospects of vortex electron as a probe of atom and molecule structure
were presented in \cite{tolstikhin2019strong}.

A systematic theoretical study of the inelastic electron scattering on atoms 
was presented in \cite{VanBoxem:2015lta}. Inelastic scattering may lead to excitation of the electronic structure
of the target atom or ion or to its ionization.
OAM conservation and transfer to excited atomic sublevels and the angular distribution
of the scattered electron were studied for incident electrons in Bessel states.
The authors conclude that, upon such collisions, specific states with the magnetic 
quantum number change $\Delta m$ can be excited only if one uses pre- and postselection of the incident 
and scattered electron. In other words, it is not enough to collide a twisted electron
with atom to transfer its OAM to its internal degrees of freedom. 
One must also project the scattered electron onto a specific OAM state or on a subspace of possible states
in order to filter the desired internal transition in the atom.
The necessity of a postselection protocol will be mentioned in Section~\ref{subsection-schemes-scattering}
when discussing
possible ways to generate high-energy vortex states of particles through free-space collisions.

A process in which transfer of the incident electron OAM to the target ion
is unavoidable is twisted electron capture followed by a photon emission.
This radiative recombination process was theoretically explored in \cite{matula2014radiative} 
within the non-relativistic approach and in \cite{Zaytsev:2016gqp} 
in a fully relativistic framework to all orders in $\alpha_{\rm em}Z$.
As expected, the angular distributions and polarization properties of the emitted photon
depended on the vortex state cone opening angle but not on the value of the OAM;
the latter sensitivity arises only for vortex electrons in superposition of different OAM states.

A fully relativistic theory of bremsstrahlung from twisted electrons in the field of heavy nuclei 
was developed in \cite{Groshev:2019uqa}.
The electron-nucleus interaction was, once again, treated to all orders 
in the nuclear binding strength parameter $\alpha_{\rm em}Z$.
The angular distributions of the final-state electron and photon, which were described as plane waves,
as well as the polarization properties were studied.

Inelastic collision of twisted electrons with atoms or molecules can also lead to
impact ionization of atoms and molecules \cite{dhankhar2020electron,mandal2021semirelativistic,singh2021twisted}.
In these $(e,2e)$ processes, the incident vortex electron knocks out
a secondary electron of much lower energy. The calculations relied on distorted wave Born approximation
and suggest that the angular distribution of this secondary electron is a more sensitive probe of the initial vortex state
parameters than the distribution of the primary electron.
This observation may inspire novel ways to explore vortex states in inelastic high energy collisions.

\subsection{Vortex states colliding with ion beams}\label{subsection-ion-beams}

Interaction of vortex electrons and photons with ions
is a new promising tool for ion storage rings
and can enhance the existing physics program of the 
FAIR facility at GSI Helmholtzzentrum for Heavy-Ion Research in Darmstadt \cite{Durante:2019hzd}.
Vortex beam collisions can be realized as the recently proposed Gamma Factory at CERN \cite{Krasny:2015ffb}, 
a facility which will generate a beam of relativistic partially stripped heavy ions 
and which can significantly expand the experimental opportunities
in atomic \cite{Budker:2020zer} and nuclear physics \cite{Budker:2021fts}.

With these prospects in mind, 
the process of radiative recombination of twisted electrons into bound states 
of free propagating, initially bare ions was studied in \cite{maiorova2020structured,Maiorova:2021tvy}
using the relativistic formalism of \cite{Zaytsev:2016gqp}.
Since the ions propagate as a beam, one can ask how the spatial structure of this ion
beam is modified upon capturing an electron which itself had a non-trivial distribution.
This question was explored in detail in \cite{maiorova2020structured} with the example of bare high-$Z$ ions.
The calculations show that population of ionic substates strongly
depends on the ion impact parameter with respect to the incident twisted electron axis. 
This effect produces a structured ion beam, that is, a beam of hydrogen-like ions 
which exhibits a multiple-ring intensity profile 
and a complex spatially-dependent spin and sublevel population patterns.

\begin{figure}[!h]
	\centering
	\includegraphics[width=0.8\textwidth]{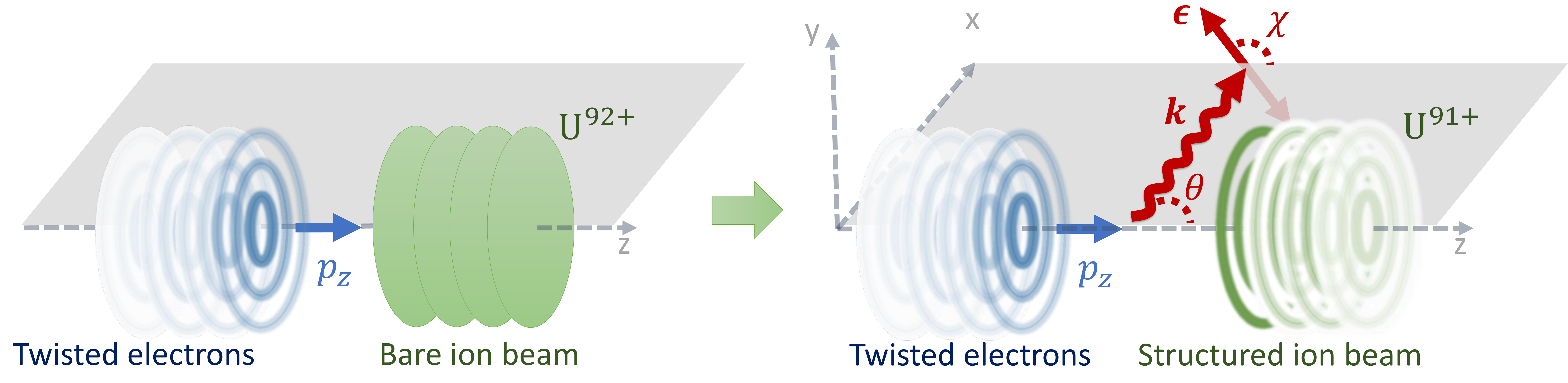}
	\caption{Sequential two-electron radiative recombination of initially bare (finally helium-like) uranium ions. 
		Left: capture of the Bessel electron by a bare uranium ion leads to formation of the structured U$^{91+}$ beam 
		with an inhomogeneous intensity profile and vortex-like spin pattern. 
		Right: capture of a vortex electron from the second bunch affects the linear polarization of the emitted photon.
		Reproduced from \cite{Maiorova:2021tvy}  with permission.}\label{fig-sequential-RR}
\end{figure}

In the follow-up paper \cite{Maiorova:2021tvy}, the authors considered the next step of this process.
The structured hydrogen-like ion beam emerging from the process just described 
can collide with the second bunch of vortex electrons, Fig.~\ref{fig-sequential-RR}.
Radiative recombination can take place again, leading to a helium-like ion with two $1s^2$ electrons.
However this time both initial states have non-trivial spatial and polarization distributions,
which resembles the double-vortex collision.
The first vortex electrons were taken as Bessel states with the kinetic energy 50 keV,
the cone opening angle $\theta_p = 15^\circ$ in the ion rest frame,
the total AM $+1/2$, and helicity $+1/2$. For the second vortex electron bunch of the same energy 50 keV,
several options for the opening angle and the total AM were investigated.
The key observable was the polarization state of the photon emitted at the second recombination;
its Stokes parameters were found to be remarkably sensitive to the total AM state
of the second vortex electron even for small cone opening angle.

The authors stress that experimental investigation of these phenomena
is feasible today at GSI and FAIR in Darmstadt
or in future, at the Gamma Factory at CERN \cite{Budker:2020zer}.
The beam of hydrogen-like ions can be separated from the parent bare ion beam
by bending magnets. Thus, one can explore in a clean environment
the dynamics of structured ion beams in magnetic fields, which is expected to be highly non-trivial.
One can also use it as tool in atomic and solid-state physics.

In what concerns twisted photon collisions with relativistic ion beams,
the future Gamma Factory at CERN, through light scattering on highly relativistic ions,
will offer unique opportunities for atomic physics \cite{Budker:2020zer}.
Indeed, heavy ions with one or few electrons can be accelerated to very high energies and stored in SPS or in the main LHC ring.
For example, helium-like heavy ions will possess gamma factors $\gamma = 96.3$ (SPS) and $\gamma=2800$ (LHC).
Optical or UV photons in the laboratory frame become X-ray photons in the ion rest frame
and can excite keV-range bound-state transitions.
For the helium-like ion Xe$^{52+}$, the transition from the ground state $1s^2$ to the excited state $1s2p$
requires energy 30.6 keV. If the ion is stored at the LHC, this transition can be excited by a modest 5.5 eV photon
colliding head-on. Deexcitation of this state produces a photon which,
if emitted backwards, can reach the energy of 170 MeV in the laboratory frame.
Such a high-energy backscattered photon could be also obtained by the usual Compton scattering off ultrarelativistic electron.
However ions, due to their sheer size and the resonant nature of the photon-ion collision, 
offer resonant scattering cross sections several orders of magnitude larger than the Compton cross section.
This is why laser light backscattering from ultrarelativistic ions can produce a brilliant, well collimated, well controlled beam
of gamma photons. 

A theory describing the properties of the high energy photons resonantly produced in such collisions
was developed in \cite{Tanaka:2021vpc,Serbo:2021cps}. 
Resonant electric dipole transitions $nS \to n' P \to nS$ with different sublevel configurations
were used.
For the plane wave incident photon, the angular distribution and the polarization properties of the final photons
were studied as functions of the energy and the crossing angle between the two beams.
For the Bessel vortex photons, the results also depended on the conical transverse momentum $\varkappa$, the incident photon total AM projection $m_i$,
as well as on the position $b$ of the atom with respect to the vortex photon axis.
Moreover, just as for the Compton backscattering of twisted light from high-energy electron, 
which was suggested in \cite{Jentschura:2010ap,Jentschura:2011ih} as a means of generation of high-energy vortex photons,
the authors of \cite{Serbo:2021cps} and the follow-up publication \cite{Karlovets:2021gcm}
investigated the phase structure of the outgoing radiation.
They found that, for $\varkappa b \ll 1$, the final photons produced by a single $m_i$ mode
can be well approximated by vortex states with $m_f = -m_i$ (no ionic spin-flip)
or $m_f = -m_i + \Delta M$ if the magnetic quantum number of the electron changes by $\Delta M$.
For an ion placed off-axis, at $\varkappa b \sim 1$, a few side modes are also produced,
but the modes with $|m_f + m_i| > 2$ are suppressed down to sub-percent level.
The authors conclude that resonant scattering from ultrarelativistic partially stripped ions
can become a powerful tool for production of high-intensity vortex gamma beams 
with moderate values of the total AM projection.

In a very recent work \cite{Tashiro:2022qrv},
it was suggested that the incident photons need not be twisted.
The authors considered two-photon transitions induced by usual laser photons
in partially stripped ions within the same Gamma Factory kinematics
and showed that deexcitation will lead to emission of a high energy photon carrying an OAM.

\subsection{Inverse Compton scattering}\label{subsection-Compton}

\begin{figure}[!h]
	\centering
	\includegraphics[height=2cm]{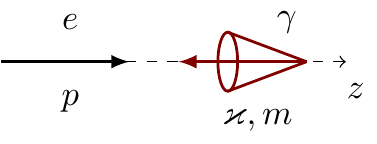}\hspace{2cm}
	\includegraphics[height=2cm]{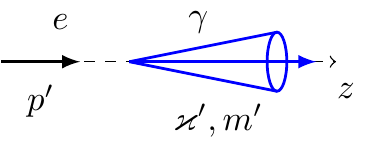}
	\caption{Schematic illustration of the Compton backscattering
		of low-energy twisted photons on ultrarelativistic electrons.
		The left and right diagrams show the configurations before and after the collision.
	}\label{fig-Compton-1}
\end{figure}

Inverse Compton scattering in various regimes, with either the incident photon or the electron taken in a vortex state,
is by far the most studied QED high-energy process with vortex particles.
This activity was kick-started in 2010 in publications
\cite{Jentschura:2010ap,Jentschura:2011ih},
where the inverse Compton scattering of twisted optical photons on GeV energy range electrons was considered, Fig.~\ref{fig-Compton-1}.
The optical twisted photon was described as a Bessel state with a non-zero OAM,
while the counterpropagating electron was taken as a plane wave. 
As is well known, for the ultrarelativistic initial electron with energy $E_e \gg m_e$, $\gamma_e \gg 1$,
the differential cross section of the Compton scattering has a sharp maximum in the backscattering region with $\theta' \lsim 1/\gamma_e$.
In the strictly backward scattering, the final photon carries a significant fraction of the initial electron energy,
$\omega' = 4\gamma_e^2\omega /(1+x) = xE_e/(1+x)$, where $x \approx 4\omega E_e/m_e^2$ and can be of order 1.
As a result, if the final electron is projected on a plane wave state propagating along the same axis $z$, 
then the final photon is up-converted into the MeV or GeV energy range while keeping its OAM and its conical transverse momentum $\varkappa$.
Thus, this process emerges as a means to generate high-energy, well collimated vortex photons, 
which then can be used in nuclear and particle physics experiments.
If scattering is not strictly backward, the final photon still retains its vortex structure due to the very small
scattering angles involved. A momentary disagreement between \cite{Jentschura:2010ap,Jentschura:2011ih} and \cite{Ivanov:2011kk}
regarding the final OAM distribution was resolved in \cite{Ivanov:2011bv} by switching from 
the exact (non-normalizable) Bessel states to more physically relevant $\varkappa$-smeared Bessel distributions.

Compton scattering of twisted photons on ultrarelativistic electrons can also be analyzed 
within classical electrodynamics, provided the parameter $x \ll 1$. 
Ref.~\cite{Petrillo:2016} reported such a study which used realistic parameters
of the electron bunch with energy of 25 MeV and of the Ti:sapphire laser pulses of picosecond duration.
The numerical results confirmed that the OAM of the laser pulse is transferred to the X-ray backscattered photons.
Concrete proposals for experimental verification of this predictions were outlined. 

Ref.~\cite{Stock:2015yha} analyzed this process for free or quasi-free electrons at rest 
(that is, in the Thomson scattering limit) and with the vortex photons described as Bessel states. 
The expressions for the angular distribution and the polarization properties of the scattered photons were derived
both for the cases of the incident photon in a pure OAM state or in a superposition of two OAM states 
which differ by $\Delta m$. In the latter case, when $\Delta m = 1$ or 2, one observes a non-trivial
azimuthal modulation of the differential cross section. Interestingly, for $\Delta m \ge 3$,
this modulation disappears as a result of the dipole selection rule which applies 
within the dipole interaction approximation. The first-order nondipole correction
was also computed, and although it generates a nonzero cross section for $\Delta m = 3$,
the result is strongly suppressed by the small recoil parameter $\omega/m_e$. 
Although these results are more of interest to atomic physics rather than high energy physics,
it is worth mentioning that the authors used density matrix formalism to describe the OAM and polarization spaces
of the photons.  

\begin{figure}[!h]
	\centering
	\includegraphics[height=2cm]{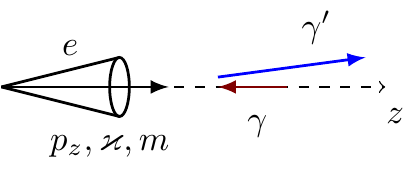}
	\caption{Compton backscattering of low-energy photons on ultrarelativistic twisted electrons.
	}\label{fig-Compton-2}
\end{figure}

A different perspective on the inverse Compton scattering was proposed in \cite{Seipt:2014bxa}. 
Here, the high-energy electron was assumed to be prepared in a vortex state with the conical opening angle $\theta_e$, 
while the optical photons were described as plane waves (a valid assumption for laser beam focal spot
much wider than the waist of the focused twisted electron beam), see Fig.~\ref{fig-Compton-2}.
The final photons were considered as plane waves.
Their spectral and angular distributions were found to depend in a remarkable way on the properties of the initial vortex electrons.
Indeed, consider first the plane-wave case in the regime $x = 4\omega E_e/m_e^2 \ll 1$.
The final photon energy is upshifted to $\omega' = 4\gamma_e^2\omega$ in the strictly backward scattering.
If the photon is observed at a small polar angle $\theta'$ with respect to axis $z$,
the photon energy is also fixed and is given by $\omega' = 4\gamma^2\omega/(1+ \gamma^2\theta^{\prime 2})$.
Now consider the inverse Compton scattering on the high-energy vortex electron with the cone opening angle $\theta_e$.
If one detects a photon at a fixed angle $\theta'$, then it may have been backscattered from any plane wave component
of the vortex electron. The scattering angle measured as the angle between the plane-wave electron and the photon detection angle
ranges from $|\theta'-\theta_e|$ to $\theta' + \theta_e$, just as discussed in Section~\ref{subsection-single-Bessel}. Therefore,
for a fixed photon detection angle $\theta'$, one observes a whole range of photon energies:
\begin{equation}
\frac{4\gamma^2\omega}{1+\gamma^2(\theta'+\theta_e)^2} \le \omega' \le \frac{4\gamma^2\omega}{1+\gamma^2(\theta'-\theta_e)^2}\,.
\label{Compton-energies}
\end{equation}
Clearly, for $\theta'$ of the same order as $\theta_e$, one can observe significantly different
photon spectra if compared with the plane wave electrons.

\begin{figure}[!h]
	\centering
	\includegraphics[width=0.6\textwidth]{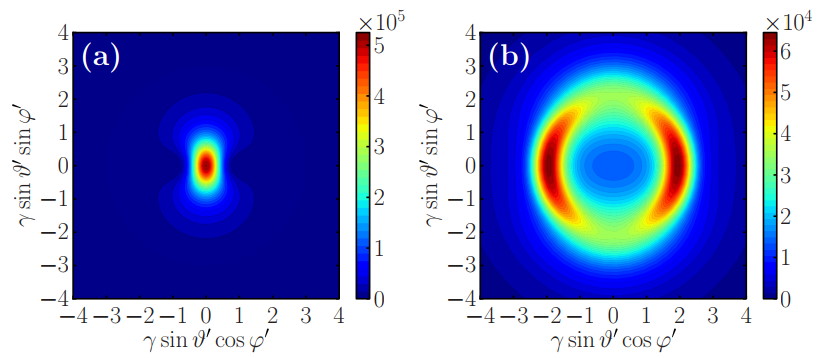}
	\caption{Angle-differential cross section for the plane-wave electrons (left panel) 
		and twisted electrons (right panel) with $\theta_e = 0.05$. 
		In both cases, the electron energy is $E_e = 20.5$ MeV ($\gamma = 40$), 
		while the laser light has frequency $\omega = 1.55$ eV and is linearly polarized along axis $x$.
		Reproduced from \cite{Seipt:2014bxa} with permission.
	}\label{fig-Compton-angular}
\end{figure}
The energy-integrated angular distribution of the X-ray photons showed the annular structure 
generated by the vortex electrons modulated by the polarization-dependent cross section, 
Fig.~\ref{fig-Compton-angular}. 
If the initial vortex electron is prepared in a superposition of different OAM states,
the azimuthal distribution of the final photons exhibited remarkable multi-petal structures.
These features explain why the process was proposed in \cite{Seipt:2014bxa} 
as a way to produce structured X-rays, which can enhance the capabilities of X-ray imaging techniques.

The series of papers \cite{Sherwin:2017wxx,Sherwin:2017vsf,Sherwin:2020vyq} addressed various regimes 
of the Compton scattering, in which the initial vortex photon already carried a significant energy
of the order of electron mass.
The initial electron was taken as a plane wave and was considered to be at rest \cite{Sherwin:2017wxx}
or moving with a (moderately) relativistic speed \cite{Sherwin:2020vyq}.
In both cases, a very detailed study of the angular distribution and the Stokes parameters 
of the final photon was presented, and differences with respect to the low-energy twisted photon case were highlighted.
Both studies also included an analysis of the case when the high energy twisted photon 
is in the superposition of different OAM states.

Ref.~\cite{Sherwin:2017vsf} theoretically explored the double Compton scattering
of a high energy initial photon with $\omega \sim m_e$ prepared in a Bessel state with the electron at rest. 
In double Compton scattering, the single photon colliding with the electron leads to production of two final photons.
This is a linear QED process, although of higher-order, which is not to be confused 
with non-linear Compton effect in intense laser driven by multiple photons absorption.
With three particles in the final state, one obtains triple-differential cross section $d\sigma/(dE_1 d\Omega_1 d\Omega_2)$,
which depends on the first photon energy and on the angles of the two final photons.
For the plane-wave case, the energy of the second electron $E_2$ is uniquely defined once these variables are fixed.
However, for the vortex electron, the situation changes in a way similar to what was described above, 
and one obtains a range of values $E_2$, which depends on the kinematical parameters.
The angular distributions were investigated in detail in \cite{Sherwin:2017vsf}; they showed 
non-trivial dependence on the cone opening angle. Also, as expected, 
when the twisted photon is assumed to be in the superposition of different OAM states, 
additional substantial changes are predicted.

Compton scattering of high-energy vortex photons with electrons at rest was also investigated
theoretically in \cite{Maruyama:2017ptl}. Compton scattering was proposed in this work
as a diagnostic tool which could help verify that the high-energy photon beam carries a non-zero OAM.
In contrast with previous works on vortex Compton scattering,
this study assumed the incident vortex photon to be in a LG mode rather than a Bessel state.
The numerical calculations were performed for the following benchmark parameters of the vortex photon beam:
$\omega = 500$ keV, the LG beam waist width is $w_0 = 25$ pm, which is ten times larger than the photon wavelength
and allows the authors to rely on the paraxial approximation.
The observables computed were the angular distributions and the energy shift $\Delta E$,
the difference between the scattered photon energies for the LG incident beam and for the plane wave photon
with the same initial energy.
The energy shift $\Delta E$ and the angular distributions show clear dependence on the two indices of the LG modes,
the OAM and the mode number. 

For very high intensities of the laser pulses, the process of Compton scattering enters the nonlinear regime \cite{DiPiazza:2011tq}.
The key parameter here is $\xi = eE_0/(m_ec \omega_0)$, where $E_0$ and $\omega_0$
are the electric field and the frequency of the incident laser light.
The works described above address the Compton scattering in the perturbative regime, $\xi \ll 1$,
in which case the assumption of a single photon absorption is valid.
In the moderately nonlinear regime, $\xi \sim 1$, known also as the higher harmonic generation regime,
the electron can absorb several photons before emitting 
a single backscattered photon. If these photons are circularly polarized,
the electron accumulates a total AM of several units of $\hbar$, which can be transferred, in the form of OAM, 
to the single MeV range gamma photon.
This idea was investigated theoretically in \cite{Taira:2017,Taira:2018onh}.
Numerical analysis confirmed the qualitative picture. 
In fact, the authors argued that the previously published experimental results at BNL \cite{Sakai:2015mra}
reporting backscattered 13 keV X-rays with ring-like intensity profile
can be readily explained as the result of the OAM generation.

When the parameter $\xi \gg 1$, the electron can absorb a large number of photons
before emitting a single backscattering gamma photons.
This regime of highly non-linear Compton scattering in ultraintense laser pulses
leading to emission of gamma photons with large OAM was investigated in \cite{Chen:2018tkb}
(the role of a non-zero incidence angle was studied in \cite{Liu:2020uhv}).
The process was calculated in the semiclassical approach: 
the electron motion in the intense laser field was treated classically
while the photon emission and the possible $e^+e^-$ pair production
were calculated via Monte Carlo QED approach. This allowed the authors,
at the same time, to take into account the radiation reaction on the electrons
and to focus on the AM of the entire electron or photon beam rather than on the OAM of individual quanta.
As for the incident laser pulse, two option were considered:
a circularly polarized Gaussian pulse or a LG pulse carrying one unit of the OAM.
The numerical results showed that backscattered gamma beam carries 
about 70\% of the total absorbed AM; the rest is retained by the electron beam. 
Moreover, if was found that the incident LG pulse generated a larger final beam OAM
than the incident circularly polarized Gaussian pulse.

The average OAM carried away by each backscattered photon was extremely large, of the order of $10^4$ to $10^5$ units of $\hbar$.
However, in the highly non-linear Compton scattering regime, the formation length of the final radiation
is $\xi$ times smaller than the laser wavelength. As the authors of \cite{Chen:2018tkb} repeatedly stress, 
the resulting gamma radiation is just an incoherent sum of individual non-vortex photons.
The large final beam OAM, defined with respect to the beam axis, is a collective effect of the beam as a whole,
not a manifestation of the vortex nature of individual photons.

The same group also explored Compton scattering of twisted light in the LG mode with $\ell = 1$ 
in the moderately non-linear regime, with $\xi = 1$ \cite{Chen:2019}. 
The electron motion was again treated classically, and two options were considered: a single electron, 
or an compact electron bunch with 20,000 electrons.
The results of the calculations show that, for a single classical electron, the spin of the incident laser pulse
can be transferred to the backscattered photon angular momentum, but not its OAM.
Indeed, the trajectory of a classical electron is determined mostly by the polarization of the driving laser field,
and not the OAM, which is a global feature of the optical vortex.
One would need to switch to the fully quantum regime and consider the electron wave packet
of the size of the optical vortex in order to probe this global structure.
However if one considers an electron bunch of the size of the optical vortex,
then, even for classical electron motion, it can collectively backscatter a vortex gamma photon 
with large energy and in a controlled OAM state.

Head-on collisions of ultrarelativistic electron bunches with the so-called Bessel-Bessel laser bullets of very high intensity 
(ultrashort and tightly focused analogues of Bessel beams) can also lead to production 
of a vortex $\gamma$ pulse via nonlinear Compton scattering. 
Semiclassical Monte Carlo simulations of this process were reported in \cite{liu2020vortex}. 

An ingenious, all-optical scheme based on non-linear Compton backscattering was recently proposed in \cite{zhang_zhao:2021}.
An extremely intense circularly polarized laser pulse irradiates a micro-channel target, drags out electrons from
the channel wall, accelerates them to high energies (hundreds of MeV) and, at the same time, transfers its spin angular momentum 
to this electron beam. At the second stage, the same driver laser pulse is reflected from a fan-foil, 
which reshapes its wave fronts and converts the Gaussian pulse into a vortex laser pulse. 
Upon reflection, this intense vortex light pulse collides with the energetic electron pulse
and gets upscattered into the MeV energy range, taking a very large OAM from the electrons.
The very detailed numerical study reported in \cite{zhang_zhao:2021} was based on
the 3D particle-in-cell simulations, with QED effects computed via a Monte Carlo algorithm,
which included radiation reaction and the feedback between the plasma and photon emission.
The parameters of the pulse were chosen to ensure a moderately non-linear regime for the Compton scattering,
with $\xi = 0.55$. As a result, a high brilliance, multi-MeV gamma beam was observed in the simulations,
with the total beam AM of the order $10^6\hbar$ per photon.
Although the resulting gamma beam, as a whole, possesses a very large OAM which can exert significant torque on a target,
it is not clear whether individual photons can be considered in a vortex state.
The authors of \cite{zhang_zhao:2021} carefully label the resulting AM as the beam angular momentum
and do not discuss the issue of coherence or the properties of individual gamma photons.

\subsection{Other QED processes}

$e^-_{\rm tw}e^-_{\rm tw}\to e^-e^-$.
Elastic electron-electron scattering (M\o{}ller scattering) 
with two incoming Bessel electrons investigated in \cite{Ivanov:2016oue} was 
already covered in Section~\ref{subsection-double-vortex}.
Here we just reiterate that collision of two initial particles in vortex states
induces novel interference features which can be detected in the 
angular distributions of the final particles.
These interference features are sensitive to the AM state of 
the vortex electrons and give access to quantities which cannot be measured in the plane wave scattering,
such as the scattering angle evolution of the overall phase.

$e^-_{\rm tw}\leftrightarrow e^-_{\rm tw}\gamma_{\rm tw}$.
The electron moving in constant magnetic field exhibits discrete energy levels (Landau levels)
and can jump between these levels by absorbing or emitting photons.
In \cite{Zhang-Xu:2020}, this process was theoretically studied under the assumption that the 
microwave photons are themselves in a vortex state.
Transitions between levels were calculated for non-relativistic and relativistic electrons,
both in empty space and inside a matching microwave cavity, which can selectively enhance
or suppress certain OAM transitions.
The method was proposed as yet another means of generating vortex electrons.

Transitions between electron Landau levels was also analyzed in \cite{maruyama2022generation}
as a potential source of vortex photons. The authors had in mind extremely strong magnetic fields
above $10^{12}$~G generated by neutrons stars and magnetars, and asked whether
electron motion in these fields can generate vortex X or gamma rays.
Although the calculations confirm that the emitted photons can carry the OAM,
it is not clear how it could be detected.

$e^+_{\rm tw}e^- \to \gamma\gamma$.
Two-photon annihilation of the twisted positron with a low-energy plane-wave electron
was examined in \cite{Sherwin:2018dah}. The incident positron was treated as a Bessel state,
and two benchmark kinetic energies were considered: 1 keV and 300 keV.
The angular distributions of the final photons
were calculated as functions of the cone opening angle $\theta_0$.
They exhibited the expected cone-generated departures from the usual plane-wave process 
but became significant only for sufficiently large opening angles, above 10$^\circ$.
However, it was found that polarization-resolved detection of the photons can strongly improve
the visibility of the vortex cone related effects.
Indeed, if one considers the plane-wave $e^+e^-$ annihilation into photons in linear polarization states,
then, in the low-energy limit, the outgoing photons must be in the orthogonal polarization states \cite{Peskin:1995ev}.
For the vortex positron, the cone structure modifies this condition,
and the cross section with the parallel polarizations becomes non-zero and large,
in line with our discussion in Section~\ref{subsection-single-Bessel}. 
For low-energy positrons, modifications are limited to the small forward and backward cone with the polar angle of 
at most a few degrees, while for 300 keV twisted positrons the changes are significant
up to twice the vortex cone opening angle.

$\gamma \gamma \to \gamma_{\rm tw}\gamma$.
Yet another mechanism to emit twisted light was proposed in \cite{Mendonca:2017tdw}.
An intense focused laser pulse propagating in vacuum along the axis of a static magnetic wiggler
was shown to be a source of vortex photons in well-defined Laguerre-Gaussian states, 
in similarity with the vortex undulator radiation emitted by electron beams, see Section~\ref{subsection-undulator}.
The OAM structure of the emitted photons is determined by the wiggler itself,
while their frequency depends on the group velocity of the laser pulse and can be much
higher than that of the incident photons.

$e_{\rm tw}A \to e^-\gamma A$.
A fully relativistic theory of bremsstrahlung from twisted electrons in the field of heavy nuclei 
was developed in \cite{Groshev:2019uqa}.
The electron-nucleus interaction was treated to all orders 
in the nuclear binding strength parameter $\alpha_{\rm em}Z$, as outlined in Section~\ref{subsection-vortex-e}.
The angular distributions of the final-state electron and photon, which were described as plane waves,
as well as the polarization properties were studied.

$\gamma_{\rm tw}\gamma_{\rm tw}\to e_{\rm tw}^+e_{\rm tw}^-$.
The Breit-Wheeler process of $e^+e^-$ pair production in collision of two energetic vortex photons
was studied in \cite{Bu:2021ebc}. 
Motivated by the continuing progress towards production of vortex gamma photons in the MeV range,
the authors asked whether collision of such photons can lead to the OAM transfer to
the final electron-positron pair.
When exploring this question, the authors took a rather unique approach and described all four particles
in this process via the Bessel vortex states. As a result, instead of the angular distribution, 
the cross section was presented as a distribution over the longitudinal momenta $p'_{1z} = -p'_{2z}$, 
the conical momenta $\varkappa'_1 =  |\bp'_{1,\perp}|$,
$\varkappa'_2 =  |\bp'_{2,\perp}|$ (or, alternatively, the final cone opening angles $\theta'_1$, $\theta'_2$) 
as well as the values of the final particle angular momentum.
As a benchmark example, the authors took the initial photons with energies $\omega_1 = \omega_2 = 5$ MeV,
longitudinal momenta $k_{1z} = -k_{2z} = 4$ MeV, the total AM $z$-projections $m_1 = 3$, $m_2 = -2$,
with various photon helicity choices. The numerical results displayed regularities
in $p'_{iz}$ and $\varkappa'_i$ consistent with the interference fringes discussed in Section~\ref{subsection-double-vortex},
as well as non-trivial distributions in $\theta'_i$ and the final particles OAM.
The authors did not discuss the fact that the two final particles are momentum-entangled,
preferring instead to focus on the distributions of only one final particle,
implicitly assuming integration over the kinematics of the second final particle.
The very tight focusing of the vortex photons assumed in this work, with $\theta_0 = \arccos(4/5) = 37^\circ$,
and the unavoidable sub-picometer alignment of their focal points
seem to be beyond the possibilities not only of the present-day technology 
but also of the schemes for MeV range vortex photon generation being now discussed.

$\gamma_{\rm tw}A \to e_{\rm tw}^+e_{\rm tw}^-A$ and $e_{\rm (tw)}A \to e^-_{\rm tw}\gamma_{\rm tw}A$.
The same research group also investigated the processes of high-energy vortex photon splitting 
into an $e^+e^-$ pair \cite{Lei:2021eqe} and bremsstrahlung in the Coulomb field of a heavy nucleus
from an energetic plane wave \cite{Wang:2022hef} or a vortex electron \cite{wang2022triple}.
The calculations were done within the same formalism, with both final state particles described as vortex states.
For the $e^+e^-$ production \cite{Lei:2021eqe}, the same benchmark photon parameters were as above. 
The nucleus was assumed to be positioned exactly on the vortex photon axis,
which will certainly represent an additional challenge to a possible future experiment.
The numerical results confirmed the overall conclusion that, under these conditions,
the initial OAM is transferred to the produced $e^+e^-$ pair, with a helicity-dependent
distribution over the final particle OAM.
This process was proposed by the authors as a way to generate vortex positrons, 
but with all the challenges mentioned above, it is unclear whether it can indeed be practical.
For bremsstrahlung from plane the wave electron with energies up to 1 GeV \cite{Wang:2022hef},
the nucleus was again considered on axis, which allowed to observe regularities
in the final particles OAM and cone opening angle distributions driven by the total AM conservation.
In contrast, in \cite{wang2022triple}, the incident multi-MeV electron was assumed
to be a vortex wave packet rather than a pure Bessel state.
In addition, a non-zero transverse impact parameter $\bb$ of the nucleus with respect
to the vortex axis was introduced, and the cross section was averaged over a range of $\bb$.
This addition renders the calculations more realistic.

\subsection{Vortex light for plasma-wakefield acceleration}

When an intense vortex laser pulse propagates in plasma, it can induce several effects
of interest to particle physics, either as a means to generate
high-energy vortex photons or as a tool for laser-plasma wakefield acceleration of charged particles.

For example, when a very intense laser pulse, with intensity above $10^{23}$ W/cm$^2$,
interacts with a dense plasma target, not only does it accelerate the electrons but it also 
leads to multiple gamma photon emission, with energies extending to tens of MeV, 
as well as $e^+e^-$ pair production processes \cite{ridgers2012dense}. 
If the driving laser pulse is in a circularly polarized LG mode,
then both its spin and the OAM can be transferred to the electrons, which in turn
transfer a part of their angular momentum to the emitted gamma photons.
In Ref.~\cite{Liu-Shen:2016}, this process was theoretically analyzed using 3D QED particle-in-cell simulations
for various values of intensity and the OAM of the driving laser pulse.
A short intense pulse of gamma radiation with an adjustable OAM was predicted.

One can also employ two ultra-intense 
circularly polarized laser pulses counter-propagating in plasma with near critical density \cite{Zhu:2018und}.
Through nonlinear Compton scattering and multiphoton Breit-Wheeler process,
one can generate a pulse of gamma photons and a dense bunch of GeV scale positrons.
Numerical simulations show that both pulses carry large beam angular momentum, up to $10^7\hbar$ per quantum,
which can be controlled through laser-plasma interaction.

A variation of this all-optical scheme relying on a LG and a Gaussian ultra-intense laser pulses was very recently analyzed in \cite{zhao2022all}.
Due to the unique structure of the twisted laser pulse, the positrons generated through the non-linear processes
are confined by the radial electric fields and experience phase-locked acceleration by the longitudinal electric field.
3D simulations revealed generation of dense sub-femtosecond quasi-monoenergetic GeV positron bunches,
with tunable energy and the beam angular momentum. Thus, twisted-laser-plasma acceleration scheme emerges 
as a versatile, highly controllable high-energy positron source.

Plasma is also capable of dramatically amplifying the intensity of an OAM pulse \cite{vieira2016amplification,vieira2016high}.
A long pump laser pulse and a counterpropagating seed OAM laser pulse can interact inside a plasma cell
via stimulated Raman backscattering, leading to the energy transfer from the pump to the seed pulse. 
Theoretical analysis and particle-in-cell simulations presented in \cite{vieira2016amplification,vieira2016high}
show that a vortex light beam with petawatt peak power can emerge from this process.

Such an intense vortex laser pulse can be shot an another plasma cell and, thanks to its hollow structure, 
can lead to doughnut-shaped wakefield configuration and to charged particle acceleration in novel regimes.
Self-injecting of electrons and strong focusing forces for positrons reaching 1000 GV/m were revealed 
by theoretical analysis and particle-in-cell simulations \cite{Vieira:2014nta}.
An electron beam accelerated in this way can also be polarized \cite{Wu:2019suy}.
Efficient trapping in the center of the LG pulse and subsequent acceleration 
of protons and ions \cite{zhang2014proton,wang2015hollow} and positive muons \cite{wang2022prompt} 
was also predicted.
It is worth noting that the LG laser pulse as a driver of the plasma wakefield acceleration can be replaced 
by a hollow electron beam \cite{Jain:2014mod};
however this hollow electron beam does not need to be vortex.

An interesting question emerging from this activity is whether an electron beam delivered 
by a plasma wakefield accelerator with a vortex driver can itself be twisted.
The calculations and simulations reported in \cite{Ju:2018vhh,Vieira:2018qqp} 
show that certain control over the topological structure of the accelerated electrons
can indeed be achieved via the helical shape of the wakefield:
at least, the electron pulse can be organized into helical bunches possessing a beam OAM.
Although these are not single-electron vortex states with definite OAM,
production electron beams with desired topologies expands 
the capacities of laser-plasma wakefield accelerators. 

In summary, intense vortex light emerges as a promising opportunity for future laser-plasma wakefield accelerator facilities \cite{Assmann:2022utc};
in fact, first experimental results have already been reported \cite{Wang:2020ifi}.

\subsection{Nuclear and hadronic processes}\label{subsection-nuclear-hadronic}

Since low-energy vortex neutrons were already demonstrated in experiment \cite{clark2015controlling,sarenac2018methods,sarenac2019generation},
one can ask how vortex neutron collision with nuclei differs from the plane-wave neutron case \cite{Afanasev:2017jdf,Afanasev:2019rlo,Afanasev:2021uth}.
In this collision setting, one usually assumes that the nucleus belongs to a fixed target, not a beam.
One can then perform calculations either treating the nucleus as an individual scattering or absorption center
placed at a fixed distance $\bb$ from the neutron phase singularity axis,
or by averaging over a range of $\bb$ for mesoscopic or a macroscopic target.

The first study of this kind \cite{Afanasev:2017jdf} explored two related processes: 
radiative capture of cold neutrons by protons $n_{\rm tw}+p \to d+\gamma$, and
deuteron photodisintegration by twisted photons $\gamma_{\rm tw} + d \to p + n$.
For low energy neutrons achieved in experiment ($\lambda = 0.27$ nm, $E_K = 11$ meV),
the capture process is dominated by the $S$-wave. As a result, the probability of the twisted neutron
capture per unit time is proportional to the plane-wave probability weighted with the spatial density of the neutron beam,
which for the Bessel neutron is
$d\dot{W}_{\rm tw}(\bb)/d\Omega_\gamma = 
(d\dot{W}_{\rm PW}/d\Omega_\gamma) \, J^2_m(\varkappa b)$,
where as usual $m$ is the total AM of the Bessel neutron state and $\varkappa$ is its conical momentum.
If the proton location can be well controlled, one can use the capture event rate to measure the neutron beam profile.
For a mesoscopic target with size of the order of $1/\varkappa$, the strong profile oscillations
get washed out, and for a macroscopic target
the cross section for twisted neutron differs from the plane-wave cross section only by the flux ratio factor $1/\cos\theta_p$
as in \eqref{single-Bes-4}.
As for the photodisintegration cross section with vortex photons, it proceeds via $^{3}S_1 \to {}^{1}S_0$ transition by M1-photons,
which leads to the selection rule near the beam axis $M_f - M_i = m_\gamma$, all quantum numbers referring to the 
total AM $z$-projections. The $b$-dependence of the event rate shows the same oscillating behavior.
However, the photon energy $E_\gamma > 2.2$ MeV, translating to sub-pm wavelength, 
makes it very difficult to observe the oscillating profile for realistic targets.

The process just considered may be a useful diagnostic tool for neutron beams but it does not offer new insights
on the dynamics of neutron-proton scattering. But other processes do.
As discovered in \cite{Afanasev:2019rlo,Afanasev:2021uth}, elastic (Schwinger) scattering of cold vortex neutrons on heavy nuclei
leads to phenomena which cannot be observed in plane-wave neutron scattering, at least in the Born approximation.
In the description below, we follow \cite{Afanasev:2021uth}. 

The neutron possesses a significant magnetic moment $\mu_n = -1.91$ nuclear magnetons.
As a result, when it elastically collides with a nucleus, not only does it scatter via strong interaction,
but it also probes the electric field of the nucleus via its magnetic moment \cite{Schwinger:1948zz}.
For plane wave neutrons, 
the corresponding scattering amplitude is well known, see e.g. \cite{Landau4}, Section 42.
For a spin-zero nucleus, it can be written as 
\begin{equation}
f_{\lambda\lambda'}(\bn,\bn') = w^{\prime\dagger}_{\lambda'}(a + i {\bm \sigma} \cdot {\bf B}) w_{\lambda}\,,\quad
{\bf B} = \beta \frac{[\bn \times \bn']}{(\bn-\bn')^2}\,, \quad \beta = \frac{\mu_n Ze^2}{m_pc^2} = -Z\times 2.94\times 10^{-16}\, \mbox{cm}.
\end{equation}
Here, $a$ is the strong nuclear amplitude, $\bn$ and $\bn'$ are the unit vectors along the initial and final neutron directions, $\lambda$ and $\lambda'$
are the initial and final helicities.
If the initial neutron is polarized along direction ${\bm \zeta}$, the standard cross section summed over final polarizations is 
\begin{equation}
\frac{d\sigma^{(st)}(\bn,\bn',{\bm \zeta})}{d\Omega'} = |a|^2 + |{\bf B}|^2 + 2 ({\bf B}\cdot {\bm \zeta})\,\Im\, a\,.
\end{equation}
Alternatively, if the initial neutron is unpolarized but we measure the final neutron polarization,
we will observe self-polarization of the neutron induced by the scattering process:
${\bm \zeta}^f =  2{\bf B}\,\Im\, a/(|a|^2 + |{\bf B}|^2)$.
Thus, for the standard Schwinger scattering, we observe that (1) the electromagnetic contribution
exhibits a strong peak in the forward direction, (2) the interference is proportional to the imaginary (absorptive) part
of the nuclear amplitude $a$ and to the transverse, not longitudinal, initial polarization.

The calculations of \cite{Afanasev:2019rlo,Afanasev:2021uth} showed that, 
for Schwinger scattering of vortex neutrons and for a fixed impact parameter $\bb$, 
all of these features change. 
The peak in polar angles shifts to the vortex neutron cone opening angle, which is to be expected from the discussion
in Section~\ref{subsection-single-Bessel}. A non-trivial result is that, for the vortex beam in superposition
of two $m$ states differing by one unit, there arises a non-zero helicity asymmetry 
$A_\lambda = (W_{\lambda=+1/2} - W_{\lambda=-1/2})/(W_{\lambda=+1/2} + W_{\lambda=-1/2})$, 
which was absent in the plane-wave case due to parity conservation.
Moreover, this helicity asymmetry can reach tens of percent for a wide range of parameters.
Finally, the interference term becomes sensitive to the real part of the nuclear amplitude $a$,
and therefore can probe its phase. This happens already at the Born approximation,
while in the standard cross section this sensitivity appears only at higher order.

Very recently, elastic scattering of slow neutrons on molecular hydrogen in gas phase was investigated in \cite{sherwin2022scattering}.
The cases of para-hydrogen and ortho-hydrogen were considered, as well as the super-elastic scattering
leading to the ortho- to para-hydrogen conversion. The incident neutrons were described 
as Bessel vortex states with the energy of 11 meV and 2.6 meV and with various options for the cone opening angle
and the OAM. The results on the angular distribution and on sensitivity to the vortex neutron parameters
agree with the expectations for single-vortex scattering. The angular distribution 
for the ortho- to para-H$_2$ transition was found to be quite different from the purely elastic case.

In summary, all these features offer new perspectives on neutron-nucleus scattering.
These observables survive for mesoscopic targets of the size of $1/\varkappa$ 
and can be readily checked in experiment, 
provided the experiments \cite{clark2015controlling,sarenac2018methods,sarenac2019generation}
can increase the flux of vortex neutrons.

\begin{figure}[!h]
	\centering
	\includegraphics[width=0.9\textwidth]{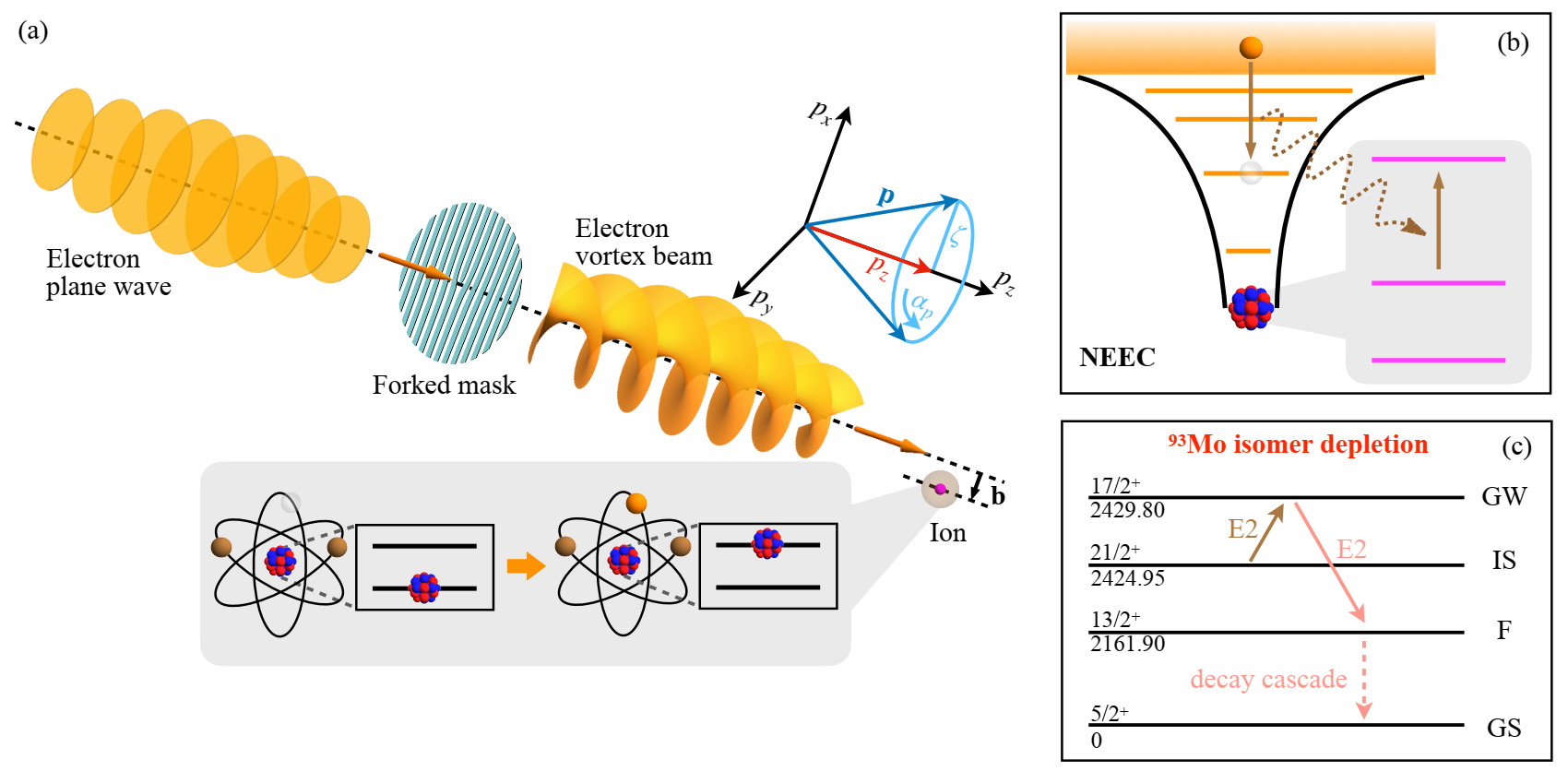}
	\caption{(a) Generation of a vortex electron beam via fork diffraction grating with subsequent collision with an atom with the impact parameter $b$
		followed by its recombination into an atomic vacancy. (b) At the resonant continuum electron energy, electron
		recombination can lead to a nuclear excitation. (c) Partial level scheme (not to scale) of $^{93}$Mo. 
		The nuclear isomeric (IS), gateway (GW), intermediate (F) and ground state (GS)
		levels are labeled by their spin, parity, and energy in keV. 
		Reproduced from \cite{Wu:2021trm} with permission.}\label{fig-isomer}
\end{figure}

Remarkable opportunities offered by vortex electrons for nuclear physics 
were discovered and discussed in \cite{Gargiulo:2021ega,Wu:2021trm}.
Ref.~\cite{Gargiulo:2021ega} theoretically investigated the resonant nuclear excitation 
which can take place, via coupling between the nuclear and electronic degrees of freedom, 
after electron recombination into an empty orbital of an ion 
(NEEC process, nuclear excitation on electron capture). 
If the electron energy matches the resonance condition, one can excite
a nucleus to an isomeric state or, through an excitation to a gateway nuclear state, 
de-excite the isomeric state in a controlled way.
This process was recently observed in experiment in $^{93}$Mo \cite{chiara2018isomer},
although that report drew certain polemics \cite{guo2021possible,reply:guo2021}.
When calculating such processes, one usually assumes that the ion was in its electronic ground state 
prior to the capture. However, as found in Ref.~\cite{Gargiulo:2021ega}, 
electron capture on a vacant orbital of an excited $^{73}$Ge ion
may increase the resonance strength by several orders of magnitude.
In particular, the authors mentioned the possibility to prepare the incident electrons in vortex states 
with the energy and the OAM tuned to the innermost shells,
arguing that vortex states offer higher control over nuclear excitation through electron capture, see Fig.~\ref{fig-isomer}.

This opportunity offered by vortex electrons was put under theoretical scrutiny in \cite{Wu:2021trm}.
Using isomers $^{93m}$Mo and $^{152m}$Eu, the authors found that the vortex
electrons with a suitable energy and cone opening angle, with the OAM equal to $m=3,4,5$, and with a controlled
distance between the ion and the vortex axis could modify the decay rate of the isomers 
by up to four orders of magnitude. For the specific case of $^{152}$Eu, it was also found that
the tailored electron vortex beam could strongly modify, with respect to plane-wave electrons, 
to which orbital ($2p_{1/2}$ vs $2p_{3/2}$) the electron is captured.
The authors are optimistic about future benefits to
nuclear physics and engineering arising from vortex beams with intense flux and dynamical control of beam parameters. 
However, it must be kept in mind that high-precision positioning of the isomer nucleus with respect to 
the vortex electron axis, required for this proposal to work, will represent a major experimenal challenge.

The sensitivity of the double-vortex scattering to the overall phase of the scattering amplitude as discussed
in Section~\ref{subsection-phase} for the elastic electron scattering applies also to scattering of hadrons.
In fact, the first proposal of this idea \cite{Ivanov:2012na} was stimulated by the 
overall phase behavior in hadron photoproduction processes, which is not measurable in traditional experiment.
However, no quantitative analysis of this new tool in hadronic physics has been published so far.
The only work which presented numerical estimates for the phase sensitivity in hadronic scattering
was \cite{Karlovets:2016jrd} with an analysis of the high energy $pp$ scattering 
with the asymmetry of the order of $10^{-10}$, or the lower-energy $ep \to X$ and $\gamma p \to X$ processes,
where it can be of the order of $10^{-4}$.
It must be stressed that this smallness is not due to the small physics effect
but is an unavoidable consequence of the small ratio $\varkappa/|\bp'_\perp|$
for vortex beams which seem to be realistic.
If there emerges a method to prepare vortex states with the conical momenta $\varkappa$ in the MeV range,
all these predictions will become much more optimistic.

The first calculation of a hadron excitation by a vortex photon
was presented in \cite{Afanasev:2021fda}.
Following the earlier works \cite{Afanasev:2017jdf,Afanasev:2020nur}, where vortex photodisintegration of the deuteron
was studied, the authors applied the same method to investigate $\Delta(1232)$ excitation in $\gamma_{\rm tw}p \to \Delta^+\to N\pi$.
The incident photon was described with the Bessel states. The initial proton was considered at rest 
and well localized with respect to the Bessel photon axis.
In the plane wave case, the photoproduction of $\Delta(1232)$ from the proton
can proceed via a spin flip with a large M1 transition amplitude,
or via the small E2 transition which requires two units of the OAM 
and must involve the small $D$-wave component of the $\Delta$ \cite{Pascalutsa:2006up}.
Accurate knowledge of the E2 transition strength would help elucidate 
the $\Delta$ composition.
In \cite{Afanasev:2021fda}, the authors show a way to isolate the small E2 amplitude
by studying the $b$-dependence of the cross section. 
In addition, intriguing polarization effects were predicted.
However, in order to have a chance to observe them,
one would need to localize and align the target proton with femtometer precision
and to focus the incident vortex photons to a focal spot of a similar size.
Since this level of control is beyond the present-day experimental possibilities,
verification of these predictions will require new experimental breakthroughs.

\subsection{Muon decay}

Decay of the vortex muon $\mu \to e \bar \nu_e \nu_\mu$ was explored in \cite{Zhao:2021joa,Zhao:2022}.
Vortex muons have not yet been experimentally demonstrated but, anticipating future progress, 
Ref.~\cite{Zhao:2021joa} considered Bessel vortex muon decay in flight. 
Muons had the energy $E_\mu$, velocity $\beta_\mu$, 
the total angular momentum $m$ and the cone opening angle $\theta_0$.
Several polarization states of the vortex muon were considered, such as the longitudinal, 
radial and azimuthal polarization, as well as spin-orbit coupled states
the polarization parameters of which exhibited $Z_n$ symmetry \cite{Zhao:2022}.
Since neutrinos are not observed, one can only measure the spectrum and the angular distribution
of the emitted electron, $d\Gamma/dE_e d\Omega$.
The calculations of \cite{Zhao:2021joa} show that the electron energy spectrum 
is a much more sensitive probe of the vortex muon properties than the $E_e$-integrated angular distribution.
The arguments resemble the discussion of Section~\ref{subsection-single-Bessel} 
and the inverse Compton backscattering on high-energy vortex electron described in Section~\ref{subsection-Compton}, 
taking into account the electron energy cut-off existing for any fixed polar angle.
Indeed, for the plane wave muon moving in the direction $\bn_\mu$, the maximal energy of the electron
detected in the direction $\bn_e$ is $E_{e\, \max} = m^2/[2E_\mu(1-\beta_\mu \bn_e \cdot \bn_\mu)]$.
However the Bessel vortex muon includes plane wave components with $\bn_\mu$ forming the vortex cone.
Thus, for any electron detection polar angle $\theta_e$ measured with respect to axis $z$,
one can define two threshold electron energies:
\begin{equation}
E_{e1} = \frac{m^2}{2E_\mu[1-\beta_\mu\cos(\theta_e+\theta_0)]}\,, \quad
E_{e2} = \frac{m^2}{2E_\mu[1-\beta_\mu\cos(\theta_e-\theta_0)]}\,,
\end{equation}
which are reminiscent of Eq.~\ref{Compton-energies}.
In the low energy region, $0 \le E_e \le E_{e1}$, the electron can emerge from all the plane wave components
of the vortex muon. In the high energy region, $E_{e1} \le E_e \le E_{e2}$, 
only a limited range of the plane wave components can emit such an electron. As a result, one expects to see
a break at $E_e = E_{e1}$ in the electron spectral distribution at fixed polar angle.
Notice that the plane wave muon corresponds to electron spectrum end-point $E_{e\, \max}$
which lies between the two thresholds $E_{e1} \le E_{e\, \max} \le E_{e2}$ for $\theta_e > \theta_0$.
Thus, the decay of the vortex muon is capable of producing higher electron energies at a given angle $\theta_e$
than of the plane wave muon moving in the same average direction.
However, when ``looking inside the cone'', that is, detecting electrons emitted at $\theta < \theta_0/2$,
one obtains $E_{e2} \le E_{e\, \max}$, so that the electron spectrum from the vortex muon decay spectrum 
shrinks with respect to the plane wave case.

\begin{figure}[!h]
	\centering
	\includegraphics[width=0.5\textwidth]{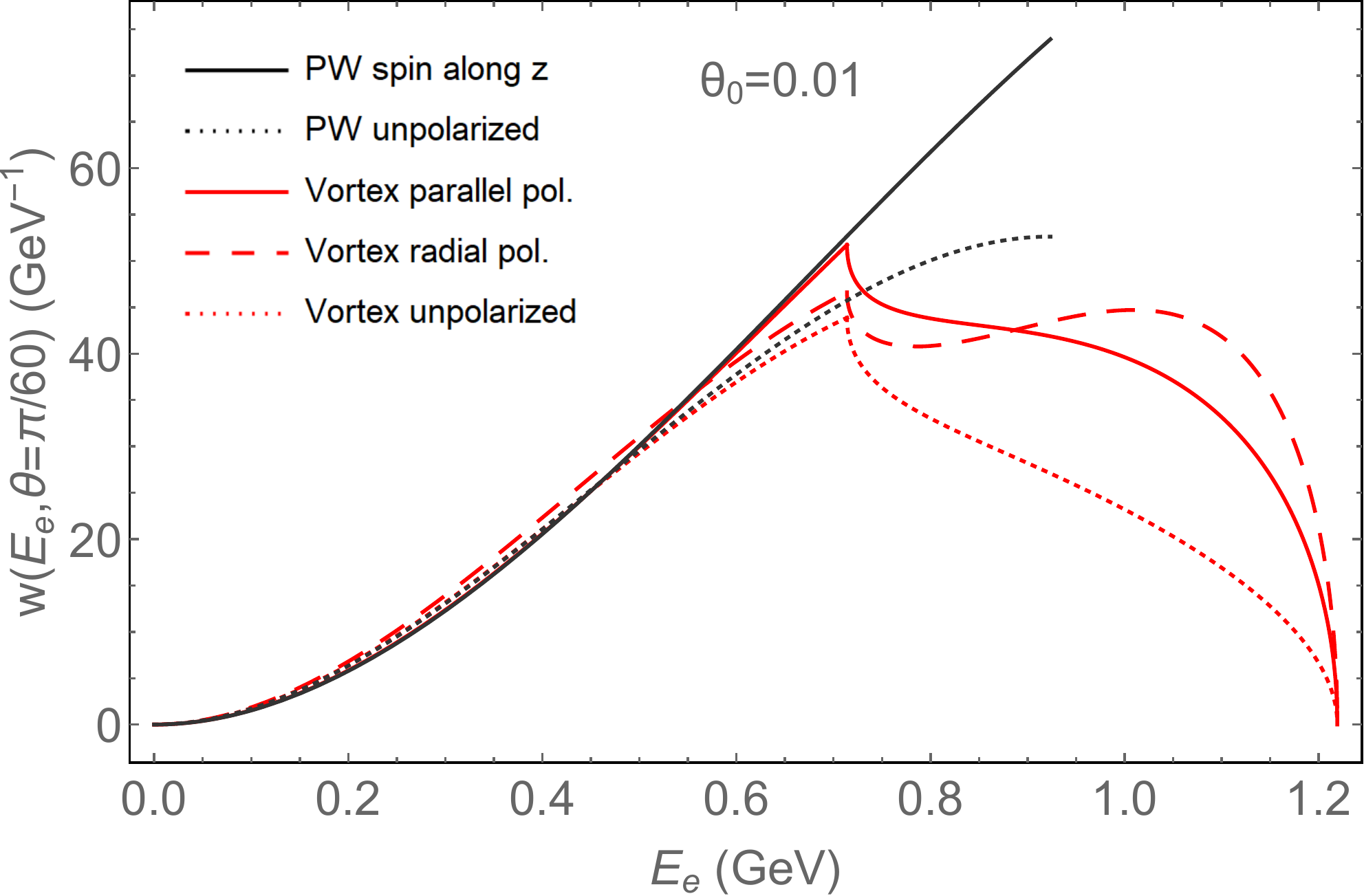}
	\caption{The electron spectra at the observation angle $\theta_e = 3^\circ$
		for the plane wave (black) and vortex (red) muons in various polarization cases:
		unpolarized (dotted lines), parallel polarization (solid lines), radial polarization (dashed line).
		The muon energy is $E_\mu = 3.1$ GeV, the cone opening angle is $\theta_0 = 0.01$. 
		Reproduced from \cite{Zhao:2021joa} with permission.}\label{fig-muon-decay}
\end{figure}

These expectations were confirmed by the numerical study \cite{Zhao:2021joa}. 
The predicted electron spectra from few GeV vortex muon decays 
showed a break at $E_{e1}$, which was clearly visible even for small cone opening angles $\theta_0 = 0.01$ provided 
the detection angle $\theta_e$ is not too large. Moreover, there were clear distinctions of the electron spectrum shapes
not only between the plane wave and the vortex muon cases but also when comparing
different polarization states of the vortex muon, Fig.~\ref{fig-muon-decay}.

All these spectral modifications depend on $E_{e2}-E_{e1}$, which becomes small
for low energy muons. Thus, in order to clearly observe the predicted effects, one should aim
to generate at least moderately relativistic vortex muons. 
As expected for single-Bessel processes with azimuthally symmetric polarization state, 
these distributions do not depend on the angular momentum state of the Bessel muon nor do they exhibit
$\varphi_e$ dependence. 
However, if the vortex muons are produced in spin-orbit coupled states with the polarization
parameters exhibiting only a discrete muon azimuthal angle invariance, 
the angular distribution will demonstrate a characteristic periodic pattern in $\varphi_e$ \cite{Zhao:2022}.

	\newpage
	\section{The experimental status}\label{section-experimental}

\subsection{Generation of vortex states: experimental results and proposals}

Traditional particle sources and injection systems are not designed to prepare 
vortex states of high energy particles.
Even if low energy vortex states are prepared and injected in accelerators and storage rings,
it is not obvious whether they will preserve their state in acceleration process and
can be brought in collisions. All these stages require novel instrumentation.
The final particles emerging from collisions of vortex states can, in most cases, be detected and analyzed 
with traditional detectors, provided one measures the angular distributions with sufficient resolution.

The future experimental program of high energy vortex states physics can benefit 
from the experience at low energies. Up to now, vortex states were experimentally demonstrated for 
photons, electron, neutrons and neutral atoms. 
Below, we overview the situation for low and moderate energies, and then discuss 
the proposals of how high energy particles can be put in vortex states.

\subsubsection{Vortex photons}\label{subsection-undulator}

Optical photons carrying OAM were demonstrated in 1990's \cite{Allen:1992zz} and are now routinely used
in many fields of fundamental and applied optics \cite{Torres-applications,andrews2012angular,Paggett:2017,photonics-review-2017,Knyazev-Serbo:2018,babiker2018atoms}.
The most popular method to producing vortex optical beams is to let a non-vortex beam
pass through a fork hologram, a diffraction grating with $\ell_0$ dislocations at the center, Fig.~\ref{fig-vortex-production}.
The incoming beam splits into several diffraction peaks, $k$-th order peak carrying the OAM of $k\ell_0 \hbar$.
Spiral phase plates or spiral mirrors can also be used to impart OAM on the incident light;
in this way, generation of photons in well controlled OAM states as high as 10,010 was reported \cite{fickler2016quantum}.

Beyond optical range, vortex photons in the extreme UV range were 
experimentally demonstrated in 2017 with the aid of the high harmonic generation technique 
\cite{gauthier2017tunable,photonics-review-2017}.
At even higher energies, the first report of experimental generation of X-ray vortices
dates back to 2002 \cite{peele2002observation,peele2004x}.
It relied on a spiral phase plate which imparted azimuthally dependent phase shift 
on 9 keV X-ray photons.
In 2006, X-ray photons with energy of 0.7 keV and the OAM up to 64$\hbar$ were produce
with carefully fabricated diffractive optical elements \cite{cojoc2006x}.
Recently, an impressive control over generation and characterization of soft X rays 
carrying OAM up to 30$\hbar$ was demonstrated in \cite{Nature-Phot-2019} 
with the aid of superimposed fork diffraction gratings.

A more versatile path to generation of vortex photons from optical to X ray domain and beyond is to employ 
electron accelerators equipped with undulators.
Already in classical electrodynamics, the wavefront of the EM radiation emitted by a charge in circular or helical motion 
contains a phase vortex with respect to the axis of circulation \cite{Katoh:2016aww};
for the angular momentum balance calculations, see \cite{Epp:2019jqk}.
In a helical undulator, the off-axis radiation can be described via LG modes \cite{Sasaki:2008,Afanasev:2011nz}.
The vortex state of this radiation was demonstrated in 2013 in a dedicated experiment
at the BESSY-II accelerator \cite{Bahrdt:2013eoa}. Soft X-ray photons with energy of 99 eV were radiated,
and their vortex states were confirmed through interference with the non-vortex reference beam.
Later, production of intense, femtosecond, coherent optical vortices in the extreme UV range 
from the FERMI light source at the Elettra Sincrotrone Trieste was reported in \cite{ribivc2017extreme}.
In 2019, production of LG modes of soft X-rays with energies up to 1 keV 
and the OAM up to 30$\hbar$ were reported \cite{Nature-Phot-2019}.
Vortex light in a variety of LG modes was also experimentally demonstrated 
inside a free-electron laser (FEL) oscillator \cite{Liu:2020wqd},
that is, a helical undulator system at an electron storage ring equipped 
with a resonant laser cavity. Unlike the single-pass generation of the OAM light mentioned above, 
a FEL oscillator allows for multi-pass self-seeded OAM light amplification with tunable wavelength.
Extension to higher photon energies should be possible.
For a recent review of other results and prospects, see \cite{photonics-review-2017}.

A different regime of generating vortex photons in undulators was proposed in \cite{Hemsing:2009,Hemsing:2011}
and experimentally verified in \cite{Hemsing:2013is}.
A high-brightness electron beam traversing a helical undulator and interacting with a linearly polarized laser pulse
experiences microbunching and organizes itself into a helically modulated structure. 
Passing through a planar undulator, it emits radiation which carries a well controlled OAM.
In the experiments at the SLAC Next Linear Collider Test Accelerator reported in \cite{Hemsing:2013is},
an initially unmodulated 120 MeV electron bunch of duration of 0.5 ps interacted with a 1 ps Gaussian laser pulse 
($\lambda = 800$ nm), which induced the required helical microbunching, and later emitted vortex light 
at the same wavelength. The approach can be extended to higher beam intensities and shorter wavelengths.

In the very recent work \cite{Morgan:2022vuj}, it was suggested that electron beam microbunching 
could be used to obtain vortex X rays with an ultrafast, sub-femtosecond scale flip of the OAM. 
An electron beam with strong intensity modulation acquired in the main FEL lasing section
can be injected into the helical undulator afterburner, where it emits vortex X rays.
As the simulations of \cite{Morgan:2022vuj} show, the electron microbunching structure
leads to temporal, hundred-attosecond scale modulation not only of the vortex light intensity but also
of individual OAM components, leading to an ultrafast OAM flip.
Such a temporal control of the vortex X ray properties, when demonstrated experimentally, will nicely match 
the similar ultrafast manipulation of the vortex electrons parameters which is already achieved in experiment \cite{vanacore2019ultrafast}.

The recent series of papers \cite{Bogdanov:2017ife,Bogdanov:2018mik,Bogdanov:2019atw,Bogdanov:2019pzp,Bogdanov:2019amj,Bogdanov:2020qtm} 
addressed twisted photon radiation by classical currents in undulators under various regimes, 
including the cases of an undulator filled with a dispersive medium.
Also, channeling radiation produced by electrons traversing a silicon crystal was suggested in \cite{Abdrashitov:2018dyi}
as yet another source of twisted photons.

In Section~\ref{subsection-Compton} we also described schemes \cite{Chen:2018tkb,Chen:2019,zhang_zhao:2021} 
based on non-linear Compton backscattering to generate gamma beams carrying large OAM,
with the average value of the OAM up to $10^6\hbar$ per photon \cite{zhang_zhao:2021}.
However since these simulations are intrinsically semiclassical, 
it is not clear whether individual photons with such a large value of OAM can exhibit a phase vortex.

\subsubsection{Vortex electrons}

\begin{figure}[!h]
	\centering
	\includegraphics[width=0.8\textwidth]{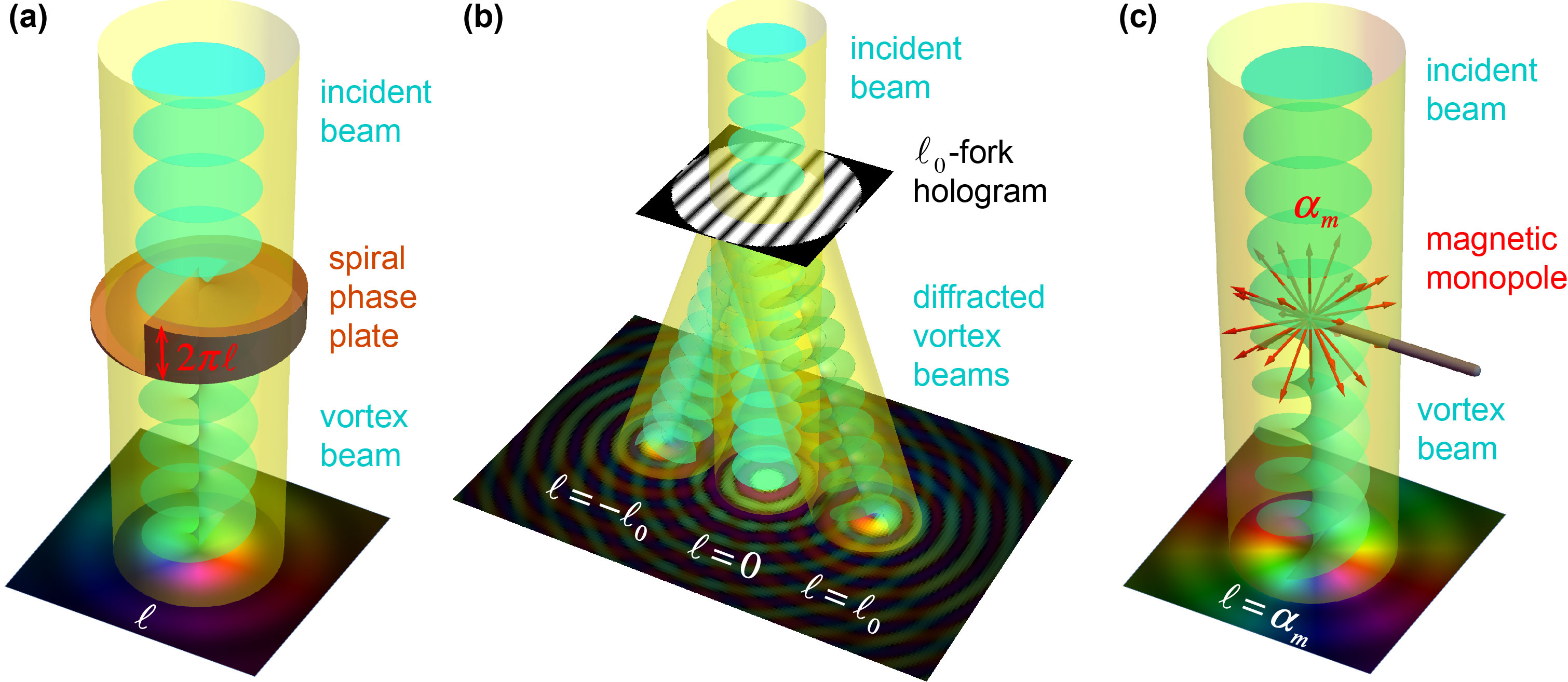}
	\caption{Basic methods for the generation of electron vortex beams: (a) a spiral phase plate; (b) a diffraction grating with a fork-like edge dislocation of order $\ell_0$ (here $\ell_0=1$); and (c) a magnetic monopole at the tip of a magnetic needle. 
		Reproduced from \cite{Bliokh:2017uvr} with permission.}\label{fig-vortex-production}
\end{figure}

Experimental work on vortex electrons started in 2010--2011 when three experimental groups, independently,
reported production of electron vortices with kinetic energy up to 300 keV using phase plates, fork diffraction gratings, or spiral zone plates
\cite{Uchida:2010,Verbeeck:2010,McMorran:2011,Verbeeck:2011-atomic,Saitoh:2012,Grillo:2014,Grillo:2014:efficient}, see Fig.~\ref{fig-vortex-production}. 
The electrons were produced in commercially available electron microscopes 
with customized electron optics elements.
A very compact electron source together with wavefront correcting elements were the key to achieving 
sufficiently large transverse coherence length in the aperture plane. 
Carefully fabricated fork holograms or more elaborate binary masks shaped the electron wavefront,
which produced diffraction peaks with well defined OAM or their well-controlled superpositions.
These vortex electron beam could be focused to $\sim 1$\AA~focal spot \cite{Verbeeck:2011-atomic}, 
which demonstrated exceptional stability
and was later used as an atomic-scale probe of magnetic properties and other material studies.
Non-diffracting Bessel states propagating over distance of 0.6~m without measurable spreading
was reported in \cite{Grillo:2014}.
Propagation of electron vortices in longitudinal magnetic fields led to new insights 
in electron wave interactions with external fields.
Details of the experimental technique can be found in \cite{Bliokh:2017uvr,Lloyd:2017,Guzzinati-PhD}.

Already in the first works on vortex electrons, the OAM as large as $100\hbar$ was achieved \cite{McMorran:2011}.
Later, holographic pitchfork masks with 200 dislocations at the center were manufactured 
and used to produce highly twisted electron beams \cite{Grillo:2014ksg}.
In \cite{mafakheri2017realization}, electron beam lithography technique was used to produce
nanofabricated masks with $\ell = 1000$ at the first diffraction order.
However, in all these experiments, only moderately relativistic electrons from the electron microscopes were used.
Ultrarelativistic vortex electron have not yet been demonstrated in experiment.

In addition to phase plates and fork holograms, electron waves can also be twisted in a contactless manner.
Since electrons are electrically charged, their wavefronts can be manipulated with external electric and magnetic fields,
which allows for novel schemes for vortex electron production.
Very recently, generation of vortex electrons with an electrostatic device was reported in \cite{Tavabi:2022uiw}.
This technique introduces no absorption or parasitic intensity loss, and allows to generate vortex beams with tunable $\ell$ up to 1000.

Another idea for vortex electron generation is to use an artificial magnetic monopole, 
which effectively exists at the tip of a thin magnetized needle placed at the center 
of a round aperture \cite{Beche:2013wua,beche2017efficient},
see Fig.~\ref{fig-vortex-production}, right.
Within classical electrodynamics, the magnetic field of a static magnetic monopole is written as
${\bf B}(\br) = (g/r^2)\bn_r$, where $g$ is the magnetic charge and $\bn_r$ is the radial unit vector.
A pointlike electric charge $e$ moving in this field experiences a non-central Lorentz force,
so that its OAM $\bL$ is not conserved. However, there appears a different integral of motion \cite{Shnir_book}:
$\bL' = \bL - (eg/c) \bn_r$. Since $\bn_r$ is the unit vector pointing from the monopole towards the charge,
this extra term does not vanish even for particles asymptotically far away for the monopole. 
If the charge arrives from $z = -\infty$ and experiences a small angle deflection,
then the vector $\bn_r$ effectively flips from $-\bn_z$ to $+ \bn_z$, so that the charge acquires 
a non-zero $z$ component of the OAM $L_z = 2eg/c$.
The quantum-mechanical description, which was first worked out in \cite{tamm1931}, supports this conclusion.
The new integral of motion is defined by the operator
\begin{equation}
\hat{\bL}' = \hat{\bL} - \frac{\hbar\mu}{2} \hat{\bn}_r\,, 
\end{equation}
where $\mu = 2eg/(\hbar c)$ is dimensionless. 
If the initial electron wave function is $\psi_0(\br) \propto \exp(i kz)$, then the small angle deflection in the monopole field 
produces an azimuthally dependent phase: $\psi(\br) \propto \exp(i kz) \exp(i \mu \varphi_r)$,
so that the initial plane wave electron acquires an OAM $L_z = \hbar \mu$.
Since the vector potential of a pointlike magnetic monopole is not uniquely defined but depends
on the auxiliary construction called the Dirac string \cite{Dirac:1931kp,Dirac:1948um,Shnir_book},
one usually requires that this Dirac string in unobservable (that is, it can be rotated by a pure gauge transformation).
This requirement leads to the famous Dirac quantization condition that $\mu = 2eg/(\hbar c)$ is integer.
In this way, the final electron wave function is uniquely defined.

In the experiments reported in \cite{Beche:2013wua,beche2017efficient}, 
the incident non-vortex electron wave passed through a round aperture with radius 10 $\mu$m 
with a thin ferromagnetic needle in a single-domain magnetic state.
The magnetic field near the tip of the needle could be approximated by the magnetic monopole field,
and the magnetic flux through the needle was chosen to correspond to $\mu \approx 2$,
leading to the final vortex-like electron wave with $L_z \approx 2\hbar$.
Since this method relies only on electromagnetic interactions, it can be used to twist any charged particle
with a sufficiently large transverse coherence length. 

An impressive technique of producing ultrafast pulses of vortex electrons
with the aid of chiral plasmonic near fields was demonstrated in \cite{vanacore2019ultrafast}.
An incident non-vortex ultrashort electron pulse with kinetic energy of 200 keV
was passing through a nanofabricated hole of diameter 0.8 $\mu$m in a perforated 
Ag/Si$_3$N$_4$ film. Optical pulses coincident with the electron pulse
illuminated the hole and excited plasmons near its edge, the near fields of which 
affected the propagation of the electron pulse and imparted an OAM on it.
Remarkably, by controlling plasmonic fields with sub-femtosecond precision,
the authors could dynamically change the properties of the vortex electron pulse
within its femtosecond scale envelop.

\subsubsection{Vortex neutrons and atoms}

Production of neutrons carrying a non-zero, controlled OAM was first reported in 2015 \cite{clark2015controlling}.
Reactor neutrons were cooled by a cryogenic moderator to 20 K and transported
through neutron guides to a neutron interferometer. A spiral phase plate inserted 
in one arm of the interferometer imparted a phase vortex with a topological charge of one or several units. 
Its subsequent interference with the non-vortex part of the neutron wave function 
revealed the azimuthally dependent pattern expected from the vortex neutron.
The transverse coherence length of the reactor neutrons was estimated to be of the order of nanometers to a few micrometers,
which was significantly smaller than the macroscopic, centimeter-sized phase plate. 
As a result, each individual neutron passing through the interferometer 
probed only a small off-centered patch of the phase plate 
and did not exhibit a well-formed phase vortex.
These observations raised doubts as to whether the results of this experiment
were at all indicative of vortex neutron generation \cite{cappelletti2018intrinsic,cappelletti2021photons}.

Later, a completely different approach, suitable to electromagnetic and matter waves, was put forth in \cite{sarenac2018generation}.
The phase plate was replaced with specially designed phase-controlling elements,
which could entangle the spin and the OAM degrees of freedom and produce a lattice of beamlets, 
each with a ring-shaped intensity profile and carrying an OAM with respect to its own phase singularity axis.
This method was verified for laser light in \cite{sarenac2018generation} and 
then applied to neutrons \cite{sarenac2018methods,sarenac2019generation} with the aid of magnetic prism pairs.
The illustrations in Fig.~\ref{fig-spin-orbit} refer to the spin-orbit coupled neutron states produced with these elements.
The neutrons were still of low energy, with the wavelength of $\lambda = 0.41$ nm \cite{sarenac2019generation}.
Since the vortex ring size imaged bythe detector is still significantly larger than the transverse coherence length,
this approach alleviates, but does not completely solve, the challenges posed by insufficient coherence.
A different development of the same technique, in which a lattice of smaller vortices was
produced via a nested loop neutron interferometer with orthogonal aluminium prisms,
was very recently reported in \cite{Geerits:2022xxq}.

The challenge of small coherence length was finally overcome in the very recent paper \cite{Sarenac:2022}
which reported experimental production of truly helical neutron waves.
This was achieved with the aid of a huge array of fork dislocation gratings tightly packed on a 5 mm $\times$ 5 mm plate.
Each diffraction grating measured just 1 $\mu$m in size, matching the $\mu$m scale transverse coherence length. 
With the lattice period of 2 $\mu$m, the plate carried 6.25 million nearly identical fork gratings,
which allowed to keep the total vortex neutrons flux at reasonable level.
After less than an hour of statistics accumulation, the far-field neutron detector demonstrated clear intensity rings 
at the first diffraction maximum even for the dislocation order of the fork grating as high as $q=7$.

An alternative way to produce a single vortex neutron is to let it pass through 
a high gradient quadrupole magnet \cite{Nsofini:2016dej}.
Numerical estimates show that a 10-cm-long quadrupole magnet with a gradient of 13 T/cm
should accomplish the task for the neutron with $\lambda = 0.27$ nm and the transverse coherence length of 100 nm.

Although the above methods are expected to generated genuine vortex neutrons
and the spin-dependent intensity profiles reported in \cite{sarenac2018methods,sarenac2019generation}
agreed with the simulation, it is desirable to find a method capable of directly
accessing the OAM carried by the vortex neutron.
The recent publication \cite{Jach:2021veo} proposed the standard reaction for 
thermal neutrons detection ${}^3$He(n,p)${}^3$H as a method to unambiguously
confirm that a spin-polarized neutron contains $\ell=1$ OAM state.
At present, this idea still awaits experimental realization.

Another recent paper \cite{Geerits:2020xzo} proposed a different idea of how twisted neutrons 
and other neutral spin-1/2 particles could be obtained.
When traversing a homogeneous electric field, an electrically neutral particle interacts 
with it through its magnetic moment.
As a result, spin-orbit coupled states are generated, and the incoming spin-1/2 particle can acquire 
an OAM $\ell = \pm 1$ parallel to the electric field direction.
Estimates show that, in the simplest configuration, extremely high electric fields are required 
to put neutrons in the OAM states \cite{Geerits:2020xzo}.

Free-propagating atoms can also be put in vortex states.
The progress in atom optics witnessed in the past several decades naturally led to 
the proposal, back in 2004, to generate vortex atomic beams with the aid of atom optical elements \cite{helseth2004atomic}.
After the impressive demonstration of vortex electron beam generation via fork diffraction grating,
a similar suggestion for atoms was presented in \cite{lembessis2014atom}.
In order to resolve individual diffraction peaks carrying different OAM, 
the incident atoms should be slow; the authors of \cite{lembessis2014atom} offered estimates based on $v = 1$ cm/s.
Another idea was put forth in \cite{lembessis2017atomic}: the Bose-Einstein condensate released from a trap 
and falling freely can be diffracted from a properly tailored light mask.

However, the first experimental realization of vortex atoms, reported very recently in \cite{luski2021vortex},
used a different technique. In that experiment, a supersonic beam of helium atoms 
with mean velocity of 1090 km/s (wavelength $\lambda = 0.09$ nm) and with a velocity spread of just 3\%
was leaving a valve and passing through a fork diffraction grating about two meters downstream.
On the way to the grating, the beam passed through a pair of narrow slit skimmers
just 0.15 mm wide. This simple setup produced a well-collimated beam with 
divergence angle of just 55 $\mu$rad. The resulting transverse coherence length
was about 1 $\mu$m, which matched the scale of the fork diffraction grating.
Each helium atom coherently passed through the grating and diffracted
into well-separated peaks of definite OAM up to $\ell = \pm 4$. 
In the far field, the diffracted atoms produced the characteristic ring intensity profile 
revealed by pixelized helium atom detection system.
The same experiment also found evidence that metastable helium dimers He$_{2}^{*}$ were also produced
and put into vortex states.
Although belonging to atomic and beam physics rather than particle physics,
this milestone result proves that composite freely propagating systems can be put into a vortex state
with the aid of mechanical skimmers and binary masks.

\subsubsection{Conical angle and conical transverse momentum}

When producing vortex states, one should keep under control not only the energy and the OAM or the total AM value,
but also the cone opening angle $\theta$, or alternatively, the conical transverse momentum $\varkappa$.
As we saw in the previous sections, the larger $\theta$, the stronger are the modification of the angular distributions
in vortex state scattering. Double-vortex collisions lead to structures in the transverse momentum distribution
of the final particles which scale with $\varkappa$. Thus, the full potential of vortex state collisions
can be realized only with the initial $\theta$ or $\varkappa$ larger than the final state measurement uncertainties.

Focusing of vortex electrons produced in electron microscopes to angstrom-scale focal spots has already been demonstrated
experimentally \cite{Verbeeck:2011-atomic}. This value of the focal spot translates to the conical transverse momentum 
$\varkappa \sim$ 10 keV. Since the electron kinetic energy of 300 keV corresponds to the longitudinal momentum
$p_z = 600$ keV, we get the cone opening angle of the order of one degree.
These are the realistic cone parameters with the present-day instrumentation.
Future particle physics applications of vortex states will critically depend on whether one can
achieve larger values of $\varkappa$. This is a separate challenge in addition to acceleration of vortex states to MeV and GeV energies.
Indeed, in the simplest case of Bessel electron acceleration in a uniform longitudinal electric field,
one can boost $p_z$ but not $\varkappa$. Increase of $\varkappa$ will require dedicated experimental efforts.

\subsection{Prospects of generating high energy vortex states}

In order to bring vortex states of various particle into nuclear and particle physics,
the energy of these particles must reach MeV or GeV range.
Generation of vortex states of any type of particles with these energies
has not yet been reported, but there exist proposals of how this could be achieved.
At present, one first needs a proof-of-principle experiment showing that vortex states 
of high energies can be produced, controlled, and brought into collisions.
The intensity of the vortex beams is not yet a figure of merit.

One can think of three main schemes of how high energy vortex states could be produced;
below we discuss them in some detail.
\begin{enumerate}
	\item First accelerate, then twist: 
	a high energy particle in an approximately plane wave state
	with sufficient transverse coherence length
	passes through a device which imparts a phase vortex.
	\item First twist, then accelerate:
	a low-energy vortex state is injected in a linear accelerator or a storage ring where its energy is increased 
	without destroying the phase vortex.
	\item
	Transfer the OAM and/or energy from an initial particle to the final one through a high-energy collision process itself.
\end{enumerate}

\subsubsection{Schemes involving twisting and acceleration}\label{section-production-twisting}

The first scheme requires sufficient transverse coherence of the incident wave function of a high energy particle 
and a ``twisting'' device.
The issue of transverse coherence is rarely
discussed when analyzing collisions of high energy particles \cite{Karlovets:2020odl}.
Particle dynamics in high-energy accelerators and storage rings within the transverse and longitudinal phase space 
is usually treated semiclassically, without invoking wave functions,
which can be seen an indirect indication that the transverse coherence length
of high energy particles is much smaller than the beam size.
An efficient ``twisting'' device is also a challenge at high energies.
The fork diffraction gratings used for low energy photons, electrons, and atoms,
are thin binary masks, which partially block incoming electromagnetic or matter waves.
However, high energy particles have large penetration depth, 
so that traditional diffraction gratings become impractical.

For charged particles, wavefront engineering \cite{madan2020quantum} can be achieved in a contactless manner with the aid of 
external electric \cite{vanacore2019ultrafast,Tavabi:2022uiw} and magnetic \cite{Beche:2013wua,beche2017efficient} fields. 
Application of this method to positrons, protons, and antiprotons, even of low energies,
can be an important step towards bringing vortex states into particle physics.
External electromagnetic fields can be also used to prepared vortex neutrons thanks to their magnetic moment 
\cite{Nsofini:2016dej,sarenac2018generation,sarenac2019generation,Geerits:2020xzo};
for a overview of neutron twisting techniques, see \cite{sarenac2018methods}.

Another promising idea proposed recently in \cite{Floettmann:2020uhc,Karlovets:2020tlg} is to generate 
vortex electrons of potentially high energies using the immersed cathode technique.
A cathode placed inside a solenoid can emit electrons along the solenoid axis. 
If the emitted electron wave function is azimuthally symmetric ($\ell = 0$),
then upon exiting the solenoid, it acquires the azimuthally dependent phase
characteristic for the electron in a twisted state. 
The acquired OAM depends on the radius of the cathode emission zone
and the magnetic field, which was first established almost 100 years in a semiclassical
approach (the so-called Busch theorem \cite{busch1926}).
Photoinjector experiments at FNAL \cite{Sun:2004zu} demonstrated production of OAM-dominated electron beams
with average angular momentum of about $10^8\hbar$ and a broad OAM spectrum.
Such an electron beam represents a useful tool in beam physics but it is hardly practical
for observation of quantum effects linked to the OAM.
The papers \cite{Floettmann:2020uhc,Karlovets:2020tlg} adapted the Busch theorem to the quantum regime
and described the arrangements required for emission of electrons in single OAM-model with small $\ell$.


For the second option from the above list --- first twist, then accelerate, ---
one would need to inject a low-energy charged particle prepared in a vortex state into a linear accelerator,
possibly a wakefield accelerator,
which would ramp up its energy without significantly distorting its phase vortex.
In the ideal case of a pure Bessel beam 
$\psi(\br) \propto e^{ik_zz}e^{i\ell\varphi_r}\, J_\ell(\varkappa \rho)$
propagating along a homogeneous and strictly longitudinal electric field, 
the transverse wave function remains unchanged as the longitudinal momentum grows.
It is not yet know whether in realistic situations, with stray electric and magnetic fields,
the vortex state parameters can be preserved during acceleration,
There are reasons to be optimistic, though. 
A phase vortex is a topologically protected feature of the wave function
and it cannot spontaneously dissolve \cite{dennis2009singular}.
These expectations are confirmed by experiments with vortex light propagation 
in turbulent atmosphere. For example, Ref.~\cite{Zeilinger:2016} reported
results of the information transmission experiment my sending optical vortex photons 
over the distance of 143 km. 
For electrons, the impressive level of vortex state control
in electron microscopes clearly shows that vortex electrons are not easily destroyed.
However, their behavior at higher energies 
and the exact tolerance levels to stray fields and other processes 
require dedicated numerical studies and proof-of-principle experiments.

One can also envisage injecting vortex electrons in circular storage rings, with the hope that 
the phase vortex will be preserved as an electron wave packet circulates in the transverse magnetic field.
It should be mentioned that there are numerous publications, 
both theoretical \cite{Bliokh:2011fi,bliokh2012electron,Gallatin:2012ai,greenshields2012vacuum,chowdhury2015electron,rajabi2017relativistic,van2017nonuniform,han2017classical,Silenko:2018eed,Zou:2021ray} and experimental \cite{Guzzinati:2012mb,schattschneider2014imaging,schachinger2015peculiar} 
(see also the reviews \cite{Bliokh:2017uvr,Lloyd:2017}), addressing vortex electron propagation in external magnetic fields. 
It was even proposed as a low energy, controllable testbed of the chiral magnetic effect taking place in heavy-ion collisions
\cite{Fukushima:2020ncb}.
However, most of them focus on the longitudinal magnetic field
and on the detailed spatial structure of the current, spin, and the OAM distribution of the vortex electron,
not on the dynamics of the vortex electron wave packet in storage rings as a whole. 
Although this is still an uncharted territory from the experimental point of view,
several theoretical works inspire certain optimism that vortex electrons can indeed be stored in rings.

The first theoretical paper on vortex electrons \cite{Bliokh:2007ec}, which triggered the entire experimental activity,
presented a semiclassical study of electron wave packets with phase vortices.
Describing the vortex electrons via the intrinsic OAM vector $\lr{\bL}$ as well as the extrinsic
parameters of the wave packet centroid $\lr{\br}$ and $\lr{\bp}$, 
Ref.~\cite{Bliokh:2007ec} discussed the equations of motion of such a vortex electron in external fields.
These equations exhibit spin-orbital coupling which leads to the OAM vector precession.
Treatment of the intrinsic OAM evolution in generic external electromagnetic fields, still within the semiclassical approach, 
was put on a more solid, fully relativistic ground in \cite{Silenko:2017fvf}.
The theory developed there leads to suggestions of how the intrinsic OAM of a vortex wave packet 
can be rotated, flipped, and otherwise manipulated in storage rings.
Not only does the OAM contribute to the magnetic moment of the vortex electron, 
but it also generates the electric quadrupole moment and the tensor magnetic polarizability. 
These were analyzed in \cite{Silenko:2019dfx}, and experimental methods for their detection were suggested.

Behavior of paraxial twisted electrons in the uniform transverse magnetic field, appropriate for storage rings, 
was analytically studied in \cite{Gallatin:2012ai}, with the optimistic conclusions
that orbital helicity evolves in a predictable way and can survive many rotation. 
The relativistic dynamics of twisted electron beams in arbitrary electric and magnetic fields was 
reformulated in quantum regime in \cite{Silenko:2018eed}.
This work predicted a remarkable phenomenon of spontaneous radiative orbital polarization of 
vortex electrons as they circulate in storage rings. This mechanism, however, strongly depends 
on the electron energy and is expected to be noticeable only in the GeV energy range.
One may think that this spontaneous polarization will be damped via orbital depolarization
caused, for example, by stray magnetic fields in storage rings.
However, as argued in \cite{Silenko:2019ziz}, this depolarization effect essentially cancel itself
during circulation: since the OAM precession period is twice the cyclotron period,
a perturbing element of the ring will affect the OAM evolution on two successive revolutions in the opposite directions.
All these remarkable predictions certainly need experimental verification. 

\subsubsection{Schemes involving scattering}\label{subsection-schemes-scattering}

A different approach to generating high-energy vortex states of particles is 
to use the scattering process itself,
namely to arrange for such a collision process in which one of the final particles emerges 
with high energy and a well-formed phase vortex. 
In this scheme, the energy, momentum, and angular momentum conservation laws
do all the job without the need to manually engineer the phase front of a high-energy particle
or to accelerate a low-energy vortex state. 
The price to pay is the necessity of a post-selection protocol, a rather exotic requirement for particle physics.
Although this idea was already used in the first publications on high-energy processes with vortex particles 
\cite{Jentschura:2010ap,Jentschura:2011ih,Ivanov:2011tu,VanBoxem:2015lta}, its general description was presented only very recently
\cite{Karlovets:2022evc,Karlovets:2022mhb}. Below, we follow the main arguments from these recent publications.

Consider elastic scattering of two plane waves with momenta $\bk_1$ and $\bk_2$
in a reference frame in which the total transverse momentum is zero: $\bk_{1\perp} + \bk_{2\perp} = 0$. 
Denoting the initial state as $\ket{i} = \ket{\bk_1, \bk_2}$,
one can represent the final state as $\ket{f} = \hat S \ket{i}$, where $\hat S$ is the evolution operator.
This final state is what emerges from the scattering process itself, before the reaction products
reach the detector or experience any other significant interaction with the environment.
In \cite{Karlovets:2022evc,Karlovets:2022mhb}, this final state was called pre-selected, or evolved state,
to clearly distinguish it from what a detector will eventually record.
This outgoing state may include several reaction channels which are kinematically possible
as well as a non-scattered contribution of the initial particles \cite{Ivanov:2022sco}.
The pre-selected final state can be expanded in a convenient basis, the usual choice being the plane waves:
\begin{equation}
\ket{f} = \hat S\, \ket{i} = \int \frac{d^3 \bk_1'}{(2\pi)^3} \frac{d^3 \bk_2'}{(2\pi)^3}\,
S_{fi}\, \ket{\bk_1',\bk_2'}\,, \quad S_{fi} = \bra{\bk_1',\bk_2'} \,\hat S\, \ket{\bk_1,\bk_2} \propto \delta^{(4)}(k_1+k_2-k_1'-k_2')\,.
\end{equation}
The delta function makes it clear that the two outgoing particles are momentum-entangled.
As the expanding two-particle state hits a traditional detector, the particle which interacts
with it first collapses to a state that can be approximated with a plane wave with momentum $\bk_1'$.
In this way, the detector performs post-selection in the form of strong measurement. As a result,
the second particle is projected on a similar approximately plane-wave state which follows from the momentum conservation law.
In particular, its transverse momentum is $\bk_{2\perp}' = -\bk_{1\perp}'$.

Now imagine that final particle 1 hits a 
detector which covers a narrow cone of polar angles but spans the entire $2\pi$ range of the azimuthal angles $\varphi'_{1}$.
Suppose also that the detector is capable of coherently interacting with the particle and projecting it onto 
eigenstates of the OAM $z$-projection operator $\psi_\ell(\bk_1') \propto \exp(i\ell\varphi'_{1})$.
Such a detector performs a different kind of post-selection, which is also a strong measurement.
Due to the transverse momentum delta function, final particle 2 is then automatically projected onto 
an eigenstate of the same OAM $z$-projection operator with eigenvalue $\ell_2' = -\ell$.
In this way, one puts the second high energy final particle into a vortex state when it was still
freely evolving, before reaching the detector.

Finally, suppose that one of the initial particles was in a vortex state with $\ell_1 \not = 0$ rather than plane wave. 
If, upon scattering, a non-local detector projects one final particle on the vortex state with $\ell_1'$,
then the other particle emerges in a vortex state with $\ell_2' = \ell_1 - \ell_1'$.
Introducing the spin degrees of freedom for the colliding particles leads to technical modifications
(the total AM rather than OAM conservation, with an additional contributions coming from helicities) but keeps the idea unchanged.  

This scheme was behind the first proposal of how to generate high-energy vortex photons via Compton backscattering
\cite{Jentschura:2010ap,Jentschura:2011ih}, see Section~\ref{subsection-Compton}. 
For ultrarelativistic initial electron, final photon angular distribution 
has a sharp backward peak with $\theta' \lsim 1/\gamma_e$.
If the final electron is projected on a plane wave state going approximately along axis $z$,
then the backscattered MeV or GeV photon approximately retains its OAM and its conical transverse momentum $\varkappa$.

\begin{figure}[h!]
	\centering
	\includegraphics[width=0.6\textwidth]{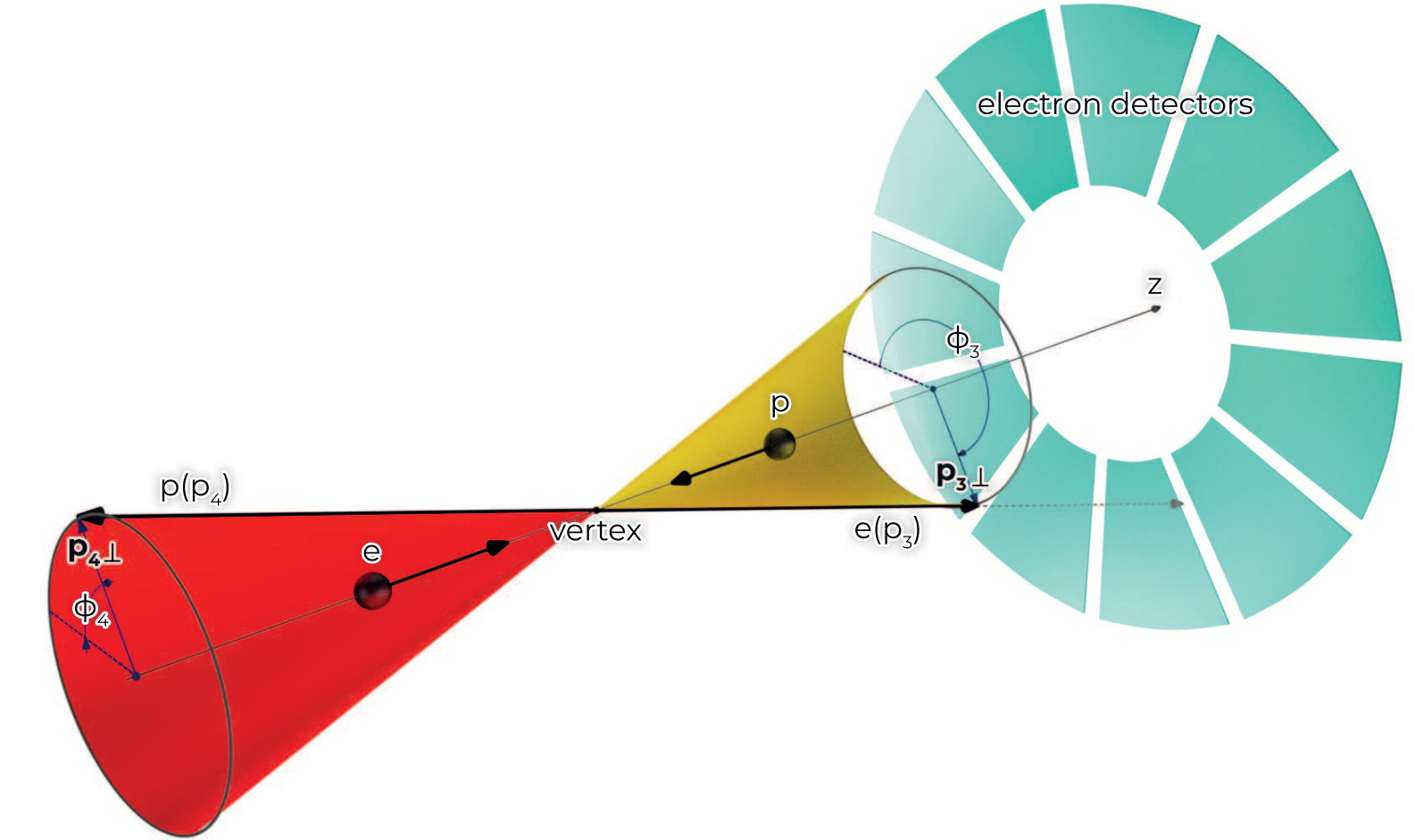}
	{\caption{\label{fig-weak-measurement} 
			An azimuthal ``which-way'' experiment with elastic $ep \to ep$ scattering. The proton becomes twisted if the electron
			azimuthal angle $\phi_3$ is measured with a large uncertainty or not measured at all. 
			Reproduced from \cite{Karlovets:2022mhb} with permission.
	}}
\end{figure}

Clearly, this scheme is not specific to Compton scattering. Entanglement in momentum space, 
which is converted to entanglement in the OAM $z$-projections of the final particles, 
is driven by momentum conservation and applies to any scattering of free-propagating beams.
In this way, it opens up opportunities to ``transfer'' a vortex state from one particle to the other
via free collision and a post-selection procedure. 
This proposal crucially relies on projection of one of the final particles onto a certain state.
Satisfying this requirement with traditional detectors is challenging or even impossible,
if projection on a definite OAM state is required.
The recent works \cite{Karlovets:2022evc,Karlovets:2022mhb} discussed different forms of postselection protocol 
which could be implemented in particle physics.
In particular, one could rely not on a strong, but on weak quantum measurement.
The authors propose not to project final particle 1 on a well-defined
plane wave nor on a state with a angular momentum but to leave its azimuthal angle $\varphi_1'$
unmeasured or, at most, measured with a significant uncertainty $\sigma_\varphi$.
This protocol should lead to final particle 2 emerging in a state with a well-controlled OAM distribution,
Fig.~\ref{fig-weak-measurement}.

Devising a post-selection protocol in high-energy scattering may seem an exotic addition.
But the authors of \cite{Karlovets:2022evc,Karlovets:2022mhb} remark
that this protocol is in fact effectively applied in scattering or emission processes
in which we cannot accurately measure the angular distribution of one of the final particles.
The specific examples considered in \cite{Karlovets:2022mhb} are Cherenkov radiation 
(the recoil of the emitting electron is not measured), non-linear Compton scattering, 
and undulator radiation. In particular, it was stated that in the semiclassical limit of radiation processes 
we do not detect nor calculate the recoil of the emitting particle, which makes the final azimuthal angle $\varphi'$
undefined and effectively leads to a weak measurement post-selection protocol.

Although specific processes for generation of vortex states via scattering
have been proposed --- such as Compton scattering on electrons \cite{Jentschura:2010ap,Jentschura:2011ih,Seipt:2014bxa,Taira:2017,Taira:2018onh,Chen:2018tkb,Chen:2019,zhang_zhao:2021}
and on relativistic partially stripped heavy ions \cite{Tanaka:2021vpc,Serbo:2021cps,Tashiro:2022qrv},
$e^+e^-$ pair production by energetic vortex photons \cite{Bu:2021ebc,Lei:2021eqe},
generic charged particle production in non-central heavy ion collisions \cite{Zou:2021pnu}, ---
a proper discussion of these processes must include a post-selection protocol.
This new ingredient of high-energy scattering analysis aimed at generating vortex states
is subtle but indispensable, as the analysis of \cite{Karlovets:2022evc,Karlovets:2022mhb} demonstrated.

	\newpage
	\section{Summary and outlook}\label{section-summary}

Vortex states, with their topologically protected coherence,
their non-zero, adjustable orbital angular momentum,
their vast freedom in preparing exotic polarization states,
represent a novel tool to probe particle structure and interactions.
The wealth of opportunities is well appreciated in the optics,
condensed matter, atomic physics, beam physics communities.
It is intriguing to see what nuclear and particle physics insights will emerge 
once vortex states of photons, electrons, and hadrons
with MeV or GeV energies become available in experiment.

In the past few years, nearly hundred of publications appeared dealing 
with possible particle physics applications of vortex states.
The objectives of the present review are not only to provide a guide 
to the literature on this emergent cross-disciplinary topic 
but also to stimulate future efforts, both experimental and theoretical.

From the experimental point of view, there is much that can be done with existing technology or
with minimal instrumentation development.
High intensity vortex light beams can already be used as a hollow driver 
for laser-plasma wakefield accelerators, forming wakefield configurations with superior properties.
Moderately relativistic vortex electrons, which are routinely obtained in electron microscopes, 
can be brought in collisions with each other or with high-energy particle beams.
Interaction of vortex neutrons with nuclei can be studied with present-day technology.
Going to higher energies, several concrete proposal for generation of keV and possibly MeV vortex photons
await experimental verification.
Proposals for MeV scale vortex electron beams can also be checked with the present day technology.
Twisting of other charged particles via electromagnetic fields
also seems possible.
Any new experimental result on vortex state preparation and scattering will represent a milestone achievement
and will point out what else needs to be done to bring vortex states to high-energy physics.
Numerical simulations of realistic vortex wave packet evolution in 
linear accelerators and storage rings are highly desirable, as they will set clear goals for instrumentation development.

Vortex state collisions can bring new physics insights in particles structure and interactions.
Novels features highlighted in this review include:
\begin{itemize}
	\item 
	modification of the spectral, angular, and polarization dependences of various processes
	driven by the cone-like kinematics of the vortex states, 
	Sections~\ref{subsection-single-Bessel},~\ref{section-processes}, and~\ref{section-experimental};
	\item
	modification of selection rules which accompany absorption or scattering of vortex particles
	by target atoms, ions, nuclei, and hadrons, Sections~\ref{subsection-atomic} and~\ref{subsection-nuclear-hadronic};
	\item
	dramatically changing the rate of metastable nuclear isomers deexcitation via capture of vortex electrons,
	Section~\ref{subsection-nuclear-hadronic};
	\item
	a new dimension in the final state kinematical distributions which allows one to probe scattering process or hadron structure via 
	interference of different plane-wave amplitudes leading to the same final state,
	Section~\ref{subsubsection-novel-kinematical};
	\item
	access to the scattering angle dependence of the overall phase of the scattering amplitude,
	Section~\ref{subsection-phase};
	\item
	tree-level access to the phase of the hadron amplitude of elastic neutron-nucleus scattering
	via Schwinger scattering of cold vortex neutrons, Section~\ref{subsection-nuclear-hadronic};
	\item
	the possibility to simultaneously produce and analyze resonances with close but different masses
	in monochromatic vortex beam collisions running at fixed energy, and to perform polarization and parity-sensitive measurements
	in fully inclusive, unpolarized collisions, Section~\ref{subsubsection-s-channel}. 
\end{itemize}
Many new theoretical developments can follow.
The formalism of vortex state collisions needs to be developed further, as different calculational
techniques are being used by different groups.
One needs to establish a reliable procedure for vortex states collision calculations
which would be convenient both for analytical results and for accurate numerical predictions.
A public code for efficient numerical computations with realistic vortex state wave packets
is also highly desirable.

With a convenient and reliable formalism at hand,
one can proceed with concrete processes, especially those involving hadrons.
Low and medium energy hadron scattering calculations should point out concrete benefits 
of vortex states with respect to approximate plane waves.
One particularly tantalizing aspect is the possibility to use single-vortex or double-vortex
electron-proton scattering to get a better understanding of the proton structure.
One can compute elastic or inelastic $ep$ scattering, or proceed directly 
to the deep inelastic scattering regime.
In vortex state scattering, new dimensions open up for the angular distributions
of the final particles, showing interference fringes and other features not available 
for the plane wave case. It is intriguing to see which particular feature
will reveal which property of the proton structure,
and what new insights the exotic polarization states of the vortex proton and electron
can bring to this field.
In the deep inelastic scattering regime, one can check whether new structure functions 
or novel unintegrated distributions are required, or whether the quantities already in use
can be accessed in a novel way.

Vortex states offer a wealth of new effects and opportunities to be discovered and explored,
and any new idea and result will pave the way to new developments.


	\section*{Acknowledgements}
	It is my pleasure to thank Ilya Ginzburg, Dmitry Karlovets, Valery Serbo, Andrey Surzhykov, and Pengming Zhang
	for many enlightening discussion on the results reported in this review. I am also grateful to Dmitry Karlovets
	and Valery Serbo for useful comments on the first version of this review.


\end{document}